\documentstyle[12pt]{article}

\author{}
\title{}
\def\permil{\%\raise.10ex\hbox{$_{\scriptstyle 0}$}}

\begin{document}

\title{Small-$x$ physics in perturbative QCD } \date{}

\author{L.N. Lipatov $^{\dagger}$\\
Petersburg Nuclear Physics\\
Institute,\\
Gatchina, 188 350, St.Petersburg, Russia}
\maketitle

\vskip 15.0pt \centerline{\bf Abstract} \noindent
We review the parton model and the Regge approach to the QCD description of
the deep-inelastic $ep$ scattering at the small Bjorken variable $x$ and
demonstrate their relation with the DGLAP and BFKL evolution equations. It
is shown, that in the leading logarithmic approximation the gluon is
reggeized and the pomeron is a compound state of two reggeized gluons. The
conformal invariance of the BFKL pomeron in the impact parameter space is
used to investigate the scattering amplitudes at high energies and fixed
momentum transfers. The remarkable properties of the Schr\"odinger equation
for compound states of an arbitrary number of reggeized gluons in the
multi-colour QCD are reviewed. The gauge-invariant effective action
describing the gluon-Reggeon interactions is constructed. The known
next-to-leading corrections to the QCD pomeron are discussed.

\vskip 3cm \hrule
\vskip 3cm \noindent
\noindent
$^{(}\dagger )$ {\it Humboldt Preistr\"ager\\ Work supported partly by INTAS
and the Russian Fund of Fundamental Investigations} \vfill

\section{Introduction}

Recent measurements of the structure functions for the deep-inelastic $ep$
scattering at HERA discovered their dramatic rise at the region of small
$x\approx 10^{-4}$ $\left[ 1\right] $. In the framework of the
Bjorken-Feynman parton model $\left[ 2\right] $ this experimental result
implies the corresponding growth of the parton distributions $n_i(x)$ inside
the rapidly moving proton as functions of the decreasing parton momentum
fraction $x$ and the increasing photon virtuality $Q^2$. In the framework of
the Dokshitzer-Gribov-Lipatov-Altarelli-Parisi (DGLAP) equation $\left[
3,4,5,6\right] $ the parton distributions grow at small $x$ as a result of
their $Q^2$-evolution. In the framework of the Balitsky-Fadin-Kuraev-Lipatov
(BFKL) equation $\left[ 7\right] $ this growth is a consequence of their $x$
-evolution. Within the double-logarithmic accuracy these equations coincide
and the increase of the structure functions at small $x$ is related with the
singularities of the anomalous dimensions for the corresponding twist-2
operators at non-physical values $j\rightarrow 1$ of the Lorentz spin $%
\left[ 2,8\right] $. The existing experimental data on structure functions
agree with the DGLAP dynamics provided that the evolution equation in $Q^2$
is applied starting from rather small $Q^2=Q_0^2$ $\left[ 9\right] $. The
growth of the structure functions at small $x$ can be also obtained with the
use of the BFKL equation $\left[ 10\right] $. In this case a large
uncertainty is related with the fact, that all next-to-leading corrections
to this equation have not been calculated yet contrary to the case of the
DGLAP equation where they are well known.

The additional information on the dynamics of the deep-inelastic scattering
at small $x$ is extracted from the study of the final state particles. An
especially clean footprint of the BFKL pomeron can be found in the processes
with the inclusive production of jets $\left[ 11\right] $. The quark-gluon
structure of the pomeron can be investigated at the hard diffractive
scattering when the hadrons are produced in the virtual photon fragmentation
region $\left[ 12\right] $. The deep-inelastic process with a large rapidity
gap for the final particle momenta was discovered at HERA $\left[ 13\right] $
. The various theoretical models for its interpretation were suggested $%
\left[ 14\right] $. The other high energy processes with the large rapidity
gaps were widely discussed to discriminate the dynamics related with the soft
and hard pomerons $\left[ 15\right] $.

In this review we consider the theory of the BFKL pomeron. Because this
theory is related closely with the parton description of the deep-inelastic
scattering in QCD and with the Regge model, we remind below the basic ideas
of these two traditional approaches to the high energy physics
(a more 
comprehensive information can be
found in Refs $\left[ 16\right] $, $\left[ 17\right] $). In the next section
the basic properties of the solution of the BFKL equation are discussed in
the framework of the impact parameter representation . The gluodynamics is
known to be a low energy limit of the super-string model of elementary
particles which includes the quantum gravity. On the contrary one can
expect that QCD at high energies could be described in terms of an effective
field theory for string-like objects. In the third section it will be
demonstrated, that in the Regge limit of large energies $\sqrt{s}$ and fixed
momentum transfers $\sqrt{-t}$ the gluon having the spin $j=1$ at $t=0$ lies
on the Regge trajectory $j=j(t)$. Such reggeization property was assumed to
be typical for hadrons which are extended objects. We consider here a
simple effective field model in which the Feynman vertices coincide with the
QCD reggeon-particle couplings. It is shown, that in the leading logarithmic
approximation the pomeron is a compound state of two reggeized gluons.
However, for restoring the $S$-matrix unitarity one should consider the
contribution of the diagrams with an arbitrary number of the reggeized
gluons in the $t$-channel. In the end of third section it will be
demonstrated, that the equations for compound states of several reggeized
gluons in the multi-colour QCD have remarkable properties: the conformal
symmetry, the holomorphic factorization of their eigen functions and the
existence of non-trivial integrals of motion in holomorphic and
anti-holomorphic subspaces. The corresponding Hamiltonian turns out to be
equivalent to the local Hamiltonian of the exactly solvable Heisenberg model
with the spins being the generators of the conformal (M\"obius) group. At
high energies it is natural to reformulate QCD as an effective field theory
for reggeized gluons. In the fourth section the gauge-invariant effective
action for the interactions between the reggeized and usual
gluons is constructed. The main results in the problem of
finding next to leading corrections to the BFKL equation are reviewed in the
fifth section. In Conclusion some unsolved problems are discussed.

\subsection{Parton model in QCD}

The deep-inelastic $ep$ scattering at large electron momentum transfers $%
q=p_e-p_{e^{\prime }}$ and a fixed Bjorken variable $x=\frac{Q^2}{2pq}$ ($%
Q^2=-q^2$) is a well investigated process for which the perturbative quantum
chromodynamics (QCD) was traditionally and successfully applied. Before the
QCD discovery the approximate scaling behaviour of the structure functions $%
W_{1,2}(x,Q^2)$ for the virtual photon-proton scattering at large $Q^2$ was
derived in the framework of the Bjorken-Feynman parton model $\left[
2\right] $. In this model the transverse momenta $k_i^{\perp }$ of partons
inside the moving proton are assumed to be independent of $Q^2$, the
cross-section $\sigma _L$ for the longitudinally polarized virtual photon is
zero and the cross-section $\sigma _T$ for the transversally polarized
photon is expressed in the impulse approximation as a sum of 
photon-quark cross-sections averaged with the distributions of quarks $%
n_q(x) $ and anti-quarks $n_{\overline{q}}(x)$ in the proton [1]:
\begin{equation}
\sigma _T=\sum_qe_q^2\frac{4\pi ^2\alpha }{Q^2}\,x\,(n_q(x)+n_{\overline{q}
}(x))\,,
\end{equation}
where $e_q$ is the quark charge measured in the units of the electron charge
$e$ and $\alpha =e^2/4\pi $. As a consequence of the charge and energy
conservation the following sum rules for the parton distributions $n_i(x)$
inside the proton
\begin{equation}
1=\sum_qe_q\int_0^1dx\,(n_q(x)-n_{\overline{q}}(x))\,,\,\,1=\sum_i\int_0^1dx
\,x\,n_i(x)
\end{equation}
are valid.

In the renormalizable field theories the Bjorken scaling for structure 
functions is violated due to
the logarithmic terms $(g^2\,\ln \,Q^2)^n$ appearing in the perturbation
theory . The leading logarithmic approximation (LLA) for the structure
functions corresponds to the sum of all such contributions [3]. The physical
reason of the scaling violation is that because of divergencies one should
introduce an ultraviolet cut-off $\Lambda $ for the integrals over the
transverse momenta $k_r^{\perp }$ of partons being quanta of bare fields
[4]:
\begin{equation}
(\overrightarrow{k_r^{\perp }})^2\,<\,\Lambda ^2\,.
\end{equation}
Nevertheless, the parton model representation (1) for the $\gamma^* p$
cross-section in terms of the quark distributions $n_q(x),\,n_{\overline{q}
}(x)$ remains to be valid in LLA [4] if we identify the ultraviolet cut-off
and the photon virtuality
\begin{equation}
\Lambda ^2\,=\,Q^2\,.
\end{equation}
Moreover, one can express the inclusive probabilities $n_i(x)$ for finding
the parton $i$ with its momentum fraction $x=\left| \overrightarrow{k}
\right| /\left| \overrightarrow{p}\right| $ inside the proton having the big
momentum $\overrightarrow{p}\rightarrow \infty$ through the partonic wave
functions $\Psi _n(k_1^{\perp },x_1;k_2^{\perp },x_2;...;k_n^{\perp },x_n)$
as follows
\begin{equation}
n_i(x)=\sum_n\prod_k\frac 1{n_k!}\int \prod_{r=1}^n\frac{d^2k_r^{\perp }}{
(2\pi )^3}\,\frac{d\,x_r}{x_r}\,\,|\Psi _n|^2\,\delta ^2(\sum_rk_r^{\perp
})\,\delta (1-\sum_rx_r)\sum_{r_i}\delta (x_{r_i}-x_i)\,.
\end{equation}
Here the index $r_i$ enumerates $n_i$ partons of the type $i$ in the state
with $n=\sum_kn_k$ partons.

In expression (5) and in the normalization condition for $\Psi _n$:

\begin{equation}
1=\sum_n\prod_i\frac 1{n_i!}\int \prod_{r=1}^n\frac{d^2k_r^{\perp }}{(2\pi
)^3}\,\frac{d\,x_r}{x_r}\,\,|\Psi _n|^2(2\pi )^3\,\delta ^2(\sum_rk_r^{\perp
})\,\delta (1-\sum_rx_r)
\end{equation}
the most essential integral contributions at large $\Lambda ^2=Q^2$ in LLA
correspond to the branching processes in which each virtual particle can
decay into two others having significantly bigger transverse momenta. The
interference terms between the decay amplitudes describing the different
branches are negligible in the physical light-cone gauge: 
\begin{equation}
n_\mu v_\mu =0\,,\,\,n_\mu ^2=0 
\end{equation}
for the gluon field $v_\mu (x)$ and therefore we can use the probabilistic
picture to calculate various characteristics of the partonic cascade [4].
From the renormalizability of the theory the $\Lambda $-dependence of $%
\left| \Psi _n\right| ^2$ is known: 
\begin{equation}
\left| \Psi _n\right| ^2\sim \prod_i\,(Z_i)^{n_i}\,\,,\,\,\sum_in_i=n\,\,, 
\end{equation}
where $Z_i$ are the renormalization constants for the corresponding parton
wave functions. Using also the above probabilistic picture for essential
contributions we obtain the following equation by differentiating the
normalization condition (6) for the wave function $\Psi _n$ [4]: 
\begin{equation}
0=\sum_iN_i\left( Z_{i\,}^{-1}\frac \partial {\partial \ln
\,Q^2}Z_i+P_i\right) \,, 
\end{equation}
where 
\begin{equation}
N_i=\int_0^1d\,x\,n_i(x) 
\end{equation}
is the averaged number of partons $i$ in the proton. The quantity $%
P_i\,\,d\,\ln \,Q^2$ is the total probability of the parton decay during the
''time'' interval $d\,\ln \,Q^2$ and it depends on the effective charge 
$g(Q^2)$. The
normalization condition (6) should be valid for the wave function $\Psi $ of
each hadron, which due to relation (9) is compatible with the
Callan-Symanzik equation [18] for the renormalization constants:

\begin{equation}
\frac \partial {\partial \,\ln \,Q^2}\,Z_i=-P_i\,Z_i\,. 
\end{equation}

Analogously by differentiating the partonic expression (5) for $n_i(x)$ one
can derive [4] the DGLAP equation [3-6]: 
\begin{equation}
\frac \partial {\partial lnQ^2}\,n_i(x)\,=\,-P_i\,n_i(x)+\sum_k\int_x^1\frac{
dx^{\prime }}{x^{^{\prime }}}\,P_{k\rightarrow i}(\frac x{x^{\prime
}})\,n_k(x^{\prime })\,,\,\,P_i=\sum_r\int_0^1d\,y\,P_{i\rightarrow r}(y) 
\end{equation}
governing the $Q^2$-dependence of the parton distributions $n_i(x)$. In the
right-hand side of this equation the first term describes the decrease of
the number of partons of the type $i$ as a result of their decay to other
partons. The second term corresponds to the increase of $n_i(x)$ due to the
decay of other partons into the states containing the partons of the type $i$
. The decay probabilities $P_{k\rightarrow i}(x/x^{\prime })$ and $%
P_k=\sum_i\int_0^1d\,y\,P_{k\rightarrow i}(y)\,$ are calculated in the form
of the perturbative expansion over the running QCD coupling constant $\alpha
(Q^2)$ at large $Q^2$

\begin{equation}
\alpha (Q^2)=\frac{g^2(Q^2)}{4\pi }=\frac{4\pi }{\beta _2\ln \,\frac{Q^2}{
\Lambda _{QCD}^2}}\,,\,\,\beta _2=\frac{11}3N_c-\frac 23n_f\,. 
\end{equation}
Here $\Lambda _{QCD}\approx 10^2$ Mev is the fundamental QCD constant, $%
N_c=3 $ is the rank of the gauge group $SU(N_c)$ for QCD and $n_f$ is the
number of the quarks with masses smaller than $\sqrt{Q^2}$. For example, in
the lowest order corresponding to LLA one obtains [3-6]:

$$
P_{k\rightarrow i}(y)=\frac{\alpha (Q^2)}{4\pi }\,w_k^i(y)\,, 
$$

$$
w_{1/2}^{1/2}(y)=\frac{N_c^2-1}{N_c}\,\frac{1+y^2}{1-y}\,,\,w_{1/2}^{-1/2}=0
\,,\,\,w_{1/2}^1(y)=\frac{N_c^2-1}{N_c}\,\frac 1{y\,}\,,\,w_{1/2}^{-1}(y)= 
\frac{N_c^2-1}{N_c}\,\frac{(1-y)^2}y\,, 
$$
$$
w_1^{1/2}(y)=w_1^{\overline{1/2}}(y)=w_1^{-1/2}(1-y)=y^2\,,\,w_1^1(y)=2N_c 
\frac{1+y^4}{y(1-y)}\,,\,w_1^{-1}(y)=2N_c\frac{(1-y)^3}y\,, 
$$
$$
w_q=w_{\overline{q}}=\frac{N_c^2-1}{N_c}\,\int_0^1d\,y\,\frac{1+y^2}{1-y}
\,,\, 
$$
\begin{equation}
w_g=\int_0^1d\,y\,\left[ y^2+(1-y)^2+2\,N_c\,\frac{1+y^4+(1-y)^4}{1-y}
\right] \,. 
\end{equation}
where $w_{\lambda _k}^{\lambda _i}(y)$ are proportional to the elementary
inclusive probabilities for the transitions $k\rightarrow i$ between the
partons with definite helicities $\lambda _k,\,\lambda _i$ ($\lambda =\pm 1$
for gluons and $\lambda =\pm 1/2$ for quarks; $\overline{\lambda }$ denotes
the helicity $\lambda $ for the anti-quark). Other inclusive probabilities
can be obtained with the use of the parity and charge conservation. In 
evolution equation (12) the infrared divergency at $y=\frac \beta {\beta
^{\prime }}\rightarrow 1$ is cancelled. The dependence from interactions in
the confinement region enters in $n_i(x)$ only through the initial condition
at some $Q^2=Q_0^2$ assumed to be sufficiently large: $Q_0^2\gg \Lambda
_{QCD}^2$. One can apply the evolution equation also for distributions
of quarks and gluons inside a virtual parton of the type $k$ substituting $%
n_i(x)\rightarrow D_k^i(x)$ and assuming, that 
\begin{equation}
D_k^i(x)=\delta _{ki}\,\delta (x-1) 
\end{equation}
for $Q^2=Q_0^2$. The parton distributions in the proton can be expressed
through $D_k^i$ using the representation: 
\begin{equation}
n_i(x)=\sum_k\int_0^1d\beta \,f_k(\beta )\,D_k^i(\frac x\beta )\,, 
\end{equation}
where the initial parton densities $f_k(\beta )$ are not calculated in the
framework of the perturbation theory. The solution of the DGLAP equation for 
$D_k^i(x)$ can be written in the following form: 
\begin{equation}
D_k^i(x)=\int_{\sigma -i\infty }^{\sigma +i\infty }\frac{d\,j}{2\pi i}
\,\left( \frac 1x\right) ^j\sum_rM_k^r(j)M_i^r(j)\,\exp \left( \xi \,\gamma
_r(j)\right) \,, 
\end{equation}
where in LLA 
\begin{equation}
\xi =\int_{Q_0^2}^{Q^2}\frac{d\,q^2}{q^2}\,\frac{\alpha (q^2)}{4\pi }=\frac
1{\beta _2}\,\ln \left( \frac{\alpha (Q_0^2)}{\alpha (Q^2)}\right)
\,,\,\,\,\alpha (Q^2)=\frac{4\pi }{\beta _2\,\ln (Q^2/\Lambda _{QCD}^2)}
\,\,. 
\end{equation}
The integration contour $L=(\sigma -i\infty \,,\,\sigma +i\infty )$ in the
complex plane of the $t$-channel angular momentum $j$ is situated to the
right of the point $j=1$. The anomalous dimensions $\gamma =\gamma _r(j)$
and couplings $M_i=M_i^r(j)$ are determined from the secular equation: 
\begin{equation}
\gamma \,M_i=\gamma _k^i(j)\,M_k\,, 
\end{equation}
where the matrix $\widehat{\gamma }(j)$ is defined as follows:

\begin{equation}
\gamma _k^i(j)=\int_0^1d\,y\,\,\left[ y^{j-1}\,\,w_k^i(y)-\delta
_{ki}\sum_rw_k^r(y)\right] 
\end{equation}
and its matrix elements are given below $\left[ 19\right] $:

$$
\gamma _{1/2}^{1/2}(j)=\frac{N_c^2-1}{N_c}\,\left[ \frac 32+\frac
1{j(j+1)}-2S_j\right] ,\,\,\,\,\gamma _{1/2}^{-1}(j)=\frac{N_c^2-1}{N_c}
\frac 2{j(j^2-1)}, 
$$
$$
\gamma _{1/2}^1(j)=\frac{N_c^2-1}{N_c}\frac 1{j-1}\,,\,\,\,\,\gamma
_1^{1/2}(j)=\frac 1{j+2}\,,\,\,\,\,\gamma _1^{-1/2}(j)=\frac
2{j(j+1)(j+2)}\,, 
$$
$$
\gamma _1^1(j)=2N_c\left[ \frac{11}6+\frac{4j^2+4j-2}{(j-1)j(j+1)(j+2)}
-2S_j\right] -\frac 23n_f\,, 
$$
\begin{equation}
\,\gamma _1^{-1}(j)=2N_c\,\frac 6{(j-1)j(j+1)(j+2)}\,. 
\end{equation}
where 
\begin{equation}
S_j=\psi (j+1)-\psi (1)\,,\,\,\,\psi (x)=\Gamma ^{^{\prime }}(x)/\Gamma (x) 
\end{equation}
and $\Gamma (x)$ is the Euler $\Gamma $-function. Due to the charge and
energy conservation the following sum rules are valid: 
$$
\gamma _{1/2}^{1/2}(1)=0\,,\,\,\gamma _{1/2}^{1/2}(2)+\gamma
_{1/2}^1(2)+\gamma _{1/2}^{-1}(2)=0\,, 
$$
\begin{equation}
\gamma _1^1(2)+\gamma _1^{-1}(2)+2\,n_f\left( \gamma _1^{1/2}(2)+\gamma
_1^{-1/2}(2)\right) =0\,. 
\end{equation}

Note, that for the $e^{+}e^{-}$ annihilation the inclusive cross-section of
the hadron production can be written in the framework of the parton model in
terms of the fragmentation function $\widetilde{D}_i^h(x)$ which is the
inclusive probability to find the hadron $h$ with the energy fraction $x$
inside the parton $i$. These fragmentation functions are expressed through
the distributions $\widetilde{D}_i^k(x)$ of dressed partons $k$ inside the
bare partons $i$. In the leading logarithmic approximation the functions $%
D_i^k(x)$ and $\widetilde{D}_i^k(x)$ coincide due to the Gribov-Lipatov
relation $\left[ 3\right] $:

\begin{equation}
\widetilde{D}_i^k(x)=D_i^k(x)\,. 
\end{equation}
Using the arguments based on the crossing symmetry and analyticity one can
expect that these two functions are related also by the Drell-Levy-Yan
relation

\begin{equation}
\widetilde{D}_i^k(x)=\pm xD_k^i(\frac 1x)\,, 
\end{equation}
where the sign in its right hand side depend on the parton types $i$ and $k$
. In the gauge theory the point $x=1$ is singular due to the Sudakov
suppression of the quasi-elastic scattering. Nevertheless, this relation is
fulfilled in a generalized form if in each order of the perturbation theory
one would continue around $x=1$ in $D(x)$ the polynomials in $\ln (1-x)$ as
the polynomials in $\ln (x-1)$ $\left[ 4,20\right] $.

As it is seen from above formulas, for $j\rightarrow 1$ the anomalous
dimensions $\gamma _q^g(j)=\gamma _{1/2}^1(j)+\gamma _{1/2}^{-1}(j)$ and $%
\gamma _g^g(j)=\gamma _1^1(j)+\gamma _1^{-1}(j)$ have the pole
singularities: 
\begin{equation}
\gamma _q^g(j)\rightarrow 2\,\frac{N_c^2-1}{N_c}\,\frac 1{j-1}\,,\,\,\gamma
_g^g(j)\rightarrow 4\,N_c\,\frac 1{j-1}\,. 
\end{equation}
Therefore for $x\rightarrow 0$ in the integrand of expression (17) for the
distributions of gluons $n_g(x)$ and quarks $n_g(x)$ one obtains a saddle
point at 
\begin{equation}
j-1=2\,\sqrt{\frac{N_c\,\xi }{\ln \,\frac 1x}}\,, 
\end{equation}
which leads after calculating the integral over $j$ to the rapid growth of
the total cross-section for the $\gamma ^{*}p$-scattering $\left[ 3,8\right] 
$: 
\begin{equation}
\sigma _T\sim \frac 1{Q^2}\,\,\exp \,\sqrt{16\,N_c\,\xi \,\ln \,\frac 1x}\,. 
\end{equation}
In next sections we shall show, that for sufficiently small $x$ this
asymptotics should be modified as follows 
\begin{equation}
\sigma _T\sim \frac 1{\sqrt{Q^2}}\,(\frac 1x)^{\omega _{BFKL}}\,,\,\,\omega
_{BFKL}=\frac{4\alpha }\pi \,N_c\,\ln \,2\, 
\end{equation}
in accordance with the BFKL equation [7]: 
\begin{equation}
\frac \partial {\partial \ln \frac 1x}\,n_g(x,k_{\perp })=2\,\omega
(k_{\perp }^2)\,n_g(x,k_{\perp })+\int d^2k_{\perp }^{\prime }\,K(k_{\perp
},k_{\perp }^{\prime })\,n_g(x,k_{\perp }^{\prime }) 
\end{equation}
written for more general distributions $n_g(x,k_{\perp })$ depending on the
longitudinal ($x$) and transverse ($k_{\perp }$) gluon momenta.

The next to leading corrections to the anomalous dimension matrix and to the
splitting kernels of the DGLAP evolution equation in QCD were calculated in
Refs $\left[ 21\right] $ and $\left[ 22\right] $. It turns out $\left[
22\right] $, that with taking into account these corrections the
Gribov-Lipatov relation is violated.

Another approach to the deep-inelastic $ep$ scattering is based on the
Wilson operator product expansion of two electromagnetic currents $\left[
19\right] $.\thinspace In this approach the $\gamma ^{*}p$ scattering
amplitude is presented at large $Q^2$ as a series over $1/x$ with the
coefficients which are the moments of the structure functions. These moments
can be expressed as a result of the operator product expansion through the
product of the coefficient functions and the matrix elements of local
operators $O^i$ between the hadron states. The coefficient functions depend
on $Q^2$ and on an intermediate parameter $\mu $ being the normalization
point for the running coupling constant $\alpha (\mu ^2)$. The matrix
elements of $O^i$ also depend on the normalization point $\mu $ playing role
of an ultraviolet cut-off in the corresponding Feynman integrals. In the
product of the coefficient functions and the matrix elements the dependence
from $\mu ^2$ is cancelled. In LLA we can chose $\mu ^2=Q^2$. Then
one should calculate the coefficient functions only in the  lowest
order of perturbation theory. 
In this case for the polarized particles only the operators 
$O_q^{V\,\alpha _1\alpha _2...\alpha _j}$ and $\,O_q^{A\,\alpha _1\alpha
_2...\alpha _j}$ having the Lorentz spin $j$ and constructed from the quark
fields $\psi (x)$ appear in the Wilson operator product expansion:

\begin{equation}
O_q^{I\,\alpha _1\alpha _2...\alpha _j}=\widehat{Sym}\,\,\overline{\psi }
(x)\gamma _I^{\alpha _1}D^{\alpha _2}D^{\alpha _3}...D^{\alpha _j}\psi
(x)\,,\,\,\,\gamma _V^\alpha =\gamma ^\alpha \,,\,\,\gamma _A^\alpha =\gamma
^\alpha \gamma _5\,, 
\end{equation}
where $\widehat{Sym}$ means the symmetrization over the indices $\alpha
_1,\alpha _2...\alpha _j$ and the subtraction of traces. The covariant
derivative $D_\alpha $ equals 
\begin{equation}
D_\alpha =\frac \partial {\partial x^\alpha }+g\,v_\alpha (x) 
\end{equation}
where the anti-hermitian matrix $v_\alpha $ is expressed in terms of the
gluon fields $v_\alpha ^a$ belonging to an adjoint representation of the
gauge group $SU(N_c)$ 
\begin{equation}
v_\alpha (x)=t^a\,v_\alpha ^a\,,\,\,\left[ t^a,t^b\right] =f^{abc}t^c. 
\end{equation}
Here $t^a$ are anti-hermitian Gell-Mann matrices and $f^{abc}$ are the
structure functions of the gauge group. The commutator of the covariant
derivatives is proportional to the strength $G_{\alpha \beta }^a$ of the
gluon field: 
\begin{equation}
\left[ D_\alpha ,D_\beta \right] =g\,G_{\alpha \beta }\,,\,\,G_{\alpha \beta
}=t^aG_{\alpha \beta }^a=\partial _\alpha v_\beta -\partial _\beta v_\alpha
+g\left[ v_\alpha ,v_\beta \right] . 
\end{equation}

Operators (31) constructed from the quark fields have the canonical
dimension $d_q=j+2$ and therefore their twist $t=d-j$ equals $t_q=2$. There
are other twist-2 operators $O_G^V$ and $O_G^A$ constructed from the gluon
fields: 
\begin{equation}
O_G^{I\,\alpha _1\alpha _2...\alpha _j}=\widehat{Sym}\,\,tr\,G^{\alpha
_1\sigma }D^{\alpha _2}D^{\alpha _3}...D^{\alpha _{j-1}}G_I^{\alpha _j\sigma
}\,,\,\,G_V^{\alpha _1\alpha _2}=G^{\alpha _1\alpha _2}\,,\,\,G_A^{\alpha
_1\alpha _2}=\epsilon ^{\alpha _1\alpha _2\alpha _3\alpha _4}G_{\alpha
_3\alpha _4}\,, 
\end{equation}
where $tr$ implies the trace for the subsequent colour matrix. Under the
renormalization the quark and gluon operators mix each with another and 
only
certain linear combinations of them are multiplicatively renormalizable and
diagonalize the evolution equations. It is convenient to use the light-cone
gauge for the gluon field $v$: 
\begin{equation}
v_{.}\equiv v_\alpha n_\alpha =0,\,\,n=\frac{q+xp}{2pq}\,,\,\,n_\alpha ^2=0. 
\end{equation}
In this gauge the light-cone components of the operators $O^{\alpha _1\alpha
_2...\alpha _j}$ appearing in the operator product expansion are the
simplest examples of the quasi-partonic operators $\left[ 23\right] $ which
have two properties: a) they are universal for all possible interactions,
which means in particular that the quasi-partonic operators do not depend on
the QCD coupling constant $g$ ; b) their matrix element between hadronic
states are expressed through the integrals from their partonic matrix
elements on the mass shell averaged with the partonic correlators.

Let us consider the twist-2 operators for a general case, when their matrix
element is calculated between the initial and final hadron states with
different longitudinal momenta [24]. Then we can construct non-equivalent
operators as bilinear combinations of fields with a different number of
derivatives: 
$$
O_{q\underbrace{....}}^I(x)\equiv n_{.}^{\alpha _1}n_{.}^{\alpha
_2}...n_{.}^{\alpha _j}O_q^{I\,\alpha _1\alpha _2...\alpha
_j}=\sum_{k=0}^{j-1}c_k\,\overline{\psi }(i\overleftarrow{\partial }%
_{.})^k\gamma _{.}^I(i\partial _{.})^{j-k-1}\psi \,,\,\,\gamma _\alpha
^V=\gamma _\alpha \,,\,\,\gamma _\alpha ^A=\gamma _\alpha \gamma _5\,, 
$$
\begin{equation}
O_{G\underbrace{....}}^I=\sum_{k=1}^{j-1}d_k\,v_\sigma ^{a\,\perp }(i 
\overleftarrow{\partial }_{.})^k(i\partial _{.})^{j-k}\Omega _I^{\sigma
\sigma ^{^{\prime }}}v_{\sigma ^{\prime }}^{a\,\perp }\,,\,\,\Omega
_V^{\sigma \sigma ^{\prime }}=\delta _{\perp }^{\sigma \sigma ^{\prime
}}\,,\,\,\Omega _A^{\sigma \sigma ^{\prime }}=\epsilon _{\perp }^{\sigma
\sigma ^{\prime }} 
\end{equation}
where $\overleftarrow{\partial }$ means, that the derivative acts at the
function situated in the left hand side from the differential operator. The
coefficients $c_r,d_r$ can be fixed up to a common factor: 
\begin{equation}
c_k=\frac{(-1)^k}{k!(k+1)!(j-k-1)!(j-k)!}\,,\,\,d_k=\frac{(-1)^k}{%
(k-1)!(k+1)!(j-k+1)!(j-k-1)!}\,. 
\end{equation}
from the requirement, that the operators $O_q$ and $O_G$ are irreducible
under the conformal transformations $\left[ 25\right] $. It is enough to
verify their covariant properties under the inversion $x_\alpha \rightarrow
x_\alpha ^{\prime }=x_\alpha /x^2$ with the simultaneous substitution of the
fields:

\begin{equation}
\psi (x)\rightarrow \psi ^{\prime }(x^{\prime })=x^2\widehat{x}\,\psi
(x)\,,\,\,v_\mu (x)\rightarrow v_\mu ^{\prime }(x^{\prime })=x^2\Lambda
_{\mu \nu }(x)\,v_\nu (x)\,, 
\end{equation}
where 
\begin{equation}
\Lambda _{\mu \nu }(x)=\delta _{\mu \nu }-2\,\frac{x_\mu x_\nu }{x^2}\,. 
\end{equation}
The free Green functions for the fields $\psi (x)$ and $F_{\alpha \beta
}(x)=\partial _\alpha v_\beta (x)-\partial _\beta v_\alpha (x)$:

$$
\langle 0\left| T\psi (x)\overline{\psi }(y)\right| 0\rangle \sim \frac{ 
\widehat{x}-\widehat{y}}{(x-y)^4}\,, 
$$
\begin{equation}
\langle 0\left| TF_{\mu \nu }(x)F_{\alpha \beta }(y)\right| 0\rangle \sim 
\frac{\Lambda _{\mu \alpha }(x-y)\Lambda _{\nu \beta }(x-y)-\Lambda _{\mu
\beta }(x-y)\Lambda _{\nu \alpha }(x-y)}{(x-y)^4}\,. 
\end{equation}
are transformed properly under the inversion. Analogously the matrix
elements of the twist-2 operators (37) with coefficients (38) can be found
in the free theory:

$$
\langle 0\left| T\psi (x)O_{q\underbrace{....}}^I(z)\overline{\psi }
(y)\right| 0\rangle \sim \left( \frac{x_{.}-z_{.}}{(x-z)^2}-\frac{
y_{.}-z_{.} }{(y-z)^2}\right) ^{j-1}\frac{\widehat{x}-\widehat{z}}{(x-z)^4}
\,\gamma _{.}^I\,\frac{\widehat{z}-\widehat{y}}{(z-y)^4}\,, 
$$
\begin{equation}
\langle 0\left| TF_{\mu \,.}^{\perp }(x)O_{G\underbrace{....}}^I(z)F_{\nu
\,.}^{\perp }(y)\right| 0\rangle \sim \left( \frac{x_{.}-z_{.}}{(x-z)^2}
\,-\, \frac{y_{.}-z_{.}}{(y-z)^2}\right) ^{j-2}\,\delta _{\mu \nu }^{\perp
}\, \frac{(x_{.}-z_{.})^2}{(x-z)^6}\,\frac{(y_{.}-z_{.})^2}{(y-z)^6} 
\end{equation}
and are transformed also correctly $\left[ 25\right] $.

Conformally-covariant operators have simple multiplicative properties under
the renormalization in LLA in accordance with the fact, that their matrix
elements between the parton states in the momentum representation are 
proportional to the Gegenbauer
polynomials which diagonalize partly the Brodsky-Lepage evolution equations
generalizing the DGLAP equations for a non-zero momentum transfer $\left[
24\right] $. These equations are invariant under a sub-group of the
conformal group which includes the transformation

\begin{equation}
x_\alpha \rightarrow \frac{x_\alpha +\epsilon n_\alpha x^2}{1+2\,\epsilon
\,x_\sigma n_\sigma }\,\,,\,\,\,n_\sigma ^2=0 
\end{equation}
where $n_\sigma $ is the above light-cone vector and $\epsilon $ is the
group parameter. One can verify, that the gauge condition $v_\sigma n_\sigma
=0$ is compatible with this conformal subgroup.

For the simplest supersymmetric Yang-Mills model containing apart from the
gluon also a gluino being the Majorana particle belonging to the adjoint
representation of the gauge group the anomalous dimensions can be easily
obtained from the above QCD expressions. If we introduce the quantities

\begin{equation}
\gamma _i^{k\pm }(j)=\gamma _{\lambda _i}^{\lambda _k}(j)\pm \gamma
_{\lambda _i}^{\lambda _k}(j) 
\end{equation}
related to the anomalous dimensions of the operators $O^V$ and $O^A$ the
corresponding matrices can be parametrized by two functions $\gamma _1(j)$
and $\gamma _2(j)$:

$$
\left( 
\begin{array}{cc}
\gamma _q^{q+}(j) & \gamma _g^{q+}(j) \\ 
\gamma _q^{g+}(j) & \gamma _g^{g+}(j) 
\end{array}
\right) =\frac{\gamma _1(j)}{2j+1}\,\left( 
\begin{array}{cc}
j-1 & j-1 \\ 
j+2 & j+2 
\end{array}
\right) +\frac{\gamma _2(j+1)}{2j+1}\,\left( 
\begin{array}{cc}
j+2 & -j+1 \\ 
-j-2 & j-1 
\end{array}
\right) \,, 
$$

\begin{equation}
\left( 
\begin{array}{cc}
\gamma _q^{q-}(j) & \gamma _g^{q-}(j) \\ 
\gamma _q^{g-}(j) & \gamma _g^{g-}(j) 
\end{array}
\right) =\frac{\gamma _2(j)}{2j+1}\,\left( 
\begin{array}{cc}
j-1 & j-1 \\ 
j+2 & j+2 
\end{array}
\right) +\frac{\gamma _1(j+1)}{2j+1}\,\left( 
\begin{array}{cc}
j+2 & -j+1 \\ 
-j-2 & j-1 
\end{array}
\right) \,. 
\end{equation}

This parametrization is a consequence of the fact $\left[ 20\right] $, that
there are four irreducible representations of the super-conformal group for
the corresponding twist-2 operators with the degenerate anomalous dimensions
proportional to $\gamma _1$ and $\gamma _2$: 
\begin{equation}
\gamma _1(j)\,=\,N_c\left[ 2\,S_{j-2}-\frac 32+\frac 3j\right] \,,\,\,\gamma
_2(j)\,=\,N_c\left[ 2\,S_{j+1}-\frac 32\,-\frac 3j\right] \,. 
\end{equation}

In particular from the above formulas we obtain the Dokshitzer relation [6]
for the anomalous dimension matrix elements 
\begin{equation}
\gamma _q^q(j)+\gamma _q^g(j)=\gamma _g^q(j)+\gamma _g^g(j) 
\end{equation}
corresponding to a particular case of the Shmushkevich rule stating, that
the probability of a process induced by a particle belonging to a group
representation is equal for any member of this representation provided that
all final states related by the symmetry transformation are taken into
account. Because $\gamma _q^q(j),\,\gamma _q^g(j)$ and $\gamma _g^q(j)$
coincide in QED $\left[ 4\right] $ and in QCD $\left[ 5,6\right] $ up to
simple colour factors, from the Dokshitzer relation (47) we conclude, that
the pure Yang-Mills anomalous dimension $\gamma _g^g(j)$ (and the
corresponding splitting kernel $w_g^g(x)$) could be obtained from the
QED results.

In the next-to-leading approximation the conformal invariance is violated
due to the conformal anomaly related with the non-vanishing $\beta $
-function $\left[ 26\right] $. Nevertheless the solution of the
Brodsky-Lepage equations can be found if one would take into account the
Ward identities of the conformal group $\left[ 27\right] $.

The quasi-partonic operators $\left[ 23\right] $ of an arbitrary twist seem
to be a natural generalization of the above twist-2 operators related with
the probabilistic picture. Their matrix elements can be presented only in
terms of the parton correlators being the integrals
from the product of initial and final state wave functions with a different
number of partons. They do not mix with other operators under the
renormalization. The evolution equations for the quasi-partonic operators
have the form of the Schr\"odinger equations with a pair-wise interaction.
The pair Hamiltonians are proportional to the splitting kernels for twist-2
operators with generally non-zero quantum numbers including colours,
flavors, spins and momentum transfers. All these kernels are calculated in
LLA $\left[ 23\right] $. They have a number of remarkable properties
including their conformal invariance. The evolution equations are simplified
at the small-$x$ region, which gives a possibility to find their solutions
in the double-logarithmic approximation $\left[ 28\right] $.

The higher twist contributions are important for finding screening 
corrections
responsible for restoring the $S$-matrix unitarity. This problem will be
discussed in this review later in more details. Here we want to remark only,
that one can calculate correlators of parton densities within a certain
accuracy using the probabilistic picture $\left[ 29\right] $. Let us
introduce the exclusive multi-parton densities using the following
definition:

\begin{equation}
f(x_1^{(1)},x_2^{(1)}...x_{n_1}^{(1)};x_1^{(2)}....x_{n_k}^{(k)})=\frac
1{\prod_rn_r!}\int \prod_{ri}\frac{d^2k_i^{(r)}}{x_i^{(r)}(2\pi )^3}\left|
\Psi _n\right| ^2(2\pi )^3\delta ^2(\sum k_{i\perp }^{(r)})\delta (1-\sum
x_i^{(r)})\,. 
\end{equation}
Then the generating functional for these quantities 
\begin{equation}
I(\phi )=\sum \int \prod_r\prod_id\,x_i^{(r)}\,\varphi
_r(x_i^r)\,\,f(x_1^{(1)},...x_{n_k}^{(k)}) 
\end{equation}
satisfies an evolution equation in $Q^2$ which can be obtained from the
above probabilistic picture. If we consider only the gluodynamics and take
into account the most singular part of the corresponding splitting kernel in
(14) by putting 
\begin{equation}
w_g^g(x)=\frac{4\,N_c}{x\,(1-x)}\,,\,\,w=\int_0^1\frac{4\,N_c\,\,d\,x}{
x\,(1-x)}\,\,, 
\end{equation}
these equations can be easily solved:

\begin{equation}
I(\varphi )=\int_{-i\infty }^{i\infty }\frac{d\,l}{2\pi i}\,\,\frac{\exp
\,(-l)\,\,w\,\exp \,(-w\xi )\,\left( 1-\exp (-w\xi )\right) ^{-1}}{1-\frac{
4\,N_c}w\,\left( 1-\exp (-w\xi )\right) \int_0^\infty \frac{d\,x}x\,\varphi
(x)\,\exp (-xl)}. 
\end{equation}

The inclusive multi-gluon densities are obtained from this expression by the
change of its argument: $\varphi \rightarrow 1+\chi $ and expanding the
obtained generating functional in the series over $\chi $$(x)$. In
particular for $\chi =0$ we obtain: $I(1)=1$, which corresponds to the
normalization condition for the partonic $\Psi $-function. The generating
functional $I(1+\chi )$ for the inclusive partonic correlators can be used
for finding the screening corrections at the quasi-Regge region $%
x\rightarrow 0$. A more general method taking into account
the possibility of the parton merging was developed in ref.$\left[ 30\right] 
$.

Below we survey the basic results of the Regge approach used in the BFKL
theory.

\subsection{Regge theory}

The elastic scattering amplitude $A(s,t,u)$ for the process $a+b\rightarrow
a^{\prime }+b^{\prime }$ is an analytic function of three invariants [17]: 
\begin{equation}
s=(p_a^{}+p_b^{})^2,\,u=(p_a^{}-p_{b^{\prime
}}^{})^2,\,t=(p_a^{}-p_{a^{\prime }}^{})^2, 
\end{equation}
related as follows: 
\begin{equation}
s+u+t=4m^2 
\end{equation}
for the case of spinless particles with an equal mass $m$.

The function $A(s,t,u)$ describe simultaneously three channels. In the $s$, $%
u$ and $t$ -channels one has 
$$
s>4m^2,\,\,\,u<0, \,\,t<0; 
$$
\begin{equation}
u>4m^2,\,\,\,s<0, \,\,t<0 
\end{equation}
and 
\begin{equation}
t>4m^2,\,\,u<0, \,\, s<0 
\end{equation}
correspondingly. For example, the scattering amplitude in the $s$-channel
(where $~\sqrt{s}$ is the c.m.energy of colliding particles) is obtained as
a boundary value of $A(s,t,u)$ on the upper side of the cut at $s>4m^2$ in
the complex $s$-plane.

The Regge kinematics of the scattered particles in the $s$-channel
corresponds to the following region: 
\begin{equation}
s\,\simeq -u>>m^2\approx -t=\overrightarrow{q}^2, 
\end{equation}
where $\overrightarrow{q}$ is the momentum transfer in the c.m. system ($%
q=p_a^{}-p_{a^{\prime }}^{}$). From the $S$-matrix unitarity one can obtain
the optical theorem for the total cross-section: 
\begin{equation}
\sigma _{tot}^{}(s)=\frac 1sIm_s^{}\,A(s,0), 
\end{equation}
where $Im_s\,A(s,t)$ is the imaginary part of the scattering amplitude in
the $s$-channel. The whole amplitude is expressed through its imaginary
parts in the $s$ and $u$ channels with the use of the dispersion relation
: 
\begin{equation}
A(s,t)=\frac 1\pi \int_{4m^2}^\infty ds^{\prime }\,\frac 1{s^{\prime
}-s-i\epsilon }\,Im_{s^{\prime }}^{}A+\frac 1\pi \int_{4m^2}^\infty
du^{\prime }\,\frac 1{u^{\prime }-u-i\epsilon }\,Im_{u^{\prime }}^{}A , 
\end{equation}
where we omitted the possible subtraction terms.

In the Regge kinematics (56) the essential $s$-channel angular momenta $%
l=\rho \,p$ ($\rho $ is the impact parameter and $p$ is the c.m. momentum)
are large. Therefore one can write the angular momentum expansion of $A$ in
the form of its Fourier transformation : 
\begin{equation}
A(s,t)=-2is\int_{}^{}d^2\rho \,\,[S(s,{\bf \rho })-1]\,e^{i{\bf 
\overrightarrow{q}\overrightarrow{\rho }}}. 
\end{equation}
The $s$-channel partial wave $S(s,{\bf \rho })$ can be considered in some
models as a two-dimensional $S$-matrix which is parametrized by the impact
parameter ${\bf \overrightarrow{\rho }}$ (see for example $\left[ 31\right] $
). Usually it is written in terms of the eikonal phase $\delta (s,{\bf \rho }%
)$: 
\begin{equation}
S(s,{\bf \rho })=e_{}^{i\delta (s,{\bf \rho })}. 
\end{equation}
In accordance with the $s$-channel unitarity one obtains $Im\,\delta >0$.
Because essential values of $\rho $ can not grow more rapidly than $\rho
_{max}^{}=c\,ln(s)$ the Froissart theorem is valid: 
\begin{equation}
\sigma _{tot}<4\pi c^2\,ln^2(s). 
\end{equation}
If the scattering amplitude is pure imaginary at high energies one can
derive from dispersion relations (58) the Pomeranchuck theorem for the
particle-particle and particle-anti-particle total cross sections: 
\begin{equation}
\sigma _t^{pp}=\sigma _t^{p\bar p}. 
\end{equation}
In a more general case when its real part is as big as possible this theorem
is modified as follows 
\begin{equation}
\sigma _t^{pp}/\sigma _t^{p\bar p}=1,\,Re(A)\propto Im(A)\propto s\,ln^2(s). 
\end{equation}

In the Regge model the asymptotics of the elastic scattering amplitude in
region (56) has the following factorized form 
\begin{equation}
A(s,t)=\sum_{p}^{}\xi _{j_p(t)}^ps_{}^{j_p(t)}g_1^p(t)g_2^p(t). 
\end{equation}
Here $g_{1,2}(t)$ are the Reggeon couplings with external particles, $\xi
_{j(t)}^{p}$ is the signature factor (for the signature $p=\pm 1$): 
\begin{equation}
\xi _j^p=i-\frac{\cos {\pi j}+p}{\sin {\pi j}} 
\end{equation}
and $j_p(t)$ is the Regge trajectory assumed to be linear: 
\begin{equation}
j_p(t)=j_0^p+\alpha _p^{\prime }t, 
\end{equation}
where $j_0^p$ and $\alpha _p^{\prime }$ are the reggeon intercept and slope
correspondingly.

For the case, when the trajectory $j_p(t)$ passes through the physical value 
$j=n$ (different for two signatures $p=\pm 1$) corresponding to the integer
(or half integer) spin $\sigma =n$ for an intermediate state, the amplitude
takes the form: 
\begin{equation}
A(s,t)\propto \,\frac{s^n}{t-t_n}, 
\end{equation}
where $t_n$ is the squared mass of the compound state lying on the Regge
trajectory $\left[ 17\right] $. Experimentally all hadrons constructed from
light quarks belong to the Regge families with almost linear trajectories
and an universal slope $\alpha ^{\prime }\approx 1Gev^{-2}$. As it will be
demonstrated below, in the perturbative QCD the gluons and quarks are
reggeized, which means, that they lie on the corresponding Regge
trajectories $\left[ 7\right] $.

The Regge asymptotics seems to be natural $\left[ 17\right] $ from the
decomposition of the scattering amplitude in the sum of contributions of
various angular momenta $j$ in the $t$-channel where the scattering angle $%
\theta _t$ is related with $s$ as follows 
\begin{equation}
z=\cos {\theta _t^{}}=1+\frac{2s}{t-4m^2}. 
\end{equation}
For its symmetric and anti-symmetric parts 
\begin{equation}
A(s,t)=A_{}^{+}(s,t)+A_{}^{-}(s,t),\,\,A_{}^{\pm }(-s,t)=\pm \,A_{}^{\pm
}(s,t), 
\end{equation}
this decomposition continued analytically to big $s$ and fixed negative $t$
takes the simple form: 
\begin{equation}
A_{}^p(s,t)=\int_{\sigma -i\infty }^{\sigma +i\infty }\frac{dj}{2\pi i}\,\xi
_j^p\,s_{}^j\,\phi _j^p(t) 
\end{equation}
and satisfies asymptotically the dispersion relation (58). The functions $%
\phi _j^p(t)$ are proportional to the $t$-channel partial waves $f_j^p(t)$: 
\begin{equation}
\phi _j^p(t)=c_j\,p\,(4m^2-t)^{-j}\,f_j^p(t),\,c_j=16\pi ^24^j\,\frac{\Gamma
(j+\frac 32)}{\Gamma (\frac 12)\Gamma (j+1)} 
\end{equation}
and are real in the physical region of the $s$-channel.

Thus, for the imaginary parts of the signatured amplitudes $A_{}^p$ one
obtains the simple formulae corresponding to the Mellin transformations: 
\begin{equation}
Im_sA^p(s,t)=\int_{\sigma -i\infty }^{\sigma +i\infty }\frac{dj}{2\pi i}
\,s^j\,\phi _j^p(t). 
\end{equation}
The inverse Mellin transformations 
\begin{equation}
\phi _j^p=\int_0^\infty d\xi \,e^{-j\xi }\,Im_sA^p(s,t),\,\xi =ln(s) 
\end{equation}
are simplified versions of the Gribov-Froissart formulae for $f_j^p(t)$ [17].

The $t$-channel elastic unitarity condition for partial waves analytically
continued to complex $j$ for $t>4m^2$ takes the form 
\begin{equation}
\frac{\phi _j^p(t+i\epsilon )-\phi _j^p(t-i\epsilon )}{2i}
=c_j^{-1}(t-4m^2)^{j+\frac 12}\,t^{-\frac 12}\,\phi _j^p(t+i\epsilon )\,\phi
_j^p(t-i\epsilon ). 
\end{equation}

Using the analytic continuation of this relation to complex $t$ one can
express the amplitude $\phi _j^p$ on the physical sheet of the $t$-plane
through its value in the same point on the second sheet and obtain for the $%
t $-channel partial wave $f_j^p(t)$ the Regge poles $\left[ 17\right] $. But
the $t$-channel partial wave could also have fixed square-root singularities
of the type $\phi _j(t)=a(t)+b(t)\sqrt{j-j_0}$ as it takes place in
renormalizable field theories including QCD.

Total cross-sections for hadron-hadron interactions are approximately
constant at high energies (up to possible logarithmic terms). To reproduce
such behaviour in the Regge model a special reggeon is introduced. It is
called the Pomeranchuck pole or the Pomeron. The Pomeron is compatible with
the Pomeranchuck theorem (62) because it has the positive signature and the
vacuum quantum numbers. Its trajectory is assumed to be close to $1$: 
\begin{equation}
j(t)=1+\omega (t)\,\,,\,\,\omega =\Delta +\alpha ^{\prime }t, 
\end{equation}
where $\Delta $$\approx 0.08$ and $\alpha ^{\prime }\approx 0.3$~Gev$^{-2}$
[32]. It means, that the signature factor approximately equals $i$ and the
real part of $A(s,t)$ is small in the agreement with experimental data. The
above quantities $\Delta $ and $\alpha ^{\prime }$ are the bare parameters
of the so called soft Pomeron.\thinspace The BFKL Pomeron has a big value
of\thinspace $\Delta $.

In the $j$-plane there should be other moving singularities of $\phi
_j^{+}(t)$ - the Mandelstam cuts arising as a result of multi-Pomeron
contributions to the $t$-channel unitarity equations. 
V.Gribov constructed the reggeon diagram technique in which all
possible Pomeron interactions are taken into account [33]. In the reggeon
field theory the parameters of the initial Lagrangian are renormalized. The
simple model for taking into account the contributions from the
multi-Pomeron exchanges is based on the assumption that the phase in the
eikonal representation can be calculated by the Fourier transformation of
the amplitude written in the Regge form : 
\begin{equation}
\delta (s,\rho )=\frac 12\int \frac{d^2q}{(2\pi )^2}\,e^{-i\overrightarrow{q}
\overrightarrow{\rho }}\,i\,g^2(t)\,s^{\Delta -\alpha ^{\prime }{\bf 
\overrightarrow{q}}^2}. 
\end{equation}
In this case the resulting amplitude satisfies the Froissart requirements.
The analogous unitarization procedure can be used also in other cases when
the scattering amplitude obtained in some approximation grows more rapidly
than any power of $ln\,(s)$ (see $\left[ 34\right] $).

The high energy theorems do not forbid the existence of another Regge pole
with the vacuum flavour quantum numbers - the Odderon which has the negative
signature and the negative charge parity $\left[ 35\right] $. It could be
situated also near $j=1$, which would lead to a large real part of
scattering amplitudes at high energies and to a significant difference
between proton-proton and proton-antiproton interactions. Such singularity
appears in the perturbative QCD simultaneously with the Pomeranchuck
singularity $\left[ 36\right] $ and therefore the discovery of the Odderon
effects would be very important.

Due to the optical theorem (57) the total cross-section is proportional to
the imaginary part of the scattering amplitude, for which the Regge
asymptotics is usually postulated. Therefore it is natural to ask what
production processes are most probable at high energies. In the dual
resonance model non-vacuum reggeons in the $t$-channel are obtained as a
result of summing over the $s$ channel contributions from resonances with 
growing
spins and masses and the Pomeron is dual to the $s$-channel background. In
multi-peripheral models all reggeons are constructed as $t$-channel compound
states of partons and from the $s$-channel point of view they describe the
processes of the multi-particle production. The produced particles have the
multi-peripheral kinematics: their transverse momenta $k_i^{\perp }$ are
fixed and their longitudinal momenta are ordered in such way, that the
invariant squared energies of neighbouring particles $s_i=(k_i+k_{i-1})^2$
are also fixed. Therefore the average number of particles in the
multi-peripheral models grows as $ln(s)$.

In QCD the situation looks to be simpler than in the multi-peripheral
models. Namely, here the average transverse momenta of produced particles
grow slowly with energy and therefore due to the asymptotic freedom the
effective coupling constant decreases. As a result in QCD (at least in the
perturbation theory) the most essential contribution to the total
cross-section in the reaction 
\begin{equation}
a+b\rightarrow a^{\prime }+d_1+d_2+...+d_n+b^{\prime },\,a^{\prime
}=d_0,\,b^{\prime }=d_{n+1} 
\end{equation}
arises from the multi-Regge kinematics for produced particle momenta: 
\begin{equation}
s\gg s_i\equiv (k_i+k_{i-1})^2\gg m^2,\,-t_i=\overrightarrow{q_i}^2\sim
m^2,\,(\overrightarrow{k_i^{\perp }})^2\sim m^2\,. 
\end{equation}
There is the following constraint for energy invariants $s_i$ resulting from
the reality condition $k_i^2=m^2$ for the final state particles 
\begin{equation}
\prod_{i=1}^{n+1}s_i=s\prod_{i=1}^n[m^2+(\overrightarrow{k_i^{\perp }})^2]. 
\end{equation}
The momentum transfers $q_i$ in the crossing channels $t_i$ are expressed
through external particle momenta as follows: 
\begin{equation}
q_i=p_a-p_{a^{\prime }}-\sum_{r=1}^{i-1}k_r. 
\end{equation}
In terms of the Sudakov parameters for produced particles 
\begin{equation}
k_i=\beta _ip_a+\alpha _ip_b+k_i^{\perp },\,(k_i^{\perp
},p_{a,b})=0,\,k_i^2=s\alpha _i\beta _i-(\overrightarrow{k_i^{\perp }}%
)^2=m^2 
\end{equation}
the multi-Regge kinematics looks especially simple: 
\begin{equation}
1\gg \beta _1\gg \beta _2\gg ...\gg \beta _n\gg \frac{m^2}s\,,\,\,\,\frac{m^2%
}s\ll \alpha _1\ll \alpha _2\ll ...\ll \alpha _n\ll 1, 
\end{equation}
\begin{equation}
s\alpha _i\beta _i\sim (\overrightarrow{k_i^{\perp }})^2\sim m^2. 
\end{equation}
The decomposition of the momentum transfers $q_i$ in terms of the Sudakov
parameters for produced particles is also simplified in the
multi-Regge region: 
\begin{equation}
q_i=\beta _ip_a-\alpha _{i-1}p_b+q_i^{\perp },\,q_i^2\simeq -( 
\overrightarrow{q_i^{\perp }})^2. 
\end{equation}
Thus, momentum transfers with a good accuracy are transverse vectors.
Finally the energy invariants also can be expressed through the Sudakov
variables: 
\begin{equation}
s_i=s\beta _{i-1}\alpha _i 
\end{equation}
and therefore the produced particle momenta in the multi-Regge kinematics
are essentially longitudinal.

Now we review shortly the theoretical description of the multi-Regge
processes in the framework of the Regge model. It is natural to generalize
the formulae for the elastic processes in the form: 
\begin{equation}
A_{2\rightarrow n+2}=s_1^{j(t_1)}s_2^{j(t_2)}...s_{n+1}^{j(t_{n+1})}\gamma
(q_1,q_2,...q_{n+1}). 
\end{equation}
Here the function $\gamma (q_1,...q_{n+1})$ is not real and should contain
something similar to signature factors for the elastic amplitude to satisfy
analytic properties in the direct channels $s_i$. These properties are
significantly simplified in the multi-Regge regime, as it was shown by K.
Ter-Martirosyan, H. Stapp, A. White, J.
Bartels and others. Namely, the inelastic amplitude has only physical
singularities in the channels where the real production of intermediate
particles is possible. There can be simultaneous singularities only in 
non-overlapping channels.

Let us consider the amplitude $A_{2\rightarrow 3}$ for the single particle
production in the multi-Regge kinematics. The signatures in channels $t_1$
and $t_2$ are assumed to be equal to $p_1$ and $p_2$ correspondingly. The
signatured amplitude has two contributions satisfying the double dispersion
representation in the non-overlapping channels $s_1,s$ and $s_2,s$
correspondingly and can be written with the use of the double
Watson-Sommerfeld transformation as follows: 
\begin{equation}
A_{2\rightarrow 3}=\int_{}^{}\frac{dj_1}{2\pi i}\frac{dj_2}{2\pi i}\left[
s_1^{j_1-j_2}\xi _{j_1-j_2}^{(p_1p_2)}s^{j_2}\xi _{j_2}^{p_2}\phi
_{j_1j_2}^1+s_2^{j_2-j_1}\xi _{j_2-j_1}^{(p_1p_2)}s^{j_1}\xi
_{j_1}^{p_1}\phi _{j_1j_2}^2\right] , 
\end{equation}
where two partial waves $\phi _{j_1j_2}^{1,2}$ are real functions. They are
expressed through the spectral functions $\rho
(s_1,s)=Im_{s_1}Im_sA_{2\rightarrow 3}$ and $\rho
(s_2,s)=Im_{s_2}Im_sA_{2\rightarrow 3}$ with the inverse Mellin
transformations: 
\begin{equation}
\phi _{j_1j_2}^{1,2}=\int_0^\infty d\xi _{1,2}e^{-j_{1,2}\xi
_{j_{1,2}}}\int_{\xi _{1,2}}^\infty d\xi e^{-j_{2,1}\xi }\,\rho
(s_{1,2},s),\,\xi _{1,2}=ln(s_{1,2}),\,\xi =ln(s). 
\end{equation}
Using the unitarity conditions in the direct channels the spectral functions
can be presented in a non-linear way again through inelastic amplitudes, 
which leads to a set of non-linear equations.
Such approach is convenient for finding the high energy asymptotics in
non-abelian gauge theories [7].

In the framework of the Regge theory $\phi _{j_1j_2}^{1,2}$ are given in the
factorized form: 
\begin{equation}
\phi _{j_1j_2}^{1,2}=g(t_1)\frac 1{j_1-j_{p_1}(t_1)}\,\Gamma ^{1,2}(t_1,t_2, 
\overrightarrow{k_{\perp }})\,\frac 1{j_2-j_{p_2}(t_2)}g(t_2). 
\end{equation}
Here $\Gamma $ is the reggeon-reggeon-particle vertex depending on the usual
invariants $t_{1,2}$ and the produced particle transverse momentum $k_{\perp
}$ expressed through $s_1,s_2$ with the use of the reality condition 
\begin{equation}
\overrightarrow{k_{\perp }}^2=\frac{s_1s_2}s-m^2. 
\end{equation}
Thus, the production amplitude in the Regge model have the form 
\begin{equation}
A_{2\rightarrow 3}=s_1^{j_{p_1}(t_1)}s_2^{j_{p_2}(t_2)}\gamma (q_1,q_2), 
\end{equation}
where the real and imaginary parts of $\gamma $ are expressed in terms of
the vertices $g$ and $\Gamma $.

In a general case of the amplitude (86) for $n$-particle production in the
multi-Regge kinematics one should introduce many partial waves $\phi
_{j_1...j_{n+1}}^i$ describing different dispersion contributions from all
non-overlapping channels.

\section{BFKL pomeron in the impact parameter space}

In this section we remind comparatively old results concerning the
description of the BFKL Pomeron in the impact parameter representation $%
\left[37\right] $.

The asymptotic behaviour of scattering amplitudes in the Born approximation
is governed by the spin $\sigma $ of the particle exchanged in the crossing
channel: 
\begin{equation}
A_{Born}\sim s^\sigma 
\end{equation}
and the Regge asymptotics is a generalization of this rule to continuous
values of the spin: $\sigma \rightarrow j=j(t)$. In higher orders of the
perturbation theory the scattering amplitude behaves
as $s^n$ (apart from possible logarithmic terms), where the power 
$n=1+\sum_i\,(\sigma _i-1)$
grows linearly with the spins $\sigma _i$ of the particles in the $t$
-channel intermediate states.

In QCD the gluon spin $\sigma $ is $1$ and therefore here the most important
high energy processes are caused by the gluon exchanges. For example, the
Born amplitude for the parton-parton scattering is $\left[ 7\right] $ 
\begin{equation}
\label{reg}A(s,t)=\,2s\,g\delta _{\lambda _a,\lambda _{a^{\prime
}}}\,T_{A^{\prime }A}^c\,\frac 1t\,g\delta _{\lambda _b,\lambda _{b^{\prime
}}}\,T_{B^{\prime }B}^c\,, 
\end{equation}
where $\lambda _i$ are helicities of the initial and final particles; $%
A,A^{\prime },B,B^{\prime }$ are their colour indices and $T_{ij}^c$ are
colour group generators in the corresponding representation. The $s$-channel
helicity for each colliding particle is conserved because the virtual gluon
in the $t$-channel for small $q$ interacts with the total colour charge $Q^c$
commuting with space-time transformations.

\subsection{Impact factors}

Let us consider now the high energy amplitude for the colorless particle
scattering described by the Feynman diagrams containing only two
intermediate gluons with momenta $k$ and $q-k$ in the $t$-channel. With a
good accuracy we can neglect the longitudinal momenta in their propagators
(cf. (84)):

\begin{equation}
k^2\simeq k_{\perp }^2\,,\,\,\,(q-k)^2\simeq (q-k)_{\perp }^2\,. 
\end{equation}
The polarization matrix for each gluon can be simplified at large energies $%
s=(p_a+p_b)^2\gg m^2$ as follows

\begin{equation}
\delta ^{\mu \nu }=\delta _{\parallel }^{\mu \nu }+\delta _{\perp }^{\mu \nu
}\simeq \,\delta _{\parallel }^{\mu \nu }=\frac{p_a^\mu p_b^\nu +p_a^\nu
p_b^\mu }{p_ap_b}. 
\end{equation}

The projector to the longitudinal subspace $\delta _{\parallel }^{\mu \nu }$
can combine the large initial momenta $p_a,p_b$ in a big scalar product $%
s/2=p_ap_b$. Moreover, if the indices $\mu $ and $\nu $ belong to the blobs
with incoming particles $a$ and $b$ correspondingly, then with a good
accuracy we have 
\begin{equation}
\delta ^{\mu \nu }\rightarrow \frac{p_b^\mu p_a^\nu }{p_ap_b}. 
\end{equation}

From the point of view of the $t$-channel unitarity for the partial waves $%
f_j(t)$ with complex $j$ this substitution has a rather simple
interpretation. The most important contribution in the $t$-channel appears
from the nonsense intermediate state leading to a pole singularity of $%
f_j(t) $ at $j=1$. For this nonphysical state the projection of the total
spin $\overrightarrow{S}$to the relative momentum of gluons in the c.m.
system  equals $2$ which corresponds to the opposite
sign of their helicities. It means, that here the complex polarization
vectors $e_1$ and $e_2$ of the gluons coincide each with another: $e_1=e_2=e$%
. The projector to the nonsense state is $e_\mu e_{\mu ^{\prime }}\,e_\nu
^{*}e_{\nu ^{\prime }}^{*} $ (note, that $e_\mu ^2=0$). Due to the Lorentz
and gauge conditions the vectors $e$ are orthogonal to both gluon momenta $k$
and $q-k$. Because these momenta are almost transverse at high energies (see
(94)), after the analytic continuation from the $t$-channel to the $s$%
-channel the above projector should be proportional to the product $p_b^\mu
p_b^{\mu ^{\prime }}p_a^\nu p_a^{\nu ^{\prime }}$ in accordance with (96).

Using eq.(96) and introducing the Sudakov parameters 
\begin{equation}
\alpha =-\frac{kp_a}{p_ap_b}=-s_a/s\,,\,\beta =\frac{kp_b}{p_ap_b}
=s_b/s\,,\, \overrightarrow{k}=\overrightarrow{k}_{\perp }\,,\,d^4k=d^2k\, 
\frac{d\,s_ad\,s_b}{2\,\left| s\right| } 
\end{equation}
for the virtual gluon momenta $k\,$,$\,q-k$ one obtains for the asymptotic
contribution of the diagrams with two gluon exchanges the following
factorized expression: 
\begin{equation}
A(s,t)=2i\,|s|\,\frac 1{2!}\int d^2k\,\frac 1{\overrightarrow{k}^2}\,\frac
1{(\overrightarrow{q}{\bf -}\overrightarrow{k})^2}\,\Phi ^a(\overrightarrow{%
k },\overrightarrow{q}{\bf -}\overrightarrow{k})\,\Phi ^b(\overrightarrow{k}%
, \overrightarrow{q}{\bf -}\overrightarrow{k}), 
\end{equation}
corresponding to the impact-factor representation $\left[ 34\right] $. Here
the sum over the colour  indices is implied and the factor $1/2!$ in
front of the integral is related with the gluon identity and compensates the
double number of Feynman diagrams appeared due to our definition of the
impact factors $\Phi ^{a,b}$ as integrals over the energy invariants $%
s_{a,b} $ from the photon-particle amplitudes $f_{\mu \nu }^{a,b}$:

\begin{equation}
\Phi ^{a,b}(\overrightarrow{k},\overrightarrow{q}{\bf -}\overrightarrow{k}
)=\int_{-\infty }^\infty \frac{ds_{a,b}}{(2\pi )^2i}\frac{p_{b,a}^\mu }s 
\frac{p_{b,a}^\nu }s\,f_{\mu \nu }^{a,b}(s_{a,b},\overrightarrow{k}, 
\overrightarrow{q}{\bf -}\overrightarrow{k})\,. 
\end{equation}
The impact factors describe the inner structure of colliding particles. For
large $\overrightarrow{k}$ they are proportional to the number of partons $%
N_i$ weighted with their colour group Casimir operators.

Note, that we neglected the longitudinal momenta in gluon propagators in
representation (98) in accordance with (94) because in the essential
integration region for the impact factors we have 
\begin{equation}
s_{a,b}\sim m^2,\,\,\overrightarrow{k}^2\sim (\overrightarrow{q}{\bf -} 
\overrightarrow{k})^2\sim m^2 
\end{equation}
and therefore $k_{\parallel }^2=\frac{s_as_b}s\ll k_{\perp }^2$. The
ultraviolet convergency of the integrals over $s_{a,b}$ follows from the
fact, that due to the Ward identities $k^\mu f_{\mu \nu }=(q-k)^\nu f_{\mu
\nu }=0$ for the scattering amplitudes $f=f^{a,b}$ one can perform the
substitution: 
\begin{equation}
\frac{p_{b,a}^\mu }s\,\frac{p_{b,a}^\nu }sf_{\mu \nu }^{a,b}\rightarrow 
\frac{k_{\perp }^\mu }{s_{a,b}}\,\frac{(q-k)_{\perp }^\nu }{s_{a,b}}\,f_{\mu
\nu }^{a,b}. 
\end{equation}
Indeed, it has been assumed above that the amplitudes $f$ do not contain
pure gluonic intermediate states in the $t$-channel and therefore $%
f^{a,b}\ll s_{a,b}$ at high energies, which leads to the rapid convergency
of the integrals over $s_{a,b}$. Therefore one can enclose the integration
contours around the right cuts of $f^{a,b}$ in the complex $s_{a,b}$ planes: 
\begin{equation}
\Phi ^{a,b}(\overrightarrow{k},\overrightarrow{q}{\bf -}\overrightarrow{k}
)=\int_{th}^\infty \frac{ds_{a,b}}{2\pi ^2}\frac{k_{\perp }^\mu }{s_{a,b}} 
\frac{(q_{\perp }^\nu -k_{\perp }^\nu )}{s_{a,b}}\,Im_{s_{a,b}}\,f_{\mu \nu
}(s_{a,b},\overrightarrow{k},\overrightarrow{q}{\bf -}\overrightarrow{k}). 
\end{equation}
after calculating its discontinuity. From this representation of $\Phi $ we
conclude that the impact factors are real functions of $\overrightarrow{k}
,\, \overrightarrow{q}{\bf -}\overrightarrow{k}$, vanishing for small $\mid $
$\overrightarrow{k}\mid $ and $\mid $$\overrightarrow{q}-\overrightarrow{k}
\mid $ in the case of the colorless particle scattering (e.g. photon-photon
collisions), which is a consequence of the gauge invariance and of the
absence of infrared divergencies in the integral over $s_{a,b}$ for small $%
\mid $$\overrightarrow{k}\mid $ and $\mid $$\overrightarrow{q}- 
\overrightarrow{k} \mid $.

From the physical point of view the infrared stability follows from the
fact, that for small $\mid $$\overrightarrow{k}\mid $ the virtual gluon
interacts with the total colour charge of scattered particles. If this
charge is zero, then the dipole and (generally) multipole interactions are
proportional to powers of $\overrightarrow{k}$. In particular the total
cross-section for the photon-photon interactions does not contain any
infrared divergency in the integral over $\overrightarrow{k}$. The various
impact-factors for the real and virtual photons are calculated $\left[
7,\,34\right] $, which in particular gives us a possibility to estimate the
hadron-hadron cross-sections using ideas of the QCD sum rules $\left[
38\right] $. For hadrons the impact factors can be extracted
phenomenologically for example from the inclusive lepton-hadron scattering
because apart from the colour and charge factors these quantities are the
same for the virtual gluons and photons.

\subsection{M\"obius invariance of the BFKL pomeron}

It is convenient to present eq.(98) in the form of the Mellin transformation
(70) 
\begin{equation}
A(s,t)=i\left| s\right| \int \frac{d\;\omega }{2\pi i}\,s^\omega \,f_\omega
(q^2),\,t=-q^2 
\end{equation}
and to pass to the impact parameter representation (cf. (59)) performing the
Fourier transformation $\left[ 38\right] $: 
\begin{equation}
f_\omega (q^2)\,\delta ^2(q-q^{\prime })=\,\int \prod_{r=1,2}\frac{d^2\rho
_rd^2\rho _{r^{\prime }}}{(2\pi )^4}\,\,\Phi ^a(\overrightarrow{\rho _1}, 
\overrightarrow{\rho _2},\overrightarrow{q})\,\,f_\omega (\overrightarrow{
\rho _1},\overrightarrow{\rho _2};\overrightarrow{\rho _1^{\prime }}, 
\overrightarrow{\rho _2^{\prime }})\,\,\Phi ^b(\overrightarrow{\rho
_1^{^{\prime }}},\overrightarrow{\rho _2^{^{\prime }}},\overrightarrow{
q^{\prime }})\,. 
\end{equation}
The quantity $f_\omega (\overrightarrow{\rho _1},\overrightarrow{\rho _2}; 
\overrightarrow{\rho _{1^{^{\prime }}}},\overrightarrow{\rho _{2^{^{\prime
}}}})$ can be considered as a four-point Green function:

\begin{equation}
f_\omega (\overrightarrow{\rho _1},\overrightarrow{\rho _2};\overrightarrow{
\rho _{1^{^{\prime }}}},\overrightarrow{\rho _{2^{^{\prime }}}})=\langle
0\left| \phi (\rho _1)\,\phi (\rho _2)\,\phi (\rho _{1^{^{\prime }}})\,\phi
(\rho _{2^{^{\prime }}})\right| 0\rangle 
\end{equation}
where the field $\varphi (\rho )$ describes the (reggeized) gluons. In the
Born approximation $f_\omega$  is proportional to the product of  
free Green functions:

\begin{equation}
f_\omega ^0(\overrightarrow{\rho _1},\overrightarrow{\rho _2}; 
\overrightarrow{\rho _{1^{^{\prime }}}},\overrightarrow{\rho _{2^{^{\prime
}}}})=\frac{4\pi ^2}\omega \,\ln \left| \rho _{11^{^{\prime }}}\right| \,\ln
\left| \rho _{22^{^{\prime }}}\right| \,,\,\,\overrightarrow{\rho _{ik}}
\equiv \overrightarrow{\rho _i}-\overrightarrow{\rho _k}\,, 
\end{equation}
related with the gluon propagators

\begin{equation}
\frac 1{\overrightarrow{k}^2}=-\int \frac{d^2\rho }{2\pi }\,\exp (i 
\overrightarrow{k}\overrightarrow{\rho })\,\ln \left| \rho \right| \,. 
\end{equation}
The functions $\Phi ^{a,b}(\overrightarrow{\rho _1},\overrightarrow{\rho _2}%
, \overrightarrow{q})$ are related with the impact factors by the Fourier
transformations: 
\begin{equation}
\Phi ^{a,b}(\overrightarrow{\rho _1},\overrightarrow{\rho _2}, 
\overrightarrow{q})=\int d^2k\,\,\Phi ^{a,b}(\overrightarrow{k}, 
\overrightarrow{q}-\overrightarrow{k})\,\exp (i\overrightarrow{k} 
\overrightarrow{\rho _1})\,\exp (i(\overrightarrow{q}-\overrightarrow{k}) 
\overrightarrow{\rho _2})\,. 
\end{equation}
The vanishing of $\Phi ^{a,b}(\overrightarrow{k},\overrightarrow{q}- 
\overrightarrow{k})$ (102) at $\overrightarrow{k}\rightarrow 0$ or at $
\overrightarrow{k}\rightarrow \overrightarrow{q}$ is equivalent to the
following sum rules for $\Phi ^{a,b}(\overrightarrow{\rho _1}, 
\overrightarrow{\rho _2},\overrightarrow{q})$: 
\begin{equation}
\int \Phi ^{a,b}(\overrightarrow{\rho _1},\overrightarrow{\rho _2}, 
\overrightarrow{q})\,d^2\rho _1=\int \Phi ^{a,b}(\overrightarrow{\rho _1}, 
\overrightarrow{\rho _2},\overrightarrow{q})\,d^2\rho _2=0\,. 
\end{equation}

Therefore the expression for $f_\omega (q^2)$ is not changed if we add to $%
f_\omega (\overrightarrow{\rho _1},\overrightarrow{\rho _2};\overrightarrow{
\rho _{1^{^{\prime }}}},\overrightarrow{\rho _{2^{^{\prime }}}})$ an
arbitrary function which does not depend on $\overrightarrow{\rho _1}$, $
\overrightarrow{\rho _2}$, $\overrightarrow{\rho _{1^{^{\prime }}}}$ or $
\overrightarrow{\rho _{2^{^{\prime }}}}$. We can use this freedom related
with the gauge invariance to modify $f_\omega ^0(\overrightarrow{\rho _1}, 
\overrightarrow{\rho _2};\overrightarrow{\rho _{1^{^{\prime }}}}, 
\overrightarrow{\rho _{2^{^{\prime }}}})$ in the following way:

\begin{equation}
f_\omega ^0(\overrightarrow{\rho _1},\overrightarrow{\rho _2}; 
\overrightarrow{\rho _{1^{^{\prime }}}},\overrightarrow{\rho _{2^{^{\prime
}}}})\rightarrow \frac{2\pi ^2}\omega \,\ln \left| \frac{\rho _{11^{^{\prime
}}}\rho _{22^{^{\prime }}}}{\rho _{12^{^{\prime }}}\rho _{1^{\prime }2}}
\right| \,\ln \left| \frac{\rho _{11^{^{\prime }}}\rho _{22^{^{\prime }}}}{
\rho _{12}\rho _{1^{\prime }2^{^{\prime }}}}\right| \,,\,\,\overrightarrow{
\rho _{ik}}\equiv \overrightarrow{\rho _i}-\overrightarrow{\rho _k}\,. 
\end{equation}
This expression is unique in comparison with all other physically equivalent
expressions for $f_\omega ^0$ because it depends only on two independent
anharmonic rations of the vectors $\overrightarrow{\rho _1}$, $
\overrightarrow{\rho _2}$, $\overrightarrow{\rho _{1^{^{\prime }}}}$ and $
\overrightarrow{\rho _{2^{^{\prime }}}}$ which can be chosen as follows

\begin{equation}
\alpha =\left| \frac{\rho _{11^{^{\prime }}}\rho _{22^{^{\prime }}}}{\rho
_{12^{^{\prime }}}\rho _{1^{\prime }2}}\right| \,,\,\,\beta =\left| \frac{
\rho _{11^{^{\prime }}}\rho _{22^{^{\prime }}}}{\rho _{12}\rho _{1^{\prime
}2^{^{\prime }}}}\right| \,,\,\,\gamma =\frac \beta \alpha =\left| \frac{
\rho _{12^{^{\prime }}}\rho _{1^{^{\prime }}2}}{\rho _{12}\rho _{1^{\prime
}2^{\prime }}}\right| \,\,. 
\end{equation}
Therefore $f_\omega ^0$ is invariant under the conformal (M\"obius)
transformations:

\begin{equation}
\rho _k\rightarrow \frac{a\,\rho _k+b\,\rho _k}{c\,\rho _k+d\,\rho _k}\, 
\end{equation}
for arbitrary complex $a$, $b$, $c$ and $d$ provided that we use the complex
coordinates

\begin{equation}
\rho _k=x_k+i\,y_k\,,\,\,\rho _k^{*}=x_k-i\,y_k\, 
\end{equation}
for all two-dimensional vectors $\overrightarrow{\rho _k}$ ($x_k,y_k$).

The solution of the BFKL equation obtained in the next section is also
M\"obius invariant in LLA and can be written in the form $\left[ 37\right] $:

\begin{equation}
f_\omega (\overrightarrow{\rho _1},\overrightarrow{\rho _2};\overrightarrow{
\rho _{1^{^{\prime }}}},\overrightarrow{\rho _{2^{^{\prime }}}}
)=\sum_{n=-\infty }^{+\infty }\int_{-\infty }^{+\infty }\frac{\,(\nu
^2+n^2/4)\,\,\,d\,\nu }{\left[ \nu ^2+(n-1)^2/4\right] \left[ \nu
^2+(n+1)^2/4\right] }\frac{G_{\nu n}(\overrightarrow{\rho _1}, 
\overrightarrow{\rho _2};\overrightarrow{\rho _{1^{^{\prime }}}}, 
\overrightarrow{\rho _{2^{^{\prime }}}})}{\omega -\omega (\nu ,n)}\,, 
\end{equation}
where for $n=\pm 1$ the integral in $\nu $ is regularized as follows:

\begin{equation}
\int_{-\infty }^{+\infty }\,d\,\nu \,\,\frac 1{\nu ^2}\,\,\varphi (\nu
)\equiv \lim _{\epsilon \rightarrow 0}\,\left( \int_{-\infty }^{+\infty
}\,d\,\nu \,\,\frac{\theta (\nu ^2-\epsilon ^2)}{\nu ^2}\,\,\varphi (\nu
)-2\,\frac{\varphi (0)}{\left| \epsilon \right| }\right) . 
\end{equation}
In the ''energy propagator'' $(\omega -\omega (\nu ,n))^{-1}$ the quantity $%
\omega (\nu ,n)$ is the eigen value of the BFKL equation $\left[ 7\right] $:

$$
\omega (\nu ,n)=\frac{N_cg^2}{2\pi ^2}\,\int_0^1\frac{d\,x}{1-x}\left[
x^{(\left| n\right| -1)/2}\cos (\nu \ln \,x)-1\right] 
$$
\begin{equation}
=-\frac{N_cg^2}{2\pi ^2}\,Re\,\left( \psi (\frac{1+\left| n\right| }2+i\nu
)-\psi (1)\right) . 
\end{equation}
The Green function $G_{\nu n}(\overrightarrow{\rho _1},\overrightarrow{\rho
_2};\overrightarrow{\rho _{1^{^{\prime }}}},\overrightarrow{\rho
_{2^{^{\prime }}}})$ is given below:

\begin{equation}
G_{\nu n}(\overrightarrow{\rho _1},\overrightarrow{\rho _2};\overrightarrow{
\rho _{1^{^{\prime }}}},\overrightarrow{\rho _{2^{^{\prime }}}})=\int
d^2\rho _0\,E_{\nu n}^{*}(\rho _{1^{^{\prime }}0},\,\rho _{2^{^{\prime
}}0})\,\,E_{\nu n}(\rho _{10},\,\rho _{20})\,, 
\end{equation}
where

\begin{equation}
E_{\nu n}(\rho _{10},\,\rho _{20})=\langle 0\mid \phi (\rho _1)\phi (\rho
_2)O_{\nu n}(\rho _0)\mid 0\rangle =(\frac{\rho _{12}}{\rho _{10}\rho _{20}}%
)^h\,(\frac{\rho _{12}^{*}}{\rho _{10}^{*}\rho _{20}^{*}})^{\widetilde{h}%
}\,, 
\end{equation}
are the solutions of the homogeneous BFKL equation $\left[ 37\right] $. They
are equivalent to the Polyakov three-point function $\left[ 25\right] $ for
the case when the fields $\phi (\rho _i)$ describing the reggeized gluons
have vanishing conformal quantum numbers $h$ and $\widetilde{h}$. The
composite field $O_{\nu n}$ describing the BFKL pomeron has the conformal
weights

\begin{equation}
h=\frac 12+i\nu +\frac n2,\,\,\widetilde{h}=\frac 12+i\nu -\frac n2 
\end{equation}
with real $\nu $ and the integer conformal spin $n$ in accordance with the
fact that they belong to the basic series of the irreducible unitary
representations of the conformal group. For $g\rightarrow 0$ partial wave $%
f_\omega $ (114) is reduced to (110) because the functions $E_{\nu n}(\rho
_{10},\,\rho _{20})$ have the following completeness property:

\begin{equation}
\delta ^2(\overrightarrow{\rho _1}-\overrightarrow{\rho _{1^{\prime }}}
)\,\delta ^2(\overrightarrow{\rho _2}-\overrightarrow{\rho _{2^{^{\prime }}}}
)=\sum_{n=-\infty }^{+\infty }\int_{-\infty }^{+\infty }\frac{d\,\nu }{\pi
^4 }\,\frac{\nu ^2+n^2/4}{\left| \rho _{11^{\prime }}\right| ^2\left| \rho
_{22^{\prime }}\right| ^2}\,G_{\nu n}(\overrightarrow{\rho _1}, 
\overrightarrow{\rho _2};\overrightarrow{\rho _{1^{^{\prime }}}}, 
\overrightarrow{\rho _{2^{^{\prime }}}})\,. 
\end{equation}

For various physical applications of the four-point Green function (114) it
is convenient to present $G_{\nu n}(\overrightarrow{\rho _1},\overrightarrow{
\rho _2};\overrightarrow{\rho _{1^{^{\prime }}}},\overrightarrow{\rho
_{2^{^{\prime }}}})$ (117) in terms of the hypergeometric functions taking
into account its conformal invariance:

$$
G_{\nu n}(\overrightarrow{\rho _1},\overrightarrow{\rho _2};\overrightarrow{
\rho _{1^{^{\prime }}}},\overrightarrow{\rho _{2^{^{\prime }}}}
)=c_1\,x^h\,x^{*\widetilde{h}}\,F(h,h,2h;x)\;F(\widetilde{h},\widetilde{h},2 
\widetilde{h};x^{*}) 
$$
\begin{equation}
+c_2\,x^{1-h}x^{*1-\widetilde{h}}F(1-h,1-h,2-2h;x)\;F(1-\widetilde{h},1- 
\widetilde{h},2-2\widetilde{h};x^{*})\,, 
\end{equation}
where $x$ is the complex anharmonic ratio:

\begin{equation}
x=\frac{\rho _{12}\,\rho _{1^{^{\prime }}2^{^{\prime }}}}{\rho
_{11^{^{\prime }}\,}\rho _{22^{^{\prime }}}}\; 
\end{equation}
and $F(a,b,c;x)$ is defined by the series

\begin{equation}
F(a,b,c;x)=1+\frac{a\,b}{1!\,c}\;x+\frac{a(a+1)\,b(b+1)}{2!\,c(c+1)}
\,x^2+...\,\,\,\,. 
\end{equation}
The coefficients

\begin{equation}
c_2=\frac{b_{n,\nu }}{2\pi ^2}\,,\,\,\frac{c_1}{c_2}=\frac{b_{n,-\nu }}{
b_{n,\nu }}=-\frac{\Gamma (2-2h)\Gamma (2-2\widetilde{h})}{\left[ \Gamma
(1-h)\Gamma (1-\widetilde{h})\right] ^2}\,\frac{\left[ \Gamma (h)\Gamma ( 
\widetilde{h})\right] ^2}{\Gamma (2h)\Gamma (2\widetilde{h})} 
\end{equation}
and the factor $b_{n,\nu }$

\begin{equation}
b_{n,\nu }=\pi ^3\,2^{4i\nu }\,\frac{\Gamma (-i\nu +(1+\left| n\right|
)/2)\;\Gamma (i\nu +\left| n\right| /2)}{\Gamma (i\nu +(1+\left| n\right|
)/2)\,\Gamma (1-i\nu +\left| n\right| /2)} 
\end{equation}
are obtained from ref.$\left[ 37\right] $. The ratio $c_1/c_2$ also can be
fixed from the condition, that $G$ is a single-valued function of its
arguments. With the use of the various relations among the hypergeometric
functions in Appendix we derive other representations for the Green 
function 
$G$, which gives us a possibility to continue it in the regions around
points $x=1$ and $x=\infty $ .

\subsection{Operator product expansion of the $t$-channel partial waves}

Let us consider now the properties of the solution (114) of the BFKL
equation in the various interesting cases. It is convenient to use the mixed
representation for the gluon-gluon scattering amplitude $\left[ 37\right] $:

\begin{equation}
f_\omega ^q(\overrightarrow{\rho },\overrightarrow{\rho }^{\prime })=\frac
1{(2\pi )^2}\int d^2R\,\exp (i(\overrightarrow{R}-\overrightarrow{R}^{\prime
})\overrightarrow{q})\,f_\omega (\overrightarrow{\rho _1},\overrightarrow{
\rho _2};\overrightarrow{\rho _{1^{^{\prime }}}},\overrightarrow{\rho
_{2^{^{\prime }}}})\,,\,\,\rho =\rho _{12}\,,\,\rho ^{^{\prime }}=\rho
_{1^{^{\prime }}2^{^{\prime }}}\,, 
\end{equation}
where $R=(\rho_1 +\rho_2)/2,\,
R^{\prime}=(\rho^{\prime}_1+\rho^{\prime}_2)/2 $.

Because the dependence of the impact factors $\Phi ^{a,b}(\overrightarrow{
\rho _1},\overrightarrow{\rho _2},\overrightarrow{q})$ from $\overrightarrow{
R}$ according to (108) is simple:

\begin{equation}
\Phi ^{a,b}(\overrightarrow{\rho _1},\overrightarrow{\rho _2}, 
\overrightarrow{q})=\Phi ^a(\overrightarrow{\rho _{12}},\overrightarrow{q}
)\,\exp (i\overrightarrow{R}\overrightarrow{q})\,, 
\end{equation}
we have another representation for $f_\omega (q^2)$ (104):

\begin{equation}
f_\omega (q^2)=\,\int \frac{d^2\rho \,d^2\rho ^{^{\prime }}}{(2\pi )^2}
\,\,\Phi ^a(\overrightarrow{\rho },\overrightarrow{q})\,\,f_\omega ^q( 
\overrightarrow{\rho },\overrightarrow{\rho ^{^{\prime }}})\,\,\Phi ^b( 
\overrightarrow{\rho ^{\prime }};\overrightarrow{q})\,. 
\end{equation}

Here the function $f_\omega ^q(\overrightarrow{\rho },\overrightarrow{\rho
^{\prime }})$ can be interpreted as the amplitude for the scattering of
two composite objects with the sizes $\overrightarrow{\rho }$ and $
\overrightarrow{\rho ^{^{\prime }}}$. Further, the impact factors $\Phi ^a( 
\overrightarrow{\rho },\overrightarrow{q})$ and $\Phi ^b(\overrightarrow{
\rho ^{^{\prime }}},\overrightarrow{q})$ describe the distributions over
these sizes for fixed $\overrightarrow{q}$. Note, that a simple space-time
picture of the high energy scattering in QCD was developed in the framework
of the dipole approach $\left[ 40\right] $.

For small $\overrightarrow{q}$ the essential values of $\overrightarrow{\rho 
}$ and $\overrightarrow{\rho ^{\prime }}$ do not depend on $\overrightarrow{%
q }$. As for $f_\omega ^q(\overrightarrow{\rho },\overrightarrow{\rho
^{^{\prime }}})$, it has a weak singularity at small $q^2$ due to massless
virtual gluons. Indeed, according to definition (122) for fixed $
\overrightarrow{\rho },\overrightarrow{\rho ^{^{\prime }}}$ and large $%
R-R^{^{\prime }}\sim 1/q$ the anharmonic ratio $x$ is small:

$$
x\simeq \frac{\rho \,\rho ^{^{\prime }}}{\left| R-R^{^{\prime }}\right| ^2}
\ll 1 
$$
and therefore expression (121) for $G_{\nu n}$ can be simplified:

\begin{equation}
G_{\nu n}(\overrightarrow{\rho _1},\overrightarrow{\rho _2};\overrightarrow{
\rho _{1^{^{\prime }}}},\overrightarrow{\rho _{2^{^{\prime }}}})\rightarrow
c_1\,x^h\,x^{*\widetilde{h}}\,+\,c_2\,x^{1-h}\,x^{*1-\widetilde{h}}\,. 
\end{equation}
By putting it in eq.(114) we obtain, that at large distances $R-R^{\prime }$
the function $f_\omega $ equals

\begin{equation}
f_\omega (\overrightarrow{\rho _1},\overrightarrow{\rho _2};\overrightarrow{
\rho _{1^{^{\prime }}}},\overrightarrow{\rho _{2^{^{\prime }}}})\sim \left| 
\frac{\rho \,\rho ^{^{\prime }}}{(R-R^{^{\prime }})^2}\right| ^{1+2i\nu
(\omega )}\,. 
\end{equation}
Here $\nu (\omega )$ is a solution of the equation

\begin{equation}
\omega =\omega (\nu ,0) 
\end{equation}
with $Im(\nu )\leq 0$. If analogously to (118) one will consider $f_\omega ( 
\overrightarrow{\rho _1},\overrightarrow{\rho _2};\overrightarrow{\rho
_{1^{^{\prime }}}},\overrightarrow{\rho _{2^{^{\prime }}}})$ as the
four-point function of a two-dimensional theory:

\begin{equation}
f_\omega (\overrightarrow{\rho _1},\overrightarrow{\rho _2};\overrightarrow{
\rho _{1^{^{\prime }}}},\overrightarrow{\rho _{2^{^{\prime }}}})=\langle
0\left| \phi (\overrightarrow{\rho _1})\phi (\overrightarrow{\rho _2})\,\phi
(\overrightarrow{\rho _{1^{^{\prime }}}})\phi (\overrightarrow{\rho
_{2^{^{\prime }}}})\right| 0\rangle 
\end{equation}
then its asymptotics at small $\rho $ and $\rho ^{^{\prime }}$ and fixed $%
R-R^{\prime }$ is related with the anomalous dimension

\begin{equation}
\gamma (\omega )=\frac 12(h+\widetilde{h})=\frac 12+i\,\nu (\omega ) 
\end{equation}
of the operators $O_\omega $ appearing in the operator-product expansion:

\begin{equation}
\phi (\overrightarrow{\rho _1})\phi (\overrightarrow{\rho _2})\sim \left|
\rho _{12}\right| ^{2\gamma }O_\omega (\overrightarrow{\rho _1})\,,\,\,\phi
( \overrightarrow{\rho _{1^{^{\prime }}}})\phi (\overrightarrow{\rho
_{2^{^{\prime }}}})\sim \left| \rho _{1^{^{\prime }}2^{^{\prime }}}\right|
^{2\gamma }O_\omega (\overrightarrow{\rho _{1^{^{\prime }}}})\,. 
\end{equation}
The quantity $\gamma (\omega )$ is the anomalous dimension of the twist-2
operators constructed in a bilinear form from the gluon fields $G_{\mu \nu
}(x)$. It can be calculated as an expansion in the parameter $g^2/\omega $
(cf. (26)):

\begin{equation}
\gamma (\omega )=\frac{N_cg^2}{4\pi ^2\,\omega }\,+\,\zeta (3)\,\left( \frac{
N_cg^2}{4\pi ^2\,\omega }\right) ^4+...\,\,, 
\end{equation}
where $\zeta $$(x)=\sum_{k=1}^\infty k^{-x}$ is the $\zeta $-function. In
LLA where $\omega \sim g^2$ the anomalous dimension is of the order of unity
and it has a square-root singularity

\begin{equation}
\gamma =\frac 12\,+\frac{\pi }{g}\,\sqrt{\frac{\omega -\omega _0}{%
14\,N_c\,\zeta (3)}} 
\end{equation}
at the point

\begin{equation}
\omega _0=\frac{g^2}{\pi ^2}\,N_c\,\ln \,2\,. 
\end{equation}
This value is the intercept of the Pomeranchuck singularity governing the
asymptotics of the cross-sections $\sigma _{tot}\sim s^{\omega _0}$.

In the deep-inelastic regime of large $Q^2$ where $\rho _{1^{\prime
}2^{\prime }}\ll \rho _{12}\sim \rho _{11^{\prime }}$ one can again use the
operator expansion for the product of the fields $\varphi (\rho _{1^{\prime
}})$ and $\varphi (\rho _{2^{\prime }})$ and\thinspace the asymptotics is
governed by the anomalous dimensions $\gamma (\omega )$ (133) of the twist-2
operators with matrix elements (118). As for the regime of large momentum
transfers $\left| t\right| \gg \left| \rho _{12}\right| ^{-2}\sim \left|
\rho _{1^{\prime }2^{\prime }}\right| ^{-2}$, we can use here the operator
expansion of the product of the fields $\varphi (\rho _i)$ and $\varphi
(\rho _{i^{\prime }})$, which corresponds to the asymptotic behaviour of $%
f_{\omega} $ (114) at $x\rightarrow \infty $ (see (122)). From the
representations of $G_{\nu n}$ given in Appendix one can obtain the
asymptotics of $f_{\omega} $ at large $x$:

\begin{equation}
f_\omega \sim c_\omega \,\ln \,\left| x\right| ^2\,, 
\end{equation}
where $c_\omega $ does not depend on $\rho _{12}$ and $\rho _{1^{\prime
}2^{\prime }}$. Therefore in the mixed representation we have at large $%
\left| t\right| $:

\begin{equation}
f_\omega ^q(\overrightarrow{\rho },\overrightarrow{\rho }^{\prime
})\rightarrow \frac 1t\,c_\omega \,, 
\end{equation}
which means, that in the momentum representation the function $f_\omega ( 
\overrightarrow{k},\overrightarrow{k^{\prime }})$ contains the contributions
proportional to $\delta ^2(\overrightarrow{k_1})$ and $\delta ^2( 
\overrightarrow{k_2})$. Using the normalization conditions for the wave
functions of the initial and final particles, one can verify that at large
momentum transfer the impulse approximation is valid and the hadron
scattering amplitudes is expressed as a sum of the effective parton
scattering amplitudes which do not contain infrared divergencies $\left[ {39}%
\right] $.

\subsection{Asymptotic freedom and vacuum Regge poles}

Because the singularity of $f_{\omega} (\overrightarrow{\rho _1}, 
\overrightarrow{\rho _2};\overrightarrow{\rho _{1^{^{\prime }}}}, 
\overrightarrow{\rho _{2^{^{\prime }}}})$ at large $R-R^{\prime }$ is weak
there is a finite limit for the scattering amplitude at small $q$. Indeed,
this amplitude can be written at the mixed representation in the form:

\begin{equation}
f_{\omega} ^q(\overrightarrow{\rho },\overrightarrow{\rho }^{\prime })=\frac
1{16}\,\sum_{n=-\infty }^{+\infty }\int_{-\infty }^{+\infty }\, \frac{%
(\nu^2+n^2/2)\,\,d\,\nu \,\,}{\left[ \nu ^2+(n-1)^2/4\right] \left[ \nu
^2+(n+1)^2/4\right] }\,\frac{ E_{\nu n}^{*}(\rho ,\,q)\,E_{\nu n}(\rho
^{\prime },\,q)}{\omega -\omega (\nu ,n)}\,, 
\end{equation}
where the function $E_{\nu n}(\rho ,\,q)$ is the Fourier transformation of $%
E_{\nu n}(\rho _{10},\,\rho _{20})$ in $\rho _0$. In particular,
for $\rho \,q\rightarrow 0$ we obtain:

\begin{equation}
E_{\nu 0}(\rho ,\,q)\sim \epsilon (\rho )=\left( \left| \rho \right|
^{-2i\nu }+\left| q^2\rho \right| ^{2i\nu }\exp \,i\delta (0,\nu )\right)
\,. 
\end{equation}
Here the phase $\delta (0,\,\nu )$ is the simple function of $\nu $ (given 
in ref. [37]) which for $\nu \rightarrow 0$ tends to the constant:

\begin{equation}
\delta (0,\,\nu )\rightarrow \pi \,. 
\end{equation}
The amplitude $\epsilon (\rho )$\thinspace is a solution of the homogeneous
BFKL equation at small $\rho $:

\begin{equation}
\omega \,\epsilon (\rho )=\alpha \,\chi (\frac i2\,\rho \,\partial
)\,\epsilon (\rho )\,,\,\,\,\,\chi (\nu )=\frac{2\,N_c}\pi \,Re\,\left( \psi
(1)-\psi (\frac 12+i\nu )\right) . 
\end{equation}
If we take into account the fact, that the QCD coupling constant is
decreasing in the region of the large transverse momentum\thinspace $k\sim
1/\rho \gg q$ this equation should be modified by the substitution\thinspace
(see (13))

\begin{equation}
\alpha \rightarrow \frac b{\ln \,\frac 1{\rho ^2\Lambda ^2}}\,\,,\,\,\,b= 
\frac{4\pi }{\beta _2}. 
\end{equation}
The solution of the modified equation can be found easily. At not too small $%
\rho $ where $\omega <\omega _0$ it can be written in a semi-classical
approximation as follow [37]

\begin{equation}
\epsilon (\rho )\sim \cos \left( \frac \pi 4+\overline{\nu }(\rho ,\omega
)\,\ln \,\frac 1{\left| \rho \right| ^2\Lambda ^2}-\frac b\omega \,\int_0^{
\overline{\nu }(\rho ,\omega )}d\,\nu ^{\prime }\,\chi (\nu ^{\prime
})\right) 
\end{equation}
where $\overline{\nu }(\rho ,\omega )$ satisfies the saddle point relation: 
\begin{equation}
\omega \,\ln \,\frac 1{\left| \rho \right| ^2\Lambda ^2}=b\,\chi (\overline{
\nu }). 
\end{equation}
By matching the logarithmic derivatives of two above expressions (141) 
and (145)
for $\epsilon (\rho )$ at $\rho \sim 1/q$ and using the approximate
expression for $\chi (\nu )$ at small $\nu $:

\begin{equation}
\chi (\nu )=\frac{2\,N_c}\pi \,\left( 2\,\ln \,2\,-\,7\,\zeta (3)\,\nu
^2\right) 
\end{equation}
we obtain the following quantum spectrum of $\overline{\nu }$\thinspace and $%
\omega $\thinspace at small $\nu $:

\begin{equation}
\overline{\nu }_k(-\overrightarrow{q}^2)=\left( \frac{3\,\pi \,(k+3/4)\,\ln
\,2}{7\,\ln \,(\overrightarrow{q}^2/\Lambda ^2)}\right) ^{1/3}\,\,, 
\end{equation}

\begin{equation}
\omega _k(-\overrightarrow{q}^2)=\frac{2\,N_c\,b}{\pi \,\ln \,\frac{
\overrightarrow{q}^2}{\Lambda ^2}}\,\left( 2\,\ln \,2-7\,\zeta (3)\, 
\overline{\nu }_k^2(-\overrightarrow{q}^2)\right) \,, 
\end{equation}
where $k=0,\,1,\,2,...$. Thus, as a consequence of the asymptotic freedom,
the fixed square-root cut of the $t$-channel partial wave $f_\omega ^q$
(140) situated at $\omega =\omega _0$\thinspace is substituted by an
infinite sequence of the Regge poles. It is possible, that the residues of
some poles including leading one with $k=0$ can be negligible small. For
hadrons they are determined by the strong interactions at large distances.
The dependence of the $t$-channel partial waves from the 
dynamics in the
confinement region is even more significant for the fixed momentum transfer $
\overrightarrow{q}^2\sim \Lambda ^2$. Because the intercept of the soft
pomeron is small $\Delta \simeq 0.08\ll 1$ $\left[ 32\right] $ one can
attempt to calculate its trajectory at small $\overrightarrow{q}$ in the
perturbation theory $\left[ 37\right] $. In the region where 
$\omega \ll \alpha (1/\left| \rho \right| ^2)\ll  1$ 
the BFKL equation does not depend on 
$\omega $ and its solution has the simple form [37]:

\begin{equation}
\epsilon (\rho )\sim \cos \left( \frac \pi 4+\nu _1\,\ln \,\frac 1{\left|
\rho \right| ^2\Lambda ^2}-\frac a\omega \right) \,,\,\,a=b\,\int_0^{\nu
_1}d\,\nu ^{\prime }\,\chi (\nu ^{\prime }) 
\end{equation}
where $\nu _1 \simeq 0.637$ 
is the root of the equation $\chi (\nu )=0$ and $a
\simeq 1.28$ for $n_f = 3$. It is resonable to expect, 
that one can neglect the $%
\omega $-dependence of the equation also in the confinement region. In this
case the small-$\rho $ asymptotics of the soft pomeron wave function should
have the form

\begin{equation}
\epsilon (\rho )\sim \cos \left( \frac \pi 4+\nu _1\,\ln \,\frac 1{\left|
\rho \right| ^2\Lambda ^2}\,-\,\varphi (\overrightarrow{q}^2)\right) \,. 
\end{equation}
Here $\varphi (\overrightarrow{q}^2)$ is a phase which is assumed to be a
linear function $\varphi (\overrightarrow{q}^2)=\varphi _0-c\,t$ with $c>0$
at small $t=-\overrightarrow{q}^2$. By matching above expressions for $%
\epsilon (\rho)$ we obtain the Regge trajectories [37]:

\begin{equation}
\omega _k(-\overrightarrow{q}^2)=\frac a{\pi \,k+\varphi (\overrightarrow{q}
^2)}\,. 
\end{equation}
Here without any restriction $\varphi_0 $ is assumed to be in the interval $%
\pi >\varphi_0 >0$. To obtain positive values for the pomeron intercepts we
should put $k=0,\,1,\,...\,\,$ and therefore the rightmost pole is situated
at $\Delta > a / \pi =0.4$ [37]. With increasing $t=- ( 
\overrightarrow{q})^2$ the function $\varphi (\overrightarrow{q}^2)$ is
decreasing and can reach the point $\varphi =0$ where according to (152) the 
first Regge pole
goes to the infinity. Near this point one should use more accurate 
expressions leading to a more moderate growth of 
$\omega _0 (t)$. Nevertheless, to avoid the non-physical behaviour of 
the scattering amplitude at positive $t$ the residue
of the first Regge pole should be small for all momentum transfers. 
In an analogous way the second pole with $k=1$
also could have a small residue. For the position of the next poles 
the value of
the unknown phase $\varphi $ is not essential. For example, the intercept of
the third Regge pole approximately equals to the phenomenological intercept
of the soft pomeron.

Note, that in the momentum representation at $t=0$ the solution of the BFKL
equation with a fixed QCD coupling constant can be written in the form:

\begin{equation}
f(\omega ,k,k^{\prime })=\frac 1{2\,\pi ^2}\,\int_{-\infty }^\infty d\,\nu
\,\sum_{n=-\infty }^\infty \,\frac{\left| k\right|^{-1} \left| k^{\prime
}\right|^{-1} (k/k^{\prime })^{i\nu +n/2}(k^{*}/k^{\prime *})^{i\nu -n/2}}{%
\omega -\omega (\nu ,n)}\,, 
\end{equation}
where $k$ and $k^{\prime }$ are the transverse momenta of incoming and
outgoing particles at the $t$-channel.\thinspace This amplitude is
normalized in such way, that at small $g$ we obtain:

$$
f(\omega ,k,k^{\prime })\rightarrow \frac 1\omega \,\, \delta ^2( 
\overrightarrow{k}-\overrightarrow{k}^{\prime }) 
$$
Near the leading singularity $\omega =\omega _0$ the region of small $\nu $
is essential and we obtain the well known diffusion expression for the
virtual gluon cross-section at high energies [7]

\begin{equation}
\sigma (s,k^2,k^{\prime 2})\sim \int \frac{d\,\omega }{2\pi i}\,s^\omega
\,f(\omega ,k,k^{\prime })\sim \frac{s^{\omega _0}}{\sqrt{\alpha \,\ln \,s}}
\exp \,(-\frac{\ln {}^2(\left| k\right| ^2/\left| k^{\prime }\right| ^2)}{
c\,\ln \,s}) 
\end{equation}
where 
$$
c=56\,\frac{\alpha \,N_c}\pi \,\zeta (3) 
$$
is the diffusion constant.

\section{Multi-Regge processes in QCD}

For the deep-inelastic scattering at small Bjorken variable $x$ the gluon
distribution $g(x,k_{\perp })$ depending on the longitudinal Sudakov
component $x$ of the gluon momentum $k$ and on its transverse projection $%
k_{\perp }$ in the infinite momentum frame of the proton $\left| 
\overrightarrow{p}_A\right| \rightarrow \infty $ can be expressed in terms
of the imaginary part of the gluon scattering amplitude at $t=0$ in the
Regge regime of high energies $\sqrt{s}=\sqrt{2p_Ap_B}$ and fixed momentum
transfers. The most probable process at large $s$ is the gluon production in
the multi-Regge kinematics for final state particle momenta $%
k_0=p_{A^{\prime }},\,k_1=q_1-q_2,...k_n=q_n-q_{n+1},\,k_{n+1}=p_{B^{\prime
}}$ (see Introduction):

\begin{equation}
s\gg s_i=2k_{i-1}k_i\gg
t_i=q_i^2=(p_A-\sum\limits_{r=0}^{i-1}k_r)^2,\,\,\prod
\limits_{i=1}^{n+1}s_i=s\prod\limits_{i=1}^n\overrightarrow{k_i}
^2\,\,,\,\,k_{\perp }^2=-\overrightarrow{k}^2. 
\end{equation}

In LLA the production amplitude in this kinematics has the multi-Regge form
[7]: 
\begin{equation}
A_{2\rightarrow 2+n}^{LLA}=A_{2\rightarrow
2+n}^{tree}\prod\limits_{i=1}^{n+1}s_i^{\omega (t_i)}\,\,. 
\end{equation}
Here $s_i^{\omega (t_i)}$ are the Regge-factors appearing from the radiative
corrections to the Born production amplitude $A_{2\rightarrow 2+n}^{tree}$.
The gluon Regge trajectory $j=1+\omega (t)$ is expressed in terms of the
quantity: 
\begin{equation}
\omega (t)=-\frac{g^2N_c}{16\pi ^3}\int d^2{\bf k}\frac{\overrightarrow{q}^2 
}{\overrightarrow{k}^2(\overrightarrow{q}-\overrightarrow{k})^2}
\,\,\,,\,\,t=-\overrightarrow{q}^2\,\,. 
\end{equation}

Infrared divergencies in the Regge factors cancel in $\sigma _{tot}$ with
analogous divergencies in the contributions of real gluons. The production
amplitude in the tree approximation has the following factorized form [7] 
\begin{equation}
A_{2\rightarrow 2+n}^{tree}=2gT_{A^{\prime }A}^{c_1}\Gamma _1\frac
1{t_1}gT_{c_2c_1}^{d_1}\Gamma _{2,1}^1\frac
1{t_2}....gT_{c_{n+1}c_n}^{d_n}\Gamma _{n+1,n}^n\frac
1{t_{n+1}}gT_{B^{\prime }B}^{c_{n+1}}\Gamma _2\,\,. 
\end{equation}
Here $A,B$ and $A^{\prime },B^{\prime },d_r$ ($r=1,2...n$) are colour
indices for initial and final gluons correspondingly. $T_{ab}^c=-if_{abc}$
are generators of the gauge group $SU(N_c)$ and $g$ is the Yang-Mills
coupling constant. Further, 
\begin{equation}
\Gamma _1=\frac 12e_\nu ^\lambda e_{\nu ^{\prime }}^{\lambda ^{\prime
}*}\Gamma ^{\nu \nu ^{\prime }},\,\Gamma _{r+1,r}^r=-\frac 12\Gamma _\mu
(q_{r+1},q_r)e_\mu ^{\lambda _r*}(k_r) 
\end{equation}
are the reggeon-particle-particle (RPP) and reggeon-reggeon-particle (RRP)
vertices correspondingly. The quantities $\lambda _r=\pm 1$ are the $s$
-channel gluon helicities in the c.m. system. They are conserved for each of
two colliding particles: $\Gamma _1=\delta _{\lambda ^{\prime }\lambda }$,
which is not valid in the one loop approximation [41]. The tensor $\Gamma
^{\nu \nu ^{\prime }}$ can be written as the sum of two terms: 
\begin{equation}
\Gamma ^{\nu \nu ^{\prime }}=\gamma ^{\nu \nu ^{\prime }+}-q^2(n^{+})^\nu
\frac 1{p_A^{+}}(n^{+})^{\nu ^{\prime }}, 
\end{equation}
where we introduced the light cone vectors 
\begin{equation}
n^{-}=\frac{p_A}E,\,n^{+}=\frac{p_B}E,\,E=\sqrt{s}/2,\,n^{+}n^{-}=2 
\end{equation}
and the light cone projections $k^{\pm }=k^\sigma n_\sigma ^{\pm }$ of the
Lorentz vectors $k^\sigma $. The first term is the light cone component of
the Yang-Mills vertex: 
\begin{equation}
\gamma ^{\nu \nu ^{\prime }+}=(p_A^{+}+p_{A^{\prime }}^{+})\delta ^{\nu \nu
^{\prime }}-2p_A^{\nu ^{\prime }}(n^{+})^\nu -2p_{A^{\prime }}^\nu (n^{+}). 
\end{equation}
The second (induced) term in (160) is a coherent contribution of the Feynman
diagrams in which the pole in the $t$-channel is absent. Indeed, it is
proportional to the factor $q^2$ cancelling the neighbouring propagator.

Similarly the effective RRP vertex $\Gamma $$(q_2,q_1)$ can be presented as
follows [7] 
\begin{equation}
\Gamma ^\sigma (q_2,q_1)=\gamma ^{\sigma -+}-2q_1^2\frac{(n^{-})^\sigma }{
k_1^{-}}+2q_2^2\frac{(n^{+})^\sigma }{k_1^{+}}, 
\end{equation}
where 
\begin{equation}
\gamma ^{\sigma +-}=2q_2^\sigma +2q_1^\sigma -2(n^{-})^\sigma
k_1^{+}+2(n^{+})^\sigma k_1^{-} 
\end{equation}
is the light-cone component of the Yang-Mills vertex.

Due to the gluon reggeization the above expression for the production
amplitude in LLA has the important property of the two-particle unitarity in
each of the $t_i$ channels (see (74)).

Furthermore, it satisfies approximately the unitarity conditions in the
direct channels $s_i$ with the intermediate particles being in the
multi-Regge kinematics [7, 37]. Note, that from the general analytic
properties of the inelastic amplitudes in the multi-Regge kinematics
discussed in Introduction (see (86, 87)) one can obtain dispersion relations
in a differential form. Indeed, in accordance with the fact, that in LLA the
real part of the production amplitudes for the negative signature is much
bigger than their imaginary part we can simplify the signature factors for
different dispersion contributions near the point $j=1$. Using this
simplification one can derive the following relation 
\begin{equation}
\sum_i\,Im_{s_{ai}}A_{2\rightarrow 2+n}=\pi \frac \partial {\partial \ln
(s_{a1})}Re\,A_{2\rightarrow 2+n}. 
\end{equation}
Here $Im_{s_{ai}}A_{2\rightarrow 2+n}$ means the imaginary part of the
production amplitude in the channel $s_{ai}=(p_{A^{\prime
}}+\sum_{r=1}^ik_r)^2$. Each of these imaginary parts can be calculated
through the products of {$Re\,A_{2\rightarrow 2+l}$} summed over all
intermediate states and integrated over the produced particle momenta with
the use of the unitarity conditions. Finally one can verify that the above
multi-Regge production amplitudes in LLA satisfy the obtained 
''bootstrap'' equations [7, 37].

Note, that $\Gamma ^\sigma $ has the important property: 
\begin{equation}
(k_1)^\mu \Gamma _\mu (q_2,q_1)=0,\,\,k_1=q_1-q_2\,\,, 
\end{equation}
which gives us a possibility to chose an arbitrary gauge for each of the
produced gluons. In the left ($l$) light cone gauge where $p_Ae^l(k)=0$ the
polarization vector $e^l(k)$ is parametrized in terms of the two-dimensional
vector $e_{\perp }^l$ 
\begin{equation}
e^l=e_{\perp }^l-\frac{k_{\perp }e_{\perp }^l}{kp_A}p_A 
\end{equation}
and satisfies the Lorentz condition $k\,e^l=0$. The matrix element of the
reggeon-reggeon-particle vertex $\Gamma $ takes an especially simple form
[42] 
\begin{equation}
\Gamma _{2,1}^1=Ce^{*}+C^{*}e,\,\,\,C=\frac{q_1^{*}q_2}{k_1^{*}}, 
\end{equation}
if we introduce the complex components 
\begin{equation}
e=e_x+ie_y\,,\,e^{*}=e_x-ie_y\,;\,k=k_x+ik_y\,,\,k^{*}=k_x-ik_y 
\end{equation}
for transverse vectors $\overrightarrow{e_{\perp }},\overrightarrow{k_{\perp
}}$. The factors $q_1^{*}$ and $q_2$ in expression (168) guarantee the
vanishing of the inelastic amplitude at small momentum transfers, which is a
consequence of the fact that the virtual gluons interact in this limit with
the total colour charge, whose matrix elements are zero between the states
with the different number of gluons. The singularity $1/k_1^{*}$ in $C$
reproduces correctly the bremsstrahlung factor for the production amplitude
in the soft gluon emission theorem.

The above complex representation was used in [42, 43] to construct the
effective scalar field theory for multi-Regge processes which in particular
can be applied for the derivation of the equations for compound states of
several reggeized gluons [44, 45]. The corresponding effective action was
obtained later from Yang-Mills one by integrating over the fields describing
the highly virtual particles [43].

The effective action describing multi-Regge processes can be written in the
form invariant under the abelian gauge transformations $\delta V_\mu
^a=i\partial _\mu \chi ^a$ for the physical gluon fields $V_\mu $ provided
that the fields $A_{\pm }$ corresponding to the reggeized gluons are gauge
invariant ( $\delta A_{\pm }=0$):

$$
S_{m\,R}=\int d^4x\{\frac 14(F_{\mu \nu }^a)^2+\frac 12(\partial _{\perp
\sigma }A_{+}^a)(\partial _{\perp \sigma }A_{-}^a)+ 
$$
$$
\frac 12\,g\,[-A_{+}^a(F_{-\sigma }T^a\,i\partial _{-}^{-1}F_{-\sigma
})-A_{-}^a(F_{+\sigma }T^a\,i\partial _{+}^{-1}F_{+\sigma })+(\partial
_{-}^{-1}F_{-\sigma }^a)(A_{-}T^a\,i\partial _\sigma A_{+})+ 
$$
\begin{equation}
+(\partial _{+}^{-1}F_{+\sigma }^a)(A_{+}T^a\,i\partial _\sigma
A_{-})+i(\frac 1{\partial _{+}}\frac 1{\partial _{-}}F_{+-}^a)(\partial
_\sigma A_{+})T^a(\partial _\sigma A_{-})+iF_{+-}^a(A_{-}T^aA_{+})]\}\,, 
\end{equation}
where $F_{\mu \nu }=\partial _\mu V_\nu -\partial _\nu V_\mu $ and we
introduced the light-cone components of the Lorentz tensors in accordance
with (161). The fields $A_{\pm }$ satisfy the kinematical constraints $%
\partial _{\pm }A_{\mp }=0$ equivalent to the condition that in the
multi-Regge kinematics the reggeized gluon takes the negligible part of
energy from the colliding particles. The Feynman vertices which are
generated by this action coincide on the mass shell with the effective
vertices (160, 163) for the reggeon-gluon interactions. However, for virtual
gluons these vertices are different, which leads to the inconsistency of this
theory. Such drawback is absent in a more general nonabelian effective field
theory which will be discussed in the next section.

\subsection{BFKL Pomeron}

Using the explicit expressions for production amplitudes in the multi-Regge
kinematics one can calculate the imaginary part of the elastic scattering
amplitude in LLA with the pomeron quantum numbers in the crossing channel 
related with the high energy asymptotics of the total cross-section. 
Its real
part is small in accordance with the fact, that this amplitude has the
positive signature. Due to the factorized form of the production amplitudes
one can write down the Bethe-Salpeter equation for the vacuum $t$-channel
partial wave $f_\omega$ describing the pomeron as a compound state of two 
reggeized
gluons [7]. The convenience of the $\omega$-representation is related 
with the angular momentum conservation in the $t$-channel. The 
contribution to the integral kernel of the BFKL equation from the real 
gluons is
proportional to the product of the effective vertices calculated in the
light cone gauge [42]:

\begin{equation}
C(p_{1,}p_{1^{\prime }})\,C^{*}(p_2,p_{2^{\prime }})+h.c.\,=\frac{
p_1^{*}p_2\,p_{1^{\prime }}p_{2^{\prime }}^{*}}{\mid k\mid ^2}+h.c. 
\end{equation}
where $p_1,p_2$ and $p_{1^{\prime }},p_{2^{\prime }}$ are the corresponding
complex transverse components of initial and final momenta in the $t$
-channel ($q=p_1+p_2=p_{1^{\prime }}+p_{2^{\prime }}$). In turn, the
contribution related with virtual corrections to the production amplitudes
is proportional to the sum of the Regge trajectories of two gluons:

\begin{equation}
\omega (-\overrightarrow{p_1}^2)+\omega (-\overrightarrow{p_2}^2)\sim \ln
\mid p_1\mid ^2+\ln \mid p_2\mid ^2+c, 
\end{equation}
where the constant $c$ contains the infrared divergent terms which are
cancelled with the analogous terms from the real contribution after its
integration in $k$. The final homogeneous equation for the wave
functions of pomerons being the singularities of $f_\omega$ takes 
the form [46]

\begin{equation}
E\Psi =H_{12}\Psi ,\,E=-\frac{8\omega \pi ^2}{g^2\,N_c}. 
\end{equation}
Here the ''Hamiltonian'' $H_{12}$ is [7, 47]

\begin{equation}
H_{12}=\ln \mid p_1\mid ^2+\ln \mid p_2\mid ^2+\frac 1{\mid p_1\mid ^2\mid
p_2\mid ^2}(p_1^{*}p_2\ln \mid \rho _{12}\mid ^2p_1p_2^{*}+h.c.)-4\psi (1) 
\end{equation}
where $\psi (x)=\frac d{dx}\ln \Gamma (x)$ and $\Gamma (x)$ is the Euler $%
\Gamma $-function. In the above expression $1/\overrightarrow{p_i}^2$ are
the gluon propagators. We introduced the complex components $\rho
_k=x_k+iy_k\,$ for the impact parameters canonically conjugated to the
momenta $p_k=i\frac \partial {\partial \rho _k}$ and performed the Fourier
transformation:

\begin{equation}
\frac 1{\mid k\mid ^2}\rightarrow \ln \mid \rho _{12}\mid ^2, 
\end{equation}
where $\rho _{ik}=\rho _i-\rho _k$. The expressions

\begin{equation}
\ln \mid p_i\mid ^2,\,\,\,\,\mid p_i\mid ^{-2} 
\end{equation}
are the integral operators in the impact parameter representation. The
Hamiltonian (174) has the property of the holomorphic separability [46]:

\begin{equation}
H_{12}=h_{12}+h_{12}^{*}\,,\,\,\,E=\epsilon +\widetilde{\epsilon }\,, 
\end{equation}
where $\epsilon $ and $\widetilde{\epsilon }$ are the energies
correspondingly in the holomorphic and anti-holomorphic subspaces:

\begin{equation}
\epsilon \,\psi (\rho _1,\rho _2)=h_{12}\,\psi (\rho _1,\rho _2)\,, \,\, 
\widetilde{\epsilon }\,\widetilde{\psi }(\rho _1^{*},\rho
_2^{*})=h_{12}^{*}\, \widetilde{\psi }(\rho _1^{*},\rho _2^{*})\,,\,\,\Psi ( 
\overrightarrow{\rho _1}, \overrightarrow{\rho _2})=\psi \,\widetilde{\psi }%
. 
\end{equation}

The holomorphic hamiltonian is [47] 
\begin{equation}
h_{12}=\frac 1{p_1}\ln \,(\rho _{12})\,p_1+\frac 1{p_2}\ln \,(\rho
_{12})\,p_2+\ln (p_1p_2)-2\psi (1). 
\end{equation}
One can verify the validity of another representation for $H_{12}$:

\begin{equation}
h_{12}=\rho _{12}\ln (p_1p_2)\,\rho _{12}^{-1}+2\ln (\rho _{12})-2\psi (1). 
\end{equation}
Further, using the following identities:

\begin{equation}
2\ln \,\partial \,+2\ln \,\rho \,=\psi (-\rho \partial )+\psi (\partial \rho
)=\psi (-\rho \partial )+\psi (1+\rho \partial ) 
\end{equation}
and

\begin{equation}
2\ln (\rho ^2\partial )-2\ln (\rho )=\psi (\rho \partial )+\psi (-\rho
^2\partial \rho ^{-1})=\psi (\rho \partial )+\psi (1-\rho \partial )\,, 
\end{equation}
we can derive the following formulas [47]:

\begin{equation}
h=\ln (\rho _{12}^2\,p_1)+\ln (\rho _{12}^2\,p_2)-2\ln (\rho _{12})- 2\psi
(1), 
\end{equation}

\begin{equation}
h=\frac 12\psi (\rho _{12}\partial _1)+\frac 12\psi (\rho _{21}\partial
_2)+\frac 12\psi (1+\rho _{21}\partial _1)+\frac 12\psi (1+\rho
_{12}\partial _2)-2\psi (1)\,. 
\end{equation}
Here $\psi (x)=\Gamma ^{\prime }(x)/\Gamma (x)$ and $\Gamma (x)$ is the
Euler $\Gamma $-function. From eq. (183) one can easily verify that $h$ is
invariant under the M\"obius transformations:

\begin{equation}
\rho _k\rightarrow \frac{a\rho _k+b}{c\rho _k+d} 
\end{equation}
for arbitrary complex values of $a,b,c,d$. It means, that solutions of the
homogeneous BFKL equation belong to irreducible unitary representations of
the M\"obius group. The generators of this group for a general case of $n$%
-particle interactions are

\begin{equation}
M^z=\sum_{k=1}^n\rho _k\partial _k\,,\,M^{-}=\sum_{k=1}^n\partial
_k\,,\,M^{+}=-\sum_{k=1}^n\rho _k^2\partial _k. 
\end{equation}
Its Casimir operator is

\begin{equation}
M^2=(M^z)^2-\frac 12(M^{+}M^{-}+M^{-}M^{+})=-\sum_{r<s}\rho _{rs}^2\partial
_r\partial _s.
\end{equation}
For the wave function of two particles in the holomorphic subspace we can
use the Polyakov ansatz (cf. (118) with the substitution $h\rightarrow m$):

\begin{equation}
\psi _m(\rho _{10},\rho _{20})=<0\mid \varphi (\rho _1)\varphi (\rho
_2)O_m(\rho _0)\mid 0>=(\frac{\rho _{12}}{\rho _{10}\rho _{20}})^m\,. 
\end{equation}
The conformal weight $m=\frac 12+i\nu +\frac n2$ of the composite operator $%
O_m$ is related with its anomalous dimension $d=1/2+i\nu $ and its conformal
spin $n$. This operator belongs to the basic series of the unitary
representations provided that $\nu $ is real and $n$ is integer. In turn,
the fields $\varphi (\rho _i)$ describe the reggeized gluons and have the
trivial quantum numbers $d=n=0$. The holomorphic factor $\psi _m$ is an
eigen function of the corresponding Casimir operator:

\begin{equation}
M^2\psi _m=m(m-1)\psi _m. 
\end{equation}
Simultaneously it is an eigen function of the BFKL equation in the
holomorphic subspace:

\begin{equation}
h\psi _m=\epsilon \psi _m\,,\,\,\,\epsilon =\psi (m)+\psi (1-m)-2\psi (1).
\end{equation}
The eigen value $\epsilon $ can be obtained from representation (184) for $h$
if one would integrate the both sides of the corresponding Schr\"odinger
equation in (178) over the coordinate $\rho _0$ with the use of the relation:

\begin{equation}
\int d\rho _0\,\psi (\rho _{10},\rho _{20})\sim \rho _{12}^{1-m}. 
\end{equation}
The second Casimir operator $M^{2*}$ is expressed through the conformal
weight $\widetilde{m}=\frac 12+i\nu -\frac n2$. The total energy is

\begin{equation}
E=\psi (m)+\psi (1-m)+\psi (\widetilde{m})+\psi (1-\widetilde{m})-4\psi (1). 
\end{equation}
After simple transformations one can rewrite $E$ as follows

\begin{equation}
E=4\,Re\,\psi (\frac 12+i\nu +\frac{\mid n\mid }2)-4\psi (1). 
\end{equation}

The minimum of the energy is obtained for $\nu =n=0$ and equals $E_0=-8\ln
\,2$. Therefore the total cross-section calculated in LLA grows as $s^\Delta 
$ (where $\Delta $$=(g^2N_c/\pi ^2)\ln \;2$), which violates the Froissart
bound $\sigma _{tot}<c\ln {}^2s$ [7]. One of the possible ways to improve
LLA is to use the above effective field theory for multi-Regge processes
[42, 43].

\subsection{Compound states of reggeized gluons in multi-colour QCD}

The simple method to unitarize the scattering amplitudes obtained in LLA is
related with the solution of the BKP equation [44] for compound states of $n$
reggeized gluons:

\begin{equation}
E\Psi =\sum_{i<k} H_{ik}\Psi \,. 
\end{equation}
Its eigen value $E$ is proportional to the position $\omega =j-1$ of the
singularity of the $t$-channel partial wave:

\begin{equation}
E=-\frac{8\pi ^2}{g^2N_c}\omega \,.
\end{equation}
Note, that a non-trivial example of the BKP dynamics is the equation for the
Odderon which is a compound state of three reggeized gluons [36, 45].

The pair Hamiltonians $H_{ik}$ in (194) have the property of the holomorphic
separability:

\begin{equation}
H_{ik}=-\frac{T_i^aT_k^a}{N_c}(h_{ik}+h_{ik}^{*}).
\end{equation}
The group generators $T_i^a$ act on colour indices of the gluon $i$. The
holomorphic pair hamiltonian is 
\begin{equation}
h_{ik}=\frac 1{p_i}\ln (\rho _{ik})p_i+\frac 1{p_k}\ln (\rho _{ik})p_k+\ln
(p_ip_k)-2\psi (1)\,.
\end{equation}
Similar to the Pomeron case we introduce the complex coordinates $\rho
_k=x_k+iy_k\;(k=1,2,...n)$ and their canonically conjugated momenta $%
p_k=i\frac \partial {\partial (\rho _k)}$ in the impact parameter space
(note, that $\rho _{ik}=\rho _i-\rho _k)$. The above Schr\"odinger equation
for $\Psi $ is invariant [46] under the M\"obius transformations: 
$$
\rho _k^{\prime }=\frac{a\;\rho _k+b}{c\;\rho _k+d} 
$$
for any complex values of $a,b,c,d$.

According to t'Hooft for the multi-colour QCD ($N_c\rightarrow \infty )$ only
planar diagrams in the colour space are important. Because the colour
structure of the eigen function at large $N_c$ is unique, due to eq. (196)
the total hamiltonian $H$ can be written as a sum of the mutually commuting
holomorphic and anti-holomorphic operators [46]:

\begin{equation}
H=\frac 12(h+h^{*})\,,\,\,\,\left[ h,h^{*}\right] =0.
\end{equation}
The colour factor $1/2$ appears because at large $N_c$ the neighbouring 
gluons are in the octet state. The holomorphic hamiltonian is
\begin{equation}
h=\sum_{i=1}^nh_{i,i+1}.
\end{equation}
Thus, in the multi-colour QCD the solution of the Schr\"odinger equation 
(194) has the property of the holomorphic factorization:

\begin{equation}
\Psi =\sum c_k\psi _k(\rho _1,...\rho _n)\,\widetilde{\psi }_k(\rho
_1^{*},...\rho _n^{*}).
\end{equation}
where $\psi $ and $\widetilde{\psi }$ are correspondingly the analytic and
anti-analytic functions of their arguments and the sum is performed over all
degenerate solutions of the Schr\"odinger equations in the holomorphic and
anti-holomorphic subspaces:

\begin{equation}
\epsilon \,\psi =h\,\psi ,\,\epsilon ^{*}\,\psi ^{*}=h^{*}\,\psi
^{*},\,E=\frac 12(\epsilon +\epsilon ^{*}).
\end{equation}
These equations have the nontrivial integrals of motion [47]: 
\begin{equation}
t(\theta )=tr\,T(\theta )\,,\,\,\,\left[ t(u),t(v)\right] =\left[ t(\theta
),h\right] =0,
\end{equation}
where $\theta $ is the spectral parameter
of the transfer matrix $t(\theta )$.
The monodromy matrix $T(\theta )$ is constructed from the product of the $L
$-operators 
\begin{equation}
T(\theta )=L_1(\theta )L_2(\theta )...L_n(\theta )
\end{equation}
expressed in terms of the M\"obius group generators:

\begin{equation}
L_k(\theta )=\left( 
\begin{array}{cc}
\theta +i\rho _k\partial _k & i\partial _k \\ 
-i\rho _k^2\partial _k & \theta -i\rho _k\partial _k
\end{array}
\right) .
\end{equation}
The solution of the Schr\"odinger equation (201) in the holomorphic subspace
is reduced to a pure algebraic problem of finding the representation of the
Yang-Baxter commutation relation [47]:

\begin{equation}
T_{i_1i_1^{\prime }}(u)T_{i_2i_2^{\prime
}}(v)(v-u+iP_{12})=(v-u+iP_{12})T_{i_2i_2^{\prime }}(v)T_{i_1i_1^{\prime
}}(u)\;,
\end{equation}
where the operator $P_{12}$ in its left and right hand sides transmutes
correspondingly the right and the left indices of the matrices $T(u)$ and $%
T(v)$. Moreover [48], Hamiltonian (199) coincides with the local Hamiltonian
for a completely integrable Heisenberg model with the spins belonging to an
infinite dimensional representation of the non-compact M\"obius group and
all physical quantities can be expressed in terms of the Baxter function $%
Q(\lambda )$ satisfying the equation: 
\begin{equation}
t(\lambda )Q(\lambda )=(\lambda +i)^nQ(\lambda +i)+(\lambda -i)^nQ(\lambda
-i),
\end{equation}
where $t(\lambda )$ is an eigen value of the transfer matrix. The solution
of the Baxter equation is known for $n=2$. In a general case $n>2$ one can
present it as a linear combination of the solutions for $n=2$ with a
recurrence relation for the coefficients $d_k$. For $n=3$ this relation
takes the form: 
\begin{equation}
Ad_k(A)=\frac{k(k+1)}{2(2k+1)}(k-m+1)(k+m)(d_{k+1}(A)+d_{k-1}(A))
\end{equation}
with the initial conditions $d_0=0,\,\,d_1=1$. If one will consider for
simplicity the integer values of the conformal weight $m$ the quantization
condition for eigen values $A$ is $d_{m-1}(A)=0$. Although the orthogonality
and completeness conditions for the polynomials $d_k(A)$ are known: 
\begin{equation}
\sum_{k=1}^{m-2}\frac{2(2k+1)}{k(k+1)(k-m+1)(k+m)}\,d_k(A)\,d_k(\widetilde{A}%
)=\delta _{A\widetilde{A}}\,d_{m-1}^{\prime }(A)\,d_{m-2}(A)\,,\,\,A\neq 0\,;
\end{equation}
\begin{equation}
\sum_{A\neq 0}\frac{d_k(A)\,d_{\widetilde{k}}(A)}{d_{m-1}^{\prime
}(A)\,d_{m-2}(A)}=
\delta _{k\widetilde{k}}\,\frac{k(k+1)(k-m+1)(k+m)}{2(2k+1)}%
\,,\,\,d^{\prime}_k(A)=\frac{d}{d\,A}\,d_k(A)\,,
\end{equation}
their full theory is not constructed yet. It does not allow us to calculate
analytically the intercept and wave function of the Odderon in QCD [45, 47].

\section{Effective action for small $x$ physics in QCD}

All above results are based on calculations of effective Reggeon vertices
and the gluon Regge trajectory in the first nontrivial order of perturbation
theory. Up to now we do not know the region of applicability of LLA
including the intervals of energies and momentum transfers fixing the scale
for the QCD coupling constant. One should develop also a self-consistent
approach for the unitarization of the BFKL pomeron. Therefore it is needed
to generalize the effective field theory of ref. [42] to the
quasi-multi-Regge processes in which the final state particles are
separated in several groups consisting of an arbitrary number of gluons and
quarks with a fixed invariant mass; each group is produced with respect to
others in the multi-Regge kinematics.
The production of two particles with a
fixed invariant mass is the simplest example of such processes [49].

At high energies the rapidity $y=\frac{1}{2} \ln \frac{k^{+}}{k^{-}}$ 
constructed 
from the light-cone components $k^{\pm }=k^\alpha n_\alpha ^{\pm }$ of the
particle momenta is similar to the time in quantum mechanics. The
corresponding Hamiltonian is determined by the interaction of gluons and
quarks with a nearly equal rapidity. Let us introduce the parameter 
$\eta $ much smaller than $\ln s$. The gauge-invariant effective action 
$S_{eff}$
local in a rapidity interval $(y_0-\eta ,y_0+\eta )$ was constructed
recently [51] and includes apart from the usual Yang-Mills action also the
reggeon-particle interactions (cf. (170)): 
\begin{equation}
S_{eff}(v,A_{\pm })=-\int d^4x\,tr\left[ \frac 12G_{\mu \nu
}^2(v)+(A_{-}(v)-A_{-})j_{+}^{reg}+(A_{+}(v)-A_{+})j_{-}^{reg}\right] 
\end{equation}
where the anti-hermitian $SU(N_c)$ matrices $v_\sigma $ and $A_{\pm }$
describe correspondingly the usual and reggeized gluons. Because the action
is local in the rapidity space, we omit temporally $y$ as an additional
argument of these fields. The reggeon current $j_{\pm }^{reg}$ is 
expressed in terms of $A_{\pm }$ as follows: 
\begin{equation}
j_{\pm }^{reg}\,=\,\partial _\sigma ^2\,A_{\pm }\,,
\end{equation}
which guarantees, that the interaction disappears on the mass shell $k^2=0$.

The fields $A_{\pm }$ are invariant 
\begin{equation}
\delta A_{\pm }\,=\,0 
\end{equation}
under the infinitesimal gauge transformation 
\begin{equation}
\delta v_\sigma =[D_\sigma ,\chi ] 
\end{equation}
with the gauge parameter $\chi $ decreasing at $x\rightarrow \infty $, but
they belong to the adjoint representation of the global $SU(N_c)$ group and
are transformed at constant $\chi $ as follows: 
\begin{equation}
\delta \,A_{\pm }=g\left[ A_{\pm },\chi \right] \,. 
\end{equation}
As usually, 
\begin{equation}
G_{\mu \nu }(v)=\frac 1g\left[ D_\mu ,D_\nu \right] =\partial _\mu v_\nu
-\partial _\nu v_\mu +g\left[ v_\mu ,v_\nu \right] ,\,\,D_\sigma =\partial
_\sigma +g\,v_\sigma \,. 
\end{equation}
The fields $A_{\pm }$ obey the additional kinematical constraints 
\begin{equation}
\partial _{+}A_{-}=0\,,\,\,\partial _{-}A_{+}=0 
\end{equation}
meaning that the Sudakov components $\alpha ,\,\beta $ of the reggeon
momentum $q=\alpha \,p_B+\beta \,p_A+k_{\perp }$ are negligible small in
comparison with the corresponding big components $\alpha _k,\,\beta _{k-1}$
of the neighbouring particle momenta. Such simplification takes place at the
quasi-multi-Regge kinematics where the gluons in the final and intermediate
states are separated in several clusters. The invariant mass of each 
cluster is
restricted from above by a value proportional to $exp(\eta )$ and the
neighbouring clusters are significantly different in their rapidities: 
$y_{k-1}-y_k\gg \eta $. 
Further, the Sudakov components $\alpha _k$, $\beta _k$
of their total momenta are strongly ordered: $\alpha _k\gg \alpha
_{k-1},\,\beta _k\ll \beta _{k-1}$ and the transverse
momenta $k_{\perp }$ are
restricted. The effective action describes the self-interaction of real and
virtual particles inside each cluster and their coupling with neighbouring
reggeized gluons. Note, that because small fractions of the Sudakov
parameters are transmitted from one to other clusters the constraints 
$\partial _{\pm }A_{\mp }=0$ (216) are not absolute. The composite reggeon
field $A_{\pm }(v)$ is given below 
\begin{equation}
A_{\pm }(v)=v_{\pm }-gv_{\pm }\frac 1{\partial _{\pm }}v_{\pm }+g^2v_{\pm
}\frac 1{\partial _{\pm }}v_{\pm }\frac 1{\partial _{\pm }}v_{\pm }-... 
\end{equation}
and can be written in the explicit form 
\begin{equation}
A_{\pm }(v)=v_{\pm }D_{\pm }^{-1}\partial _{\pm }=-\frac 1g\partial _{\pm
}U(v_{\pm })\,,\,U(v_{\pm })=P\exp (-\frac g2\int_{-\infty }^{x^{\pm
}}d\,x^{\prime \pm }v_{\pm }(x^{\prime }))\,, 
\end{equation}
where the integral operator $D_{\pm }^{-1}\partial _{\pm }$ is implied to
act on an unit constant matrix from the left hand side and the symbol $P$
means the ordering of the fields $v$ in the matrix product in accordance
with the increasing of their arguments $x^{\prime \pm }$. We chose for the
Wilson exponent $U(v_{\pm })$ the following boundary condition: $%
U\rightarrow 1$ at $x\rightarrow -\infty $. This exponent does not have
infrared singularities in the essential integration region in accordance
with the fact, that the operators $\partial _{\pm }^{-1}$ are restricted
from above for all particles inside the cluster. Because
$j^{reg}$ (211) in the momentum representation
contains the factor $t=q^2$ killing the pole in
the neighbouring reggeon propagator, the corresponding scattering amplitudes
satisfy the Steinman relations forbidding simultaneous
singularities in the overlapping channels $t$ and $s$. The interaction terms
describe contributions of the Feynman diagrams, in which the gluons in the
given rapidity interval $\left( y_0-\eta ,y_0+\eta \right) $ are coherently
emitted by the neighbouring particles with essentially different rapidities.

The interaction terms of the action are gauge invariant due to the following
relations 
\begin{equation}
\left[ D_{\pm },j_{\mp }^{ind}\right] =0\,,\,\,j_{\mp }^{ind}=\frac \partial
{\partial v_{\pm }}tr(A_{\pm }(v)\,\partial _\sigma ^2A_{\mp })\, 
\end{equation}
where $j_\sigma ^{ind}$ is the induced current. For the particles belonging
to the same cluster the parameter $\eta $ plays a role of the ultraviolet
cut-off in their relative rapidity.

Note, that  action (210) is similar to the Legendre transformation
for which the $A_{\pm }$-dependence of the current $j_{\pm
}^{reg}$ would be fixed by the stationary condition $\delta \,$$F=0$ for
the free energy:

\begin{equation}
\exp (-iF(j_{\pm }^{reg},A_{\pm }))\,=\,\int \prod_xd\,v(x)\,\exp
\;(-iS_{eff}(v,A_{\pm }))\, 
\end{equation}
with taking into account the Faddeev-Popov ghosts. Below some
arguments supporting our choice (211) of $j_{\pm }^{reg}$ will be given.
Because we do not use the Legendre transformation, there is a finite
renormalization of the kinetic term for the fields $A_{\pm }$ in (210)
even in the
tree approximation:

\begin{equation}
-A_{-}\,j_{+}^{reg}-A_{+}\,j_{-}^{reg}\,\rightarrow \,(\partial _\sigma
A_{-})\,(\partial _\sigma A_{+}) 
\end{equation}
related with the transition $A\rightarrow v\rightarrow A$ induced by the
bilinear terms in the gluon and reggeon fields.

\subsection{$\eta$ - independence}

The physical results should not depend on the auxiliary parameter $\eta $ 
due to
cancellations between the integrals over the invariant masses of the
produced clusters and the integrals over their relative rapidities (for
which $\eta $ plays a part of the infrared cut-off). This criterion is very
important for the self-consistency of the effective action (210).
For big values of $\eta $ the most essential polynomial
contribution from integrals over particle rapidities inside each cluster
corresponds to the production of an arbitrary number $n$ of subgroups
of real and virtual particles in the quasi-multi-Regge kinematics. We
show below,
that the interaction of the particles inside the whole cluster can be
described by the same effective action for each subgroup with a 
smaller cut-off and the
integrals over their relative rapidities reproduce correctly
the contribution
proportional to the polynomial $P_{n-1}(\eta )$.

To begin with, let us consider  only two subgroups of real and virtual
particles with the fixed invariant masses being restricted from above by a
value proportional to $\exp (\eta _0)$ where $\eta _0$ is a new parameter
which is assumed to be
much smaller than initial one $\eta $ for the whole cluster.
The rapidities $y^1$ and $%
y^2 $ of these subgroups satisfy the condition $0<$$y^1-y^2<2\eta $ leading
to the
integral contribution proportional to $\eta $.

The field $v_\alpha $ can be presented as the sum of three Fourier
components: 
\begin{equation}
v=v^1+v^2+v^{12}\,. 
\end{equation}
The fields $v^1$ and $v^2$ describe the particles inside two extracted
subgroups and $v^{12}$ is related with the strongly virtual gluons having
the relatively big Sudakov parameters $\beta \sim k^{+}$ and $\alpha \sim 
k^{-}$ of the
order of value of $\beta $$_1\sim k_1^{+}$ and $\alpha _2\sim k_2^{-}$ for
the first and second subgroups correspondingly. Because $\alpha _i$ and $%
\beta _i$ are strongly ordered in the quasi-multi-Regge kinematics: $\alpha
_2\gg \alpha _1 $, $\beta _1\gg \beta _2$, the virtuality of fields $v^{12}$
is large: $M^2=S\alpha \beta \gg k_{\perp }^2$.

Let us consider the functional integral over the heavy fields $v^{12}$ for
fixed $v^1$ and $v^2$ (cf. [43]). In expression (217) the integral operators 
$\partial _{\pm }^{-1}$ correspond to the propagators of the strongly
virtual neighbouring particles producing gluons within the chosen rapidity
interval $(y_0-\eta ,y_0+\eta )$. Because these virtualities are
significantly different for the emission of gluons inside above two
subgroups we can simplify the composite fields $A_{\pm }(v)$ by neglecting
the more virtual fields: 
\begin{equation}
A_{-}(v)\simeq -\frac{\partial _{-}}gU\,(v_{-}^2)+...\,\,,\,\,A_{+}(v)\simeq
-\frac{\partial _{+}}g\,U(v_{+}^1)+....\,\,. 
\end{equation}

The factors $U(v_{+}^1)$ and $U(v_{-}^2)$ in the above expansion lead to the
corresponding interaction terms $A_{+}(v^1)\,\partial _\alpha ^2A_{-}$ and $%
A_{-}(v^2)\,\partial _\alpha ^2A_{+}$ in the effective Lagrangians $L^1$ and 
$L^2$ for each of the above two subgroups. To get the other two terms in
these Lagrangians one should calculate the functional integral over $v^{12}$%
. It leads to the contributions depending only on $v_{-}^1$ and $v_{+}^2$
because they enter with the coefficients proportional to the big Sudakov
parameters $\alpha $$_2\sim k_2^{+}$ or $\beta $$_1\sim k_1^{-}$. Due to the
gauge invariance of the initial action these contributions should be also
gauge invariant. This property fixes their form almost completely: $\Delta
L^1=A_{-}(v_{-}^1)\,\partial _\alpha ^2a_{+}$ and $\Delta
L^2=A_{+}(v_{+}^2)\,\partial _\alpha ^2a_{-}$ up to two arbitrary functions $%
a_{\pm }(x)$ satisfying the kinematical constraints $\partial
_{-}a_{+}=\partial _{+}a_{-}=0$. Further, one can integrate the obtained
expression $\exp (iS_1^{int}+iS_2^{int})$ over $a_{\pm } $ with an arbitrary
weight depending on $a_{\pm }$ but to achieve an agreement with the
perturbation theory the total Lagrangian should contain at least the
quadratic term $-2\,(\partial _\alpha a_{+})\,(\partial _\alpha a_{-})$
leading to a correctly normalized propagator of fields $a_{\pm }$
responsible for the interaction of the particles with the relative
rapidities smaller than $\eta $ (and bigger than the new intermediate
parameter $\eta _0$).

The integration over $a_{\pm }$ with the subsequent integration over the
relative rapidity gives the result proportional to $\eta $ in accordance
with our expectations. Non-linear interactions of the fields $a_{\pm }$
would lead to the contributions having higher powers of $\eta $ and
therefore they should be absent for the two-group kinematics. Note, that 
these
higher power contributions appear if we shall consider several subgroups of
real and virtual particles in the quasi-multi-Regge kinematics inside the
given cluster. In this more general case one can also verify, that the
integration over momenta of the strongly virtual particles gives the
interaction terms linear in $a_{\pm }$ for each subgroup. It
means, that the functions $a_{\pm }$ coincide in fact with the fields $%
A_{\pm }$ and describe the interaction between the subgroups of particles
within the additional interval $(\eta _0,\eta )$ of their relative rapidity.
Thus, in the framework of the effective action approach the $\eta $
-dependence of integrals over particle momenta inside each group is
compensated by the $\eta $-dependence of integrals over the relative
rapidities of these groups. In particular for $\eta =y$ the fields $A_\pm$
are absent and we return to the initial Yang-Mills theory.

\subsection{Classical equations}

One can verify, that the composite fields $A_{\pm }(v)$ are changed under
the infinitesimal transformations of the gauge group $\delta v_\sigma
=\left[ D_\sigma ,\chi \right] $ according to the abelian law: $\delta $$%
A_{\pm }(v)=\partial _{\pm }\varphi .$ Therefore after integrating by parts
and using the kinematical constraint $\partial _{\pm }A_{\mp }=0$ we
conclude, that the effective action $S_{eff}$ is gauge invariant for $\chi $
decreasing at $\infty $ provided that the reggeon fields are fixed: $\delta
A_{\pm }=0$. For constant $\chi $ the action is also invariant because the
fields $A_{\pm }$ belong to the adjoint representation of the gauge group $%
SU(N_c)$.

The effective action $S_{eff}$ has a nontrivial stationary point $v=\bar v$
satisfying the following Euler-Lagrange equations [51]: 
\begin{equation}
\left[ D_\sigma ,G_{\sigma \pm }\right] =j_{\pm }^{ind},\,\,\left[ D_\sigma
,G_{\sigma \rho }^{\perp }\right] =0, 
\end{equation}
where the induced current $j_\sigma ^{ind}$ equals 
\begin{equation}
j_{\pm }^{ind}=\frac 1{D_{\mp }}\partial _{\mp }\,\,(\partial _{\perp \sigma
}^2A_{\pm })\,\,\partial _{\mp }\frac 1{D_{\mp }}\,,\,\,j_{\perp \sigma
}^{ind}=0 
\end{equation}
and due to (219) satisfies the covariant conservation law: 
\begin{equation}
\left[ D_\sigma ,j_\sigma ^{ind}\right] =0\,. 
\end{equation}
We remind, that the integral operators $D_{\mp }^{-1}\partial _{\mp }$ and $%
\partial _{\mp }D_{\mp }^{-1}$ are implied to act on unit constant matrices
correspondingly from the left and right hand sides (in the second case after
integrating by parts the operators $\partial _{\mp }$ should be substituted
by $-\partial _{\mp }$ acting on $A_{\pm }$). Using the following expression
for the Yang-Mills current 
$$
j_\mu ^{YM}=-\left[ D_\sigma ,G_{\sigma \mu }\right] = 
$$
\begin{equation}
\partial _\mu \partial _\sigma v_\sigma -\partial _\sigma ^2v_\mu +g\left(
\left[ v_\sigma ,\partial _\mu v_\sigma \right] -\left[ \partial _\sigma
v_\sigma ,v_\mu \right] -2\left[ v_\sigma ,\partial _\sigma v_\mu \right]
\right) -g^2\left[ v_\sigma ,\left[ v_\sigma ,v_\mu \right] \right] 
\end{equation}
we can rewrite the above equations in the form: 
\begin{equation}
\partial _\sigma G_{\sigma \mu }=j_\mu \,, 
\end{equation}
where 
\begin{equation}
j_\mu ^{\perp }=-g\left[ v_\sigma ,G_{\sigma \mu }^{\perp }\right] , 
\end{equation}
and 
\begin{equation}
j_{\pm }=-g\left[ v_\sigma ,G_{\sigma \pm }\right] +j_{\pm }^{ind}. 
\end{equation}
The above equations are self-consistent because the current $j$ is
conserved: 
\begin{equation}
\partial _\mu \,j_\mu =\frac 12\left[ D_{-},j_{+}^{ind}\right] +\frac
12\left[ D_{+},j_{-}^{ind}\right] =0\,. 
\end{equation}

We can construct a perturbative solution $v=\overline{v}$ of the classical
equations. For example in the Landau gauge where 
\begin{equation}
\partial _\sigma \bar v_\sigma =\,0 
\end{equation}
one obtains 
\begin{equation}
\bar v_\sigma =\sum_{n=0}^\infty g^nf_\sigma ^n\,,\,\,f_{\pm }^0=A_{\pm
}\,,\,\,f_{\perp \sigma }^0=0\,. 
\end{equation}
and 
\begin{equation}
\overline{j}_\sigma ^{ind}=\sum_{n=0}^\infty g^n\,\Delta _\sigma
^n\,,\,\,\Delta _{\pm }^0=\partial _\alpha ^2\,A_{\pm }\,,\,\,\Delta _{\perp
\sigma }^n=0\,. 
\end{equation}
The higher order coefficients $f_\sigma ^n\,,\,\Delta _\sigma ^n$ satisfy
the reccurence relations: 
\begin{equation}
\partial _\alpha ^2f_\sigma ^n=\Delta _\sigma ^n-\sum_{i=0}^{n-1}\left[
f_\alpha ^{n-1-i},\,2\partial _\alpha f_\sigma ^i-\partial _\sigma f_\alpha
^i+\sum_{j=0}^{i-1}\left[ f_\alpha ^{i-1-j},f_\sigma ^j\right] \right] 
\end{equation}
and 
\begin{equation}
\partial _{\mp }\Delta _{\pm }^n=-\sum_{i=0}^{n-1}\left[ f_{\mp
}^{n-1-i}\,,\,\,\Delta _{\pm }^i\right] \,, 
\end{equation}
where we used the gauge properties (219) of $j_{\pm }^{ind}$ .

The coefficients $\Delta $$_\sigma ^n$ can be presented in the explicit
form: 
\begin{equation}
\Delta _{\pm }^n=\sum_{k=0}^n(\frac 1{D_{\mp }}\partial _{\mp })_k\,\partial
_\sigma ^2A_{\pm }\,(\partial _{\mp }\frac 1{D_{\mp }})_{n-k}\,,\,\,\Delta
_{\sigma \perp }^n=0 
\end{equation}
where 
\begin{equation}
(\frac 1{D_{\mp }}\partial _{\mp })_k=\sum_{perm}\prod_{r=0}^\infty (-\frac
1{\partial _{\mp }}f_{\mp }^r)^{i_r},\,\,\sum_{r=0}^\infty (r+1)i_r=k, 
\end{equation}
\begin{equation}
(\partial _{\mp }\frac 1{D_{\mp }})_k=\sum_{perm}\prod_{r=0}^\infty (-f_{\mp
}^r\frac 1{\partial _{\mp }})^{i_r},\,\,\sum_{r=0}^\infty (r+1)i_r=k. 
\end{equation}
In these expressions the summation is performed over all permutations of
the non-commuting integral operators $(-\frac 1{\partial _{\mp }}f_{\mp }^r)$
and$(-f_{\mp }^r\frac 1{\partial _{\mp }})$ correspondingly and the
following reccurence relations are valid:

\begin{equation}
\partial _{\mp }(\frac 1{D_{\mp }}\partial _{\mp
})_n=-\sum_{i=0}^{n-1}f_{\mp }^{n-1-i}(\frac 1{D_{\mp }}\partial _{\mp
})_i\,, 
\end{equation}
\begin{equation}
\partial _{\mp }(\partial _{\mp }\frac 1{D_{\mp
}})_n=-\sum_{i=0}^{n-1}(\partial _{\mp }\frac 1{D_{\mp }})_i\,\,f_{\mp
}^{n-1-i}. 
\end{equation}

Using equations (235, 236), we can construct $f^n$ and $\Delta ^n$ in several
orders of perturbation theory taking into account that $\partial _{\mp
}A_{\pm }=0$. Namely, 
\begin{equation}
f_{\pm }^0=A_{\pm }\,,\,\,f_\sigma ^{0\perp }=0\,,\,\,\Delta _{\pm
}^0=\partial _\sigma ^2\,A_{\pm }\,,\,\Delta _\sigma ^{\perp }=0\,; 
\end{equation}
\begin{equation}
\partial _\alpha ^2f_{\pm }^1=\Delta _{\pm }^1-\frac 12\left[ A_{\mp
},\partial _{\pm }A_{\pm }\right] \,,\,\,\Lambda _{\pm }^1=\left[ \partial
_\alpha ^2A_{\pm },\partial _{\mp }^{-1}A_{\mp }\right] , 
\end{equation}
\begin{equation}
\partial _\alpha ^2f_\sigma ^{1\perp }=\frac 12\left[ A_{+},\partial _\sigma
^{\perp }A_{-}\right] +\frac 12\left[ A_{-},\partial _\sigma ^{\perp
}A_{+}\right] ; 
\end{equation}
and 
$$
\partial _\alpha ^2f_\sigma ^2=\frac 12\left( \left[ A_{+},\partial _\sigma
f_{-}^1\right] +\left[ A_{-},\partial _\sigma f_{+}^1\right] +\left[
f_{+}^1,\partial _\sigma A_{-}\right] +\left[ f_{-}^1,\partial _\sigma
A_{+}\right] \right) 
$$
\begin{equation}
-\left[ A_{+},\partial _{-}f_\sigma ^1\right] -\left[ A_{-},\partial
_{+}f_\sigma ^1\right] +\Omega _\sigma , 
\end{equation}
where $\Omega _\sigma ^{\perp }=0$ and 
\begin{equation}
\Omega _{\pm }=\Delta _{\pm }^2-2\left[ f_\alpha ^1,\partial _\alpha A_{\pm
}\right] -\frac 12\left[ A_{\pm },\left[ A_{\mp },A_{\pm }\right] \right] , 
\end{equation}
\begin{equation}
\Delta _{\pm }^2=-\partial _{\mp }^{-1}\left( \left[ f_{\mp }^1,\partial
_\sigma ^2A_{\pm }\right] +\left[ A_{\mp },\Delta _{\pm }^1\right] \right) . 
\end{equation}
These perturbative terms $f_\sigma ^n$ satisfy the Landau gauge condition $%
\partial _\alpha f_\alpha ^n=0$. For the quasi-elastic kinematics
when $A_{+}=0$ ($A_{-}=0$)
the solution of the reccurence equations is
trivial: $\bar v_{-}=A_{-}$  ($\bar v_{+}=A_{+}$).

Note, that the Euler-Lagrange equations for the effective action $S_{eff}$
can be obtained also from the usual Yang-Mills action in the form 
\begin{equation}
\left[ D_{\mp },[D_\sigma ,G_{\sigma \pm }]\right] =0\,,\,\,[D_\sigma
,G_{\sigma \rho }^{\perp }]=0 
\end{equation}
if one will restrict oneself to the quasi-gauge variations $\delta v_{\pm
}=\left[ D_{\pm },\chi _{\pm }\right] $ under which
the interaction terms are stationary. Therefore the reggeon fields
$A_{\pm }$ can be considered as some parameters
of the solutions of eqs (248) and the
effective theory can be derived from the gluodynamics by imposing certain
constraints for the initial Yang-Mills fields: $A_{\pm }(v)=\varphi _{\pm }$
where $\varphi _{\pm }$ is an abelian field.

\subsection{Reggeon action in the semi-classical approximation}

To obtain the reggeon action in the tree approximation we should substitute $%
v_\sigma \rightarrow
\overline{v}_\sigma $ in $S_{eff}$ [51]. The interaction terms in $%
S_{eff}$ can be expressed in terms of the Wilson integrals with the
contours displaced  along two light-cone lines:
$$
S_{eff}^{int}=\frac 1g\,\int d^2x_{\perp }\,tr\,\left( \int_{-\infty
}^\infty d\,x^{+}(T(v_{-})-1)\,\partial _\sigma ^2A_{+}+\int_{-\infty
}^\infty d\,x^{-}(T(v_{+})-1)\,\partial _\sigma ^2A_{-}\right) , 
$$
\begin{equation}
T(v_{\pm })=P\,\exp \,(-\frac g2\int_{-\infty }^{+\infty }dx^{\pm }v_{\pm
})=\lim _{x^{\pm }\rightarrow \infty }\frac 1{D_{\pm }}\partial _{\pm }\,, 
\end{equation}
where we used the constraints $\partial _{\pm }A_{\mp }=0$. Note, however,
that because this constraint is not absolute and $T(v_{\pm })$ contains
singular operators $\partial _{\pm }^{-1}$ there could be uncertainties. In
this case the initial representation for the action is preferable.

The expression $\partial _{\pm }D_{\pm }^{-1}$ appearing in expression (225)
has the following integral representation: 
\begin{equation}
\partial _{\pm }D_{\pm }^{-1}=P\,\exp \,(-\frac g2\int_{x^{\pm }}^\infty
dx^{\prime \pm }v_{\pm }). 
\end{equation}
Therefore it tends to $1$ at $x^{\pm }\rightarrow \infty $ and the
interaction term can be expressed in terms of the induced current $j_{\pm
}^{ind}$ (225):
\begin{equation}
S_{eff}^{int}=\frac 1g\,\int d^2x_{\perp }dx^{+}\,tr\,(j_{+}^{ind}(\infty
)-\partial _\sigma ^2A_{+})+\frac 1g\,\int d^2x_{\perp
}dx^{-}\,tr\,(j_{-}^{ind}(\infty )-\partial _\sigma ^2A_{-})\,. 
\end{equation}
For this choice of the boundary conditions for the perturbative coefficients 
$\Delta _{\pm }^n$ of $j_{\pm }^{ind}$ the factors $(\frac 1{D_{\mp
}}\partial _{\mp })^k$ and $(\partial _{\mp }\frac 1{D_{\mp }})^k$ are
defined as the coefficients of the perturbative expansion of the following
expressions: 
\begin{equation}
\frac 1{D_{\mp }}\partial _{\mp }=P\,\prod_{n=0}^\infty \exp \left( -\frac{
g^{n+1}}2\int_{-\infty }^{x^{\mp }}dx^{\prime \mp }f_{\mp }^n\right) , 
\end{equation}
\begin{equation}
\partial _{\mp }\frac 1{D_{\mp }}=P\,\prod_{n=0}^\infty \exp \left( -\frac{
g^{n+1}}2\int_{x^{\mp }}^\infty dx^{\prime \mp }f_{\mp }^n\right) . 
\end{equation}

In the particular case of the quasi-elastic process, when there are only two
clusters, after integration over the fields $A_{\pm }$ we obtain for the
effective action

\begin{equation}
S_{eff}^{qe}=-\frac 12\,\int d^4x\,tr\,G_{\mu \nu }^2(v)\,+\,\Delta \,S\,, 
\end{equation}
where $\Delta \,S$ is the action for a two-dimensional $\sigma $-model:

\begin{equation}
\Delta \,S\,=\,-\frac 2{g^2}\,\int d^2x_{\perp }\,tr\,\left( \partial
_{\perp \sigma }T(v_{-})\right) \left( \partial _{\perp \sigma
}T(v_{+})\right) \,, 
\end{equation}
introduced firstly by E. Verlinde and H. Verlinde using other arguments $%
\left[ 52\right] $. In our approach
the Yang-Mills interaction of the fields $v$ inside of two separated
clusters is also essential.

By inserting the perturbative solution $v_\sigma =\overline{v}_\sigma $, $%
j_{\pm }^{ind}=\overline{j}_{\pm }^{ind}$ of classical equations in $S_{eff}$
one can write the reggeon action in the tree approximation as follows: 
\begin{equation}
S_{eff}^{tree}=-\int d^4x\;tr\;\left( s_2+g\;s_3+g^2\;s_4+...\right) , 
\end{equation}
where 
\begin{equation}
s_2=\partial _\sigma ^{\perp }A_{+}\;\partial _\sigma ^{\perp }A_{-}\,, 
\end{equation}
\begin{equation}
s_3=-(\partial _\sigma ^2A_{-})\,A_{+}\partial _{+}^{-1}A_{+}-(\partial
_\sigma ^2A_{+})\,A_{-}\partial _{-}^{-1}A_{-}\,, 
\end{equation}
$$
s_4=-(\partial _\mu f_\nu ^1)^2-\frac 14\left[ A_{+},A_{-}\right] ^2+ 
$$
\begin{equation}
+(\partial _\sigma ^2A_{-})\,A_{+}\partial _{+}^{-1}A_{+}\partial
_{+}^{-1}A_{+}+(\partial _\sigma ^2A_{+})\,A_{-}\partial
_{-}^{-1}A_{-}\partial _{-}^{-1}A_{-}\,
\end{equation}
and $f^1_\nu$ is determined by eq. (243).

The term $s_2$ leads to the correctly normalized propagator

\begin{equation}
\langle 0\left| T\,(A_{+}^y(x))_b^c(A_{-}^{y^{\prime }}(x^{\prime
}))_{b^{\prime }}^{c^{\prime }}\right| 0\rangle =\Lambda _{bb^{\prime
}}^{cc^{\prime }}\,\theta (y-y^{\prime }-\eta )\,\delta ^2(x_{\parallel
}-x_{\parallel }^{\prime })\,\frac 1{2\pi i}\,\ln \,(x-x^{\prime })_{\perp
}^2 
\end{equation}
of the reggeon fields $(A_{\pm }^y)_b^c$ ($b,c$ are colour indices and $%
\Lambda _{bb^{\prime }}^{cc^{\prime }}=\delta _b^{c^{\prime }}\delta
_{b^{\prime }}^c-N_c^{-1}\delta _b^c\delta _{b^{\prime }}^{c^{\prime }}$ is
the projector to the adjoint representation of the colour group). Note, that
we introduced $y$ as an additional argument enumerating the
Fourier components of the fields and included the
factor $\theta (y-y^{\prime }-\eta )$ in the Green function
to remind that $A_{\pm }$ describe the
interaction of the clusters with the relative rapidity bigger than $\eta $.
One can modify the term $s_2$ to guarantee this property:

\begin{equation}
s_2\rightarrow \int_{-\infty }^{+\infty }d\,y\,\,\partial _\sigma ^{\perp
}A_{+}^y\,\frac \partial {\partial \,y}\,\partial _\sigma ^{\perp
}A_{-}^{y-\eta }. 
\end{equation}
Here we took into account
the interactions for all rapidities contrary to the case of the
initial action defined in the fixed rapidity interval $(y_0-\eta ,y_0+\eta )$%
.

The term $s_3$ corresponds to the triple reggeon coupling. This reggeon
vertex was introduced earlier by A. White $\left[ 53\right] $. Because the
signature of the gluon is negative, $s_3$ does not give any contribution to
the elastic scattering amplitude
due to the Gribov
signature conservation rule but this vertex can be essential for some
inclusive processes.

The term $s_4$ describes the four-linear interaction of the reggeons. The
first two contributions correspond to their elastic scattering in the $t$
-channel and last ones give the gluon transitions $1\Leftrightarrow 3$ which
are not suppressed by the Gribov rule. The quantity $%
f_\sigma ^1$ is conserved: $\partial _\sigma f_\sigma ^1=0$ and is directly
related with the effective RRP vertex $C_\sigma (q_1,q_2)$. Because for the
big squared mass $\kappa \sim $$\partial _\sigma ^2$ of the intermediate
state and fixed transverse momenta we have

$$
f_\nu ^1\,\partial _\sigma ^2\,f_\nu ^1\,\simeq \,\frac 14\,\left[
A_{+},A_{-}\right] ^2\,, 
$$
there is a significant cancellation between two first terms in $s_4$ leading
to a better
convergency of the corresponding integral: $\int \frac{
d\,\kappa }\kappa \,$ where $\ln \,\kappa \,<\eta $. For the colorless
state in the $t$-channel this integral is convergent in the ultraviolet
region and by taking the
residue at $\kappa =0$ one can obtain the result, corresponding to the real
contribution to the BFKL kernel.

The term $s_5$ also can be expressed in terms of the above calculated
quantities $f^1$ and $f^2$. The higher order term $s_6$ describes in
particular the reggeon transition $2\rightarrow 4$ related with the triple
pomeron vertex $\left[ 54\right] $.

In a general case to cancel the infrared divergencies one should take into
account apart from the classical expressions for the corresponding
transition vertices also the contributions from quantum fluctuations near
classical solutions. For example in LLA they are responsible for the gluon
reggeization.

Let us write down the field $v$ as a sum of its classical component $
\overline{\nu }$ and the small variation $\epsilon $ describing its
fluctuations near the classical solution: 
\begin{equation}
v\,=\,\overline{v}\,+\,\epsilon \,\,
\end{equation}
and expand the action in $\epsilon $: 
$$
\Delta S=S_{eff}-S_{eff}^{tree}\,=\,-\int d^4x\,tr\, \left\{\,[D_\mu
,\epsilon _\nu ]^2\,-\,[D_\mu ,\epsilon _\nu ][D_\nu ,\epsilon _\mu
]\,+\,g\,G_{\mu \nu }\,[\epsilon _\mu ,\epsilon _\nu ] \right. 
$$
\begin{equation}
\left. +\frac 12(\epsilon _{-}\frac \partial {\partial v_{-}})(\epsilon
_{-}\frac \partial {\partial v_{-}})j_{-}^{ind}(v_{-})A_{+}+\frac
12(\epsilon _{+}\frac \partial {\partial v_{+}})(\epsilon _{+}\frac \partial
{\partial v_{+}})j_{+}^{ind}(v_{+})A_{-}+O(\epsilon ^3)\, \right\}. 
\end{equation}
The reggeon action in the semiclassical approximation can be obtained if one
would calculate the functional integral over the quantum fluctuations $%
\epsilon $ (with taking into account the Faddeev-Popov ghosts). Here we use
the perturbative solution of the classical equations to write down $\Delta S$
only up to quadratic terms in $\epsilon $ and bilinear in $A_{\pm }$: 
$$
\Delta S=-\int d^4x \,tr \, \left\{(\partial _\mu \epsilon _\nu
)^2-(\partial _\mu \epsilon _\nu )(\partial _\nu \epsilon _\mu
)+g\{2(\partial _\mu \epsilon _\nu )[\overline{v}_\mu ,\epsilon _\nu
]-2(\partial _\nu \epsilon _\mu )[ \overline{v}_\mu ,\epsilon _\nu ] \right. 
$$
$$
\left. +2(\partial _\nu \overline{v}_\mu )[\epsilon _\nu ,\epsilon _\mu
]-(\partial _{\perp \sigma }^2A_{+})\epsilon _{-}\frac 1{\partial
_{-}}\epsilon _{-}-(\partial _{\perp \sigma }^2A_{-})\epsilon _{+}\frac
1{\partial _{+}}\epsilon _{+}\,\}\,\right. 
$$
$$
\left. +g^2\{[A_{+},\epsilon _\nu ][A_{-},\epsilon _\nu ]-\frac
12[A_{+},\epsilon _{+}][A_{-},\epsilon _{-}]-\frac 14[A_{+},\epsilon
_{-}]^2-\frac 14[A_{-},\epsilon _{+}]^2\right. 
$$
$$
\left. -\frac 12[A_{+},A_{-}][\epsilon _{+},\epsilon _{-}]+(\partial _{\perp
\sigma }^2A_{+})(\epsilon _{-}\frac 1{\partial _{-}}\epsilon _{-}\frac
1{\partial _{-}}A_{-}+\epsilon _{-}\frac 1{\partial _{-}}A_{-}\frac
1{\partial _{-}}\epsilon _{-}+A_{-}\frac 1{\partial _{-}}\epsilon _{-}\frac
1{\partial _{-}}\epsilon _{-})\,\right. 
$$
\begin{equation}
\left.+(\partial _{\perp \sigma }^2A_{-})(\epsilon _{+}\frac 1{\partial
_{+}}\epsilon _{+}\frac 1{\partial _{+}}A_{+}+\epsilon _{+}\frac 1{\partial
_{+}}A_{+}\frac 1{\partial _{+}}\epsilon _{+}+A_{+}\frac 1{\partial
_{+}}\epsilon _{+}\frac 1{\partial _{+}}\epsilon _{+})\}\right\}\,, 
\end{equation}
where one should substitute $\overline{v}$ by the classical solution. The
terms bilinear simultaneously in $A_{\pm }$ and in $\epsilon $ can be used
to find the next-to-leading corrections to the BFKL pomeron. In comparison
with the dispersion approach [49] the method based on the effective action
could provide a better infrared convergency of intermediate expressions.

We consider below only the contributions which are linear in $A_{\pm }$ and
bilinear in $\epsilon $$_\sigma $ with the singularities $\partial _{\pm
}^{-1}$. Expanding the integrand $exp\,(-iS_{eff})$
up to the order containing linearly both $A_{+}$ and $A_{-}$ and
substituting the products of $\epsilon _{+}$ and $\epsilon _{-}$ by the free
propagators in the Feynman gauge 
\begin{equation}
<\epsilon _{-}(x)\epsilon _{+}(0)>\,=\,-2i\int \frac{d^4k}{(2\pi )^4}\,
\frac{1}{k^2+i\,0}\,exp(ikx)\,\,, 
\end{equation}
we obtain effectively the box diagram because apart from
the gluon propagators $(k^2+i0)^{-1}$ and $((q-k)^2+i0)^{-1}$
there are also the
propagators $k_{+}^{-1}$ and $k_{-}^{-1}$ appearing from the corresponding
singular factors $%
\partial _{+}^{-1}$ and $\partial _{-}^{-1}$ in the vertices.
The contribution of the Faddeev-Popov ghosts is small. As usual, one
can obtain the logarithmic term $\sim \ln \,s$ as a result of integration
over $k_{+}$ and $k_{-}$ restricted by the constraint $k_{+}k_{-}\sim
k_{\perp }^2$. We interpret it as the integral over the rapidity $y$ of the 
$t$-channel gluons with momenta $k$ and $q-k$ and write down the
corresponding one-loop contribution to the effective action in the form 
\begin{equation}
S_{eff}^1=-\int d^4x_{\perp }\,\int d^2x_{\perp }^{\prime }\,\int
d\,y\,tr\,(\partial _{\perp \sigma }A_{+}^y(x_{-},x_{\perp }))\,(\partial
_{\perp \sigma }A_{-}^{y-\eta} (x_{+},x_{\perp }^{\prime }))\,\,\beta 
(x_{\perp }-x_{\perp }^{\prime }) 
\end{equation}
where $\beta (x_{\perp })$ is obtained by the Fourier transformation

\begin{equation}
\beta (x_{\perp })=\int \frac{d^2q}{(2\pi )^2}\,\omega (-\overrightarrow{q}
^2)\,\exp \,(i\,\overrightarrow{x_{\perp }}\,\overrightarrow{q}) 
\end{equation}
from the gluon Regge trajectory:

\begin{equation}
\omega (-q^2)=\frac{g^2}{16\,\pi ^2}\,N_c\,\int \frac{(-q^2)\,\,d^2\,k}{
k^2\,(q-k)^2}\,. 
\end{equation}
Note, that in LLA $\omega (t)$ does not depend on the intermediate parameter
$\eta$ because it does not contain the ultraviolet divergencies in the 
rapidity.

With taking into account apart from $s_2$ also the one loop correction $%
S_{eff}^1$ the renormalized correlation function contains the Regge factor $%
\exp (\omega (q^2)\,(y-y^{\prime }))$ in the momentum representation: 
\begin{equation}
\int d^2\,x_{\perp }\,\exp (ix_{\perp }q)\,<A_{+}^{y}(x)
A_{-}^{y^{\prime}}(0) >_{ren}\sim \theta (y-y^{\prime }-\eta )\,\exp
(\omega (q^2)\,(y-y^{\prime }))\,\,. 
\end{equation}
Analogously starting from the term $-(\partial _\mu f_\nu )^2$ in the
contribution $s_4$ (259) one can reduce the integrals over the momenta $%
k_{\pm }$ of the real intermediate gluon to the integral over its rapidity $%
y $ and derive the corresponding term of the BFKL kernel.

\subsection{Effective vertices}

Because the classical extremum of the effective action is situated at
non-zero fields $\overline{v}_{\pm }=A_{\pm }+...\,$, it is natural to
parametrize $v_{\pm }$ as follows [51]:

\begin{equation}
v_{\pm }=V_{\pm }+U(V_{\mp })A_{\pm }U^{-1}(V_{\mp })\;,\,\,U(V_{\pm
})=\frac 1{1+g\partial _{\pm }^{-1}V_{\pm }}\;, 
\end{equation}
where $V_{\pm }$ is transformed under the gauge group similar to $v_{\pm }$
and $\overline{V}_{\pm }=0$ for small $g$. For this parametrization the
expansion of $S_{eff}$ in the series over $A_{\pm }$ is gauge invariant but
the coefficients are rather complicated and some contributions to scattering
amplitudes contain the simultaneous singularities in  overlapping
channels (cancelling after the use of equations of motion). It is the 
reason, why we shall use the simpler parametrization of $v_{\pm }$:

\begin{equation}
v_{\pm }=V_{\pm }+A_{\pm }\,. 
\end{equation}
In this case one has also $\overline{V}_{\pm }=0$ at $g=0$ in the Landau
gauge $\partial _\sigma v_\sigma =0$ and the homogeneous polynomials $L^i$
of fields $A_{\pm }$ appearing in the expansion of $S_{eff}$:

\begin{equation}
S_{eff}(V,A_{\pm })=-\int d^4x\;tr\;\sum_{i=0}^\infty L^i\;, 
\end{equation}
are compatible with
the Steinman relations. The terms $L^i$ do not have simple gauge
properties but the corresponding scattering amplitudes are invariant under
the gauge transformation after using equations of motion.

In the above expansion $L^0$ describes the pure gluonic interaction: 
\begin{equation}
L^0=\frac 12G_{\mu \nu }^2(V)=\frac 12(\partial _\mu V_\nu -\partial _\nu
V_\mu )^2+2g\left[ V_\mu ,V_\nu \right] \partial _\mu V_\nu +\frac{g^2}
2\left[ V_\mu ,V_\nu \right] ^2. 
\end{equation}

The next term $L^1$ contains the reggeon fields $A_{\pm }$ linearly and in
particular it governs the quasi-elastic processes: 
\begin{equation}
L^1=j_{+}\,A_{-}\;+\;j_{-}\,A_{+}\,, 
\end{equation}
where the currents $j_{\pm }$ are given below: 
$$
j_{\pm }=-\left[ D_\sigma ,G_{\sigma \pm }\right] -\frac 1g\,\partial
_{\perp \sigma }^2\,\partial _{\pm }\frac 1{D_{\pm }}\partial _{\pm }\, 
$$
$$
=g\left( \left[ V_\sigma ,\partial _{\pm }V_\sigma \right] -\left[ \partial
_\sigma V_\sigma ,V_{\pm }\right] -2\left[ V_\sigma ,\partial _\sigma V_{\pm
}\right] -\partial _\sigma ^2\,V_{\pm }\partial _{\pm }^{-1}V_{\pm }\right) 
$$
$$
+g^2\left( \left[ V_\sigma ,\left[ V_\sigma ,V_{\pm }\right] \right]
+\partial _{\perp \sigma }^2\,V_{\pm }\partial _{\pm }^{-1}V_{\pm }\partial
_{\pm }^{-1}V_{\pm }\right) 
$$
\begin{equation}
-g^3\,\partial _{\perp \sigma }^2\,V_{\pm }\partial _{\pm }^{-1}V_{\pm
}\partial _{\pm }^{-1}V_{\pm }\partial _{\pm }^{-1}V_{\pm }+... 
\end{equation}

The quadratic term of the expansion

$$
L^2=\left[ D_\sigma ^{\perp },A_{+}\right] \left[ D_\sigma ^{\perp
},A_{-}\right] -\frac 14\left( \left[ D_{+},A_{-}\right] -\left[
D_{-},A_{+}\right] \right) ^2+\frac g2\,G_{+-}\left[ A_{-},A_{+}\right] 
$$
$$
+\left( \partial _{+}\frac 1{D_{+}}A_{+}\frac 1{D_{+}}\partial
_{+}-A_{+}\right) \partial _{\perp \sigma }^2A_{-}+\left( \partial _{-}\frac
1{D_{-}}A_{-}\frac 1{D_{-}}\partial _{-}-A_{-}\right) \partial _{\perp
\sigma }^2A_{+} 
$$
$$
=\partial _\sigma ^{\perp }A_{+}\partial _\sigma ^{\perp }A_{-}\,+\,g\left\{
\left[ V_\sigma ^{\perp },A_{+}\right] \partial _\sigma ^{\perp }A_{-}+\frac
12\left[ A_{-},A_{+}\right] \partial _{+}V_{-}-\left[ V_{+},\partial
_{+}^{-1}A_{+}\right] \partial _{\perp \sigma }^2A_{-}\right. 
$$
$$
\left. +\left[ V_\sigma ^{\perp },A_{-}\right] \partial _\sigma ^{\perp
}A_{+}+\frac 12\left[ A_{+},A_{-}\right] \partial _{-}V_{+}-\left[
V_{-},\partial _{-}^{-1}A_{-}\right] \partial _{\perp \sigma
}^2A_{+}\right\} 
$$
$$
+g^2\left\{ \left[ V_\sigma ^{\perp },A_{+}\right] \left[ V_\sigma ^{\perp
},A_{-}\right] -\frac 14\left( \left[ V_{-},A_{+}\right] -\left[
V_{+},A_{-}\right] \right) ^2+\frac 12\left[ V_{+},V_{-}\right] \left[
A_{-},A_{+}\right] \right. 
$$
$$
\left. +\left( (V_{+}\partial _{+}^{-1})^2A_{+}+A_{+}(\partial
_{+}^{-1}V_{+})^2+V_{+}\partial _{+}^{-1}A_{+}\partial _{+}^{-1}V_{+}\right)
\partial _{\perp \sigma }^2A_{-}\right. 
$$
\begin{equation}
\left. +\left( (V_{-}\partial _{-}^{-1})^2A_{-}+A_{-}(\partial
_{-}^{-1}V_{-})^2+V_{-}\partial _{-}^{-1}A_{-}\partial _{-}^{-1}V_{-}\right)
\partial _{\perp \sigma }^2A_{+}\right\} +... 
\end{equation}
describes the gluon production due to the fusion of two reggeized gluons at
the central rapidity region.

The following terms non-linear in fields $A_{\pm }$ are responsible for more
complicated processes 
$$
L^3=\frac g2\left( \left[ D_{+},A_{-}\right] -\left[ D_{-},A_{+}\right]
\right) \left[ A_{-},A_{+}\right] - 
$$
$$
-\frac g2\left( \partial _{+}\frac 1{D_{+}}A_{+}\frac 1{D_{+}}A_{+}\frac
1{D_{+}}\partial _{+}\partial _{\perp \sigma }^2A_{-}+\partial _{-}\frac
1{D_{-}}A_{-}\frac 1{D_{-}}A_{-}\frac 1{D_{-}}\partial _{-}\partial _{\perp
\sigma }^2A_{+}\right) 
$$
\begin{equation}
=-\frac g2\left( A_{+}\partial _{+}^{-1}A_{+}\partial _{\perp \sigma
}^2A_{-}+A_{-}\partial _{-}^{-1}A_{-}\partial _{\perp \sigma }^2A_{+}\right)
+O(g^2)\,. 
\end{equation}
$$
L^4=-\frac{g^2}4\left[ A_{+},A_{-}\right] ^2+\frac{g^2}6\left( \partial
_{+}\frac 1{D_{+}}A_{+}\frac 1{D_{+}}A_{+}\frac 1{D_{+}}A_{+}\frac
1{D_{+}}\partial _{+}\partial _{\perp \sigma }^2A_{-}\right. 
$$
\begin{equation}
\left. +\partial _{-}\frac 1{D_{-}}A_{-}\frac 1{D_{-}}A_{-}\frac
1{D_{-}}A_{-}\frac 1{D_{-}}\partial _{-}\partial _{\perp \sigma
}^2A_{+}\right) . 
\end{equation}
For $i>4$ we have 
\begin{equation}
L^i=(-1)^i\frac{g^{i-2}}{(i-1)!}\left( \partial _{+}\frac 1{D_{+}}\left(
A_{+}\frac 1{D_{+}}\right) ^{i-1}\partial _{+}\partial _{\perp \sigma
}^2A_{-}+\partial _{-}\frac 1{D_{-}}\left( A_{-}\frac 1{D_{-}}\right)
^{i-1}\partial _{-}\partial _{\perp \sigma }^2A_{+}\right) . 
\end{equation}

The effective vertices obtained according to the Feynman rules from the
above action are sums of the usual Yang-Mills couplings and some nonlocal
induced terms. Below we construct the gluon production amplitudes in the
quasi-multi-Regge kinematics using these effective vertices. One can find in
the framework of this approach also the perturbative expansion of the
reggeon action $S_{regg}$ defined as follows 
\begin{equation}
\exp (-iS_{regg}(A_{\pm }))=\int DV\;\exp (-iS_{eff})\, 
\end{equation}
and depending on the reggeon fields $A_{\pm }$.

The subsequent functional integration over $A_{\pm }$ corresponds to the
solution of the reggeon field theory defined in the two-dimensional impact
parameter subspace with the rapidity playing the role of time. This 
theory is obtained after the integration of the multi-reggeon couplings 
over the reggeon longitudinal momenta within the fixed rapidity interval 
$y_0-\eta , y_0+\eta $, 
which is equivalent to calculating certain limits of these 
couplings in the longitudinal subspace. It is
important, that in the above approach the $t$-channel dynamics of the
reggeon interactions turns out to be in the agreement with the $s$-channel
unitarity of the $S$-matrix in the initial Yang-Mills model. In the
Hamiltonian formulation of this reggeon calculus the wave function will
contain the components with an arbitrary number of reggeized gluons.
Nevertheless, one can hope that at least some of the remarkable properties
of the BFKL equation [46-48] will remain in the general case of the
non-conserving number of reggeized gluons.

Note, that to build the effective action for the multi-Regge kinematics one
should take into account only two first terms of the perturbative expansion
of $L$: 
$$
L_{mR}=\frac 12\,(\partial _\mu V_\nu -\partial _\nu V_\mu )^2+\partial
_\sigma ^{\perp }A_{+}\partial _\sigma ^{\perp
}A_{-}+g\,b_3(A_{+},A_{-},V)\,, 
$$
$$
b_3=-\frac 12\,A_{+}\partial _{+}^{-1}A_{+}\partial _{\perp \sigma
}^2A_{-}-\frac 12\,A_{-}\partial _{-}^{-1}A_{-}\partial _{\perp \sigma
}^2A_{+}+F_{+-}\,\left[ A_{-},A_{+}\right] 
$$
$$
-\left( \partial _{+}^{-1}\partial _{-}^{-1}F_{+-}\right) \left[ \partial
_\sigma A_{-},\partial _\sigma A_{+}\right] +\left( \partial
_{-}^{-1}F_{-\sigma }\right) \left[ A_{-},\partial _\sigma A_{+}\right]
+\left( \partial _{+}^{-1}F_{+\sigma }\right) \left[ A_{+},\partial _\sigma
A_{-}\right] 
$$
\begin{equation}
-A_{+}\,\left[ F_{-\sigma },\partial _{-}^{-1}F_{-\sigma }\right]
\,-\,A_{-\,}\left[ F_{+\sigma },\partial _{+}^{-1}F_{+\sigma }\right] \,, 
\end{equation}
where we introduced the abelian strength tensor:

\begin{equation}
F_{\mu \nu }\,=\,\partial _\mu V_\nu -\partial _\nu V_\mu 
\end{equation}
and omitted in the last line some terms containing the factors $\partial
_\sigma V_\sigma $ and $\partial _\sigma ^2V_\mu $ vanishing for the
real gluons. The Feynman vertices of this theory coincide with the effective
reggeon-particle vertices of the leading logarithmic approximation.

The obtained multi-Regge action is invariant under the abelian gauge
transformations:

\begin{equation}
V_\sigma \rightarrow V_\sigma \,+\,\partial _\sigma \,\phi \,. 
\end{equation}
Its disadvantage is that the contributions of loop diagrams can contain the
terms incompatible with the Steinman relations forbidding the simultaneous
singularities in the overlapping channels.\ Such unpleasant properties are
absent for the full effective action (210) invariant under nonabelian gauge
transformations.

\subsection{Gluon and quark pair production in the quasi-multi-Regge
kinematics}

As an example of the general approach we consider the quasi-elastic process
in which several gluons are produced with a fixed invariant mass in the
fragmentation region of the initial gluon $A$ provided that the momentum of 
the
other particle $B$ is almost conserved: 
$p_{B^{\prime }}\simeq p_B$ [49, 51]. It
is convenient to denote the colour indices of the produced gluons by $%
a_1,a_2,...a_n$ ($a_i=1,2,...N_c^2-1$) leaving the index $a_0$ for the
particle $A$. Further, the momenta of the produced gluons and of the
particle $A$ are denoted by $k_1,k_2,...k_n$ and $-k_0$ correspondingly.
The momentum transfer $q=-\sum \limits_{i=0}^nk_i$ is fixed and $%
t=q^2$. Omitting the polarization vectors $e_{{\nu }_i}(k_i)$ for the gluons
we can write the production amplitude related with the one-gluon exchange in
the $t$-channel in the factorized form: 
\begin{equation}
A_{a_0a_1...a_nB^{\prime }B}^{{\nu }_0{\nu }_1...{\nu }_n}\,=\,\,-\phi
_{a_0a_1...a_nc}^{\nu _0\nu _1...\nu _n+}\,\,\,\frac
1t\,\,g\,\,p_B^{-}T_{B^{\prime }B}^c\delta _{\lambda _{B^{\prime }},\lambda
_B}\,. 
\end{equation}
Here the form-factor $\phi $ depends on the invariants constructed from the
momenta $k_0,...k_n$.

In the simplest case of the elastic scattering the tensor $\phi $ equals

\begin{equation}
\phi _{a_0a_1c}^{\nu _0\nu _1+}=g\,\Gamma _{a_0a_1c}^{\nu _0\nu _1+} 
\end{equation}
where after extracting the colour group generators

\begin{equation}
\Gamma _{a_0a_1c}^{\nu _0\nu _1+}=T_{a_1a_0}^c\,\Gamma ^{\nu _1\nu
_0+}(k_1,-k_0)\, 
\end{equation}
the quantity $\Gamma ^{\nu _1\nu _0+}$ can be presented as the sum of two
terms: 
\begin{equation}
\Gamma ^{\nu _1\nu _0+}(k_1,-k_0)=\gamma ^{\nu _1\nu
_0+}(k_1,-k_0)\,+\,\Delta ^{\nu _1\nu _0+}(k_1,-k_0). 
\end{equation}
Here $\gamma ^{\nu _1\nu _0+}$ is the light-cone component of the Yang-Mills
vertex:

\begin{equation}
\gamma ^{\nu _1\nu _0+}(k_1,-k_0)=\delta ^{\nu _1\nu
_0}\,(k_1^{+}+k_0^{+})+(n^{+})^{\nu _0}\,(2k_0+k_1)^{\nu _1}+(n^{+})^{\nu
_1}\,(-2k_1-k_0)^{\nu _0} 
\end{equation}
and $\Delta ^{\nu _1\nu _0+}$ is the induced vertex: 
\begin{equation}
\Delta ^{\nu _1\nu _0+}(k_1,-k_0)\,=\,-t\,(n^{+})^{\nu _1}\frac
1{k_1^{+}}(n^{+})^{\nu _0}\,,\,\,k_0^{+}+k_1^{+}=0. 
\end{equation}

For the case of the production of one extra gluon the amplitude $\phi $ was
calculated several years ago [49] and in accordance with the above effective
action (272) it can be written in the gauge invariant form

$$
{\phi }_{a_0a_1a_2c}^{{\nu }_0{\nu }_1{\nu }_2+}=g^2\{\Gamma _{a_0a_1a_2c}^{{%
\ \nu }_0{\nu }_1{\nu }_2+}-T_{a_1a_0}^aT_{a_2a}^c\frac{\gamma ^{{\nu }_1{%
\nu } _0\sigma }(k_1,-k_0)\,\Gamma ^{{\nu }_2\sigma +}(k_2,k_2+q)}{%
(k_0+k_1)^2}\, 
$$
$$
-T_{a_2a_0}^aT_{a_1a}^c\frac{\gamma ^{{\nu }_2{\nu }_0\sigma
}(k_2,-k_0)\,\Gamma ^{{\nu }_1\sigma +}(k_1,k_1+q)}{(k_0+k_2)^2} 
$$
\begin{equation}
-T_{a_2a_1}^aT_{a_0a}^c\frac{\gamma ^{{\nu }_2{\nu }_1\sigma
}(k_2,-k_1)\,\Gamma ^{{\nu }_0\sigma +}(k_0,k_0+q)}{(k_1+k_2)^2}\}. 
\end{equation}

The last three terms in the brackets correspond to the Feynman diagram
contributions constructed from the gluon propagator combining the usual
Yang-Mills vertex $\gamma $ and the effective reggeon-gluon-gluon
vertex\thinspace $\Gamma $. The first term is the sum of two terms 
\begin{equation}
\Gamma _{a_0a_1a_2c}^{{\nu }_0{\nu }_1{\nu }_2+}\,=\,\gamma
_{a_0a_1a_2c}^{\nu _0\nu _1\nu _2+}\,+\,\Delta _{a_0a_1a_2c}^{\nu _0\nu
_1\nu _2+}\,\,, 
\end{equation}
where $\gamma $ is the light-cone projection of the usual quadri-linear
Yang-Mills vertex 
$$
\gamma _{a_0a_1a_2c}^{\nu _0\nu _1\nu _2+}=T_{a_1a_0}^aT_{a_2a}^c(\delta
^{\nu _1\nu _2}\delta ^{\nu _0+}-\delta ^{\nu _1+}\delta ^{\nu _0\nu _2}) 
$$
\begin{equation}
+T_{a_2a_0}^aT_{a_1a}^c(\delta ^{\nu _2\nu _1}\delta ^{\nu _0+}-\delta ^{\nu
_2+}\delta ^{\nu _0\nu _1})+T_{a_2a_1}^aT_{a_0a}^c(\delta ^{\nu _2\nu
_0}\delta ^{\nu _1+}-\delta ^{\nu _2+}\delta ^{\nu _1\nu _0}) 
\end{equation}
and $\Delta $ is the induced vertex appearing in the effective action 
(see (275)): 
\begin{equation}
\Delta _{a_0a_1a_2c}^{\nu _0\nu _1\nu
_2+}(k_0^{+},k_1^{+},k_2^{+})\,=-t\,(n^{+})^{\nu _0}(n^{+})^{\nu
_1}(n^{+})^{\nu _2}\{\frac{T_{a_2a_0}^aT_{a_1a}^c}{k_1^{+}\,\,\,k_2^{+}}+ 
\frac{T_{a_2a_1}^aT_{a_0a}^c}{k_0^{+}\,\,\,k_2^{+}}\}\,. 
\end{equation}
Note that due to the Jacobi identity

\begin{equation}
T_{a_2a_0}^aT_{a_1a}^c-T_{a_2a_1}^aT_{a_0a}^c=T_{a_1a_0}^aT_{a_2a}^c 
\end{equation}
and the momentum conservation law 
\begin{equation}
k_0^{+}\,+\,k_1^{+}\,+\,k_2^{+}\,=\,0 
\end{equation}
valid in the quasi-elastic kinematics the tensor $\Delta $ has the Bose
symmetry with respect to the simultaneous transmutation of momenta, colour
and Lorentz indices of the gluons $0,1,2$.\thinspace The above expression
for ${\phi }_{a_0a_1a_2c}^{{\nu }_0{\nu }_1{\nu }_2+}$ is significantly
simplified if we use the light-cone gauge $e^{+}(k_i)=0$ for the gluon
polarization vectors because in this gauge all induced terms disappear.

In a general case $n>2$ for the gauge invariance of $\phi $ one should take
into account apart from the usual Yang-Mills vertices $\gamma $ also an
arbitrary number of the induced vertices $\Delta $ appearing in $j^+$ 
(275) and satisfying the reccurence relation: 
$$
\Delta _{a_0a_1...a_rc}^{\nu _0\nu _1...\nu _r+}(k_0^{+},k_1^{+},...k_r^{+}) 
$$
\begin{equation}
=\frac{(n^{+})^{\nu _r}}{k_r^{+}}\sum\limits_{i=0}^{r-1}T_{a_ra_i}^a\Delta
_{a_0a_1...a_{i-1}aa_{i+1}...a_{r-1}c}^{\nu _0...\nu
_{r-1}+}(k_0^{+},...k_{i-1}^{+},k_i^{+}+k_r^{+},k_{i+1}^{+},...k_{r-1}^{+})
\,. 
\end{equation}
These induced vertices are invariant under arbitrary transmutations of
indices $i$ : 
\begin{equation}
\Delta _{a_{i_0}a_{i_1}...a_{i_r}c}^{\nu _{i_0}\nu _{i_1}...\nu
_{i_r}+}(k_{i_0}^{+},k_{i_1}^{+},...k_{i_r}^{+})=\Delta
_{a_0a_1...a_rc}^{\nu _0\nu _1...\nu _r+}(k_0^{+},k_1^{+},...k_r^{+}) 
\end{equation}
due to the Jacobi identity for the colour group generators $T$ and the
energy-momentum conservation 
\begin{equation}
\sum\limits_{i=0}^rk_i^{+}\,\,=\,\,0. 
\end{equation}
One can calculate easily also amplitudes of the quark production in the
quasi-elastic kinematics because the reggeized and usual gluons interact
with quarks in a similar way.

Let us consider now the multi-gluon production in the central rapidity
region [49, 51]. The following kinematics of the final state
particles is essential: 
the gluons $A^{\prime }$ and $B^{\prime }$ move almost along the
momenta of the initial gluons $A$ and $B$ and there is a group of produced
gluons with a fixed invariant mass in the central 
rapidity region: $y=\frac 12\ln (k^{+}/k^{-}) \sim 0$. The momentum transfers $%
q_1=p_A-p_{A^{\prime }} $ and $q_2=p_{B^{\prime }}-p_B$ in this regime have
the decomposition 
\begin{equation}
q_1=q_{1\perp }+\beta \,p_A\,,\,\,q_2=q_{2\perp }-\alpha \,p_B\, 
\end{equation}
where $\beta $ and $\alpha $ are the Sudakov parameters of the total
momentum $k=\sum\limits_{i=1}^nk_i$ of the produced gluons: 
\begin{equation}
k=k_{\perp }+\beta \,p_A+\alpha \,p_B\,\,,\,\,\kappa =k^2=s\alpha \beta
+(q_1-q_2)_{\perp }^2 
\end{equation}
and $\sqrt{\kappa }$ is their invariant mass which is assumed to be fixed at
high energies: $\kappa \ll s$.
In this kinematical region the production amplitude has the factorized form:

\begin{equation}
A_{d_1d_2....d_nA^{\prime }AB^{\prime }B}^{\nu _1\nu _2...\nu
_n+-}=-\,g\,p_A^{+}\,T_{A^{\prime }A}^{c_1}\,\delta _{\lambda _A\lambda
_{A^{\prime }}}\,\frac 1{t_1}\,\psi _{d_1d_2...d_nc_2c_1}^{\nu _1\nu
_2...\nu _n+-}\,\frac 1{t_2}\,g\,p_B^{-}\,T_{B^{\prime }B}^{c_2}\,\delta
_{\lambda _B\lambda _{B^{\prime }}}. 
\end{equation}

For the simplest case of one gluon emission we have 
\begin{equation}
\psi _{d_1c_2c_1}^{\nu _1+-}=g\,\Gamma _{d_1c_2c_1}^{\nu _1+-}, 
\end{equation}
where $\Gamma $ is the sum of two terms 
\begin{equation}
\Gamma _{d_1c_2c_1}^{\nu _1+-}=\gamma _{d_1c_2c_1}^{\nu _1+-}+\Delta
_{d_1c_2c_1}^{\nu _1+-}. 
\end{equation}
The first term is the contribution from the tri-linear Yang-Mills vertex 
\begin{equation}
\gamma _{d_1c_2c_1}^{\nu _1+-}=T_{c_2c_1}^{d_1}\gamma ^{\nu _1+-}\,,\,\gamma
=2(q_2+q_1)-2k_1^{+}n^{-}+2k_1^{-}n^{+}. 
\end{equation}
The second term is the induced one 
\begin{equation}
\Delta _{d_1c_2c_1}^{\nu _1+-}=T_{c_2c_1}^{d_1}\Delta ^{\nu _1+-}\,,\,\Delta
=-2\,t_1\,\frac{n^{-}}{k_1^{-}}\,+\,2\,t_2\,\frac{n^{+}}{k_1^{+}}\,\,. 
\end{equation}
Due to the relation 
\begin{equation}
\gamma +\Delta =-2\,C, 
\end{equation}
we obtain the known result for the multi-Regge kinematics.

For the more complicated case of the two gluon production in the central
rapidity region the amplitude was calculated also and the quantity $\psi $
is given below [49, 51]: 
$$
\psi _{d_1d_2c_2c_1}^{\nu _1\nu _2+-}=g^2\{\Gamma _{d_1d_2c_2c_1}^{\nu _1\nu
_2+-}-\frac{T_{d_2d_1}^d\gamma ^{\nu _2\nu _1\sigma }(k_2,-k_1)\,\,\Gamma
_{dc_2c_1}^{\sigma +-}(q_2,q_1)}{(k_1+k_2)^2} 
$$
\begin{equation}
-\frac{\Gamma _{d_1dc_1}^{\nu _1\sigma -}(k_1,k_1-q_1)\Gamma _{d_2dc_2}^{\nu
_2\sigma +}(k_2,k_2+q_2)}{(q_1-k_1)^2}-\frac{\Gamma _{d_2dc_1}^{\nu _2\sigma
-}(k_2,k_2-q_1)\Gamma _{d_1dc_2}^{\nu _1\sigma +}(k_1,k_1+q_2)}{(q_1-k_2)^2}
\}. 
\end{equation}

The second term in the brackets describes the production of a pair of gluons
through the decay of the virtual gluon in the direct channel. This
contribution is a product of the effective vertex $\Gamma $, the usual YM
vertex $\gamma $ and the gluon propagator. Analogously, the third and fourth
contributions are products of two effective vertices $\Gamma $ having the
light cone components $\pm $ and of the gluon propagator in the crossing
channels.

The first term in the brackets can be presented as the sum of two terms 
\begin{equation}
\Gamma _{d_1d_2c_2c_1}^{\nu _1\nu _2+-}=\gamma _{d_1d_2c_2c_1}^{\nu _1\nu
_2+-}+\Delta _{d_1d_2c_2c_1}^{\nu _1\nu _2+-}, 
\end{equation}
where the contribution $\gamma $ is the light cone component of the
quadri-linear Yang-Mills vertex 
$$
\gamma _{d_1d_2c_2c_1}^{\nu _1\nu _2+-}=T_{d_1c_1}^dT_{d_2d}^{c_2}(\delta
^{\nu _1\nu _2}\delta ^{-+}-\delta ^{\nu _1+}\delta ^{-\nu _2})\,\,\,+ 
$$
\begin{equation}
+\,\,T_{d_2c_1}^dT_{d_1d}^{c_2}(\delta ^{\nu _2\nu _1}\delta ^{-+}-\delta
^{\nu _2+}\delta ^{-\nu _1})+T_{d_2d_1}^dT_{c_1d}^{c_2}(\delta ^{\nu
_2-}\delta ^{\nu _1+}-\delta ^{\nu _2+}\delta ^{\nu _1-}) 
\end{equation}
and $\Delta $ is the new induced vertex appearing in the effective action
(see (276)):

$$
\Delta _{d_1d_2c_2c_1}^{\nu _1\nu _2+-}=-2\,t_2\,(n^{+})^{\nu
_1}(n^{+})^{\nu _2}\{\frac{T_{d_2c_1}^dT_{d_1d}^{c_2}}{k_1^{+}\,\,k_2^{+}}+ 
\frac{T_{d_2d_1}^dT_{c_1d}^{c_2}}{(-k_1^{+}-k_2^{+})\,k_2^{+}}\}\,\,\,- 
$$
\begin{equation}
-2\,t_1\,(n^{-})^{\nu _1}(n^{-})^{\nu _2}\{\frac{T_{d_1c_2}^dT_{d_2d}^{c_1}}{
k_2^{-}\,\,k_1^{-}}+\frac{T_{d_1d_2}^dT_{c_2d}^{c_1}}{(-k_1^{-}-k_2^{-})
\,k_1^{-}}\}\,. 
\end{equation}
One can verify that $\Delta $ is symmetric under simultaneous transmutations
of $n^{-},t_1,d_1$ and $n^{+},t_2,d_2$.

Note, that in the general case of multi-gluon production one should take
into account the induced vertices with an arbitrary number of external legs
which are expressed in terms of the light cone projections of the vertices
introduced above for the quasi-elastic case 
\begin{equation}
\Gamma _{d_1...d_nc_2c_1}^{\nu _1...\nu _n+-}=\Delta
_{c_1d_1...d_nc_2}^{+\nu _1...\nu _n-}(k_0^{+}k_1^{+}...k_n^{+})+\Delta
_{c_2d_1...d_nc_1}^{+\nu _1...\nu _n-}(k_0^{-}k_1^{-}...k_n^{-})\,, 
\end{equation}
where $k_0^{+},k_0^{-}$ are determined by the momentum conservation 
\begin{equation}
\sum\limits_{i=0}^nk_i^{+}\,=\,\sum\limits_{i=0}^nk_i^{-}\,=\,0. 
\end{equation}

The tensor $\psi _{d_1d_2c_2c_1}^{\alpha _1\alpha _2+-}$ appearing in the
two gluon production amplitude $A_{2\rightarrow 4}$ can be written in the
other form:

\begin{equation}
\psi _{d_1d_2c_2c_1}^{\alpha _1\alpha _2+-}=2\,g^2\left[
T_{d_1d}^{c_1}\,T_{d_2d}^{c_2}\,A^{\alpha _1\alpha
_2}(k_1,k_2)+T_{d_2d}^{c_1}\,T_{d_1d}^{c_2}\,A^{\alpha _2\alpha
_1}(k_2,k_1)\right] , 
\end{equation}
where 
$$
A^{\alpha _1\alpha _2}=-\frac{\Gamma ^{\alpha _1\beta
-}(k_1,k_1-q_1)\,\Gamma ^{\alpha _2\beta +}(k_2,k_2+q_2)}{2\,\,(q_1-k_1)^2}- 
\frac{\gamma ^{\alpha _2\alpha _1\beta }(k_2,-k_1)\,\Gamma ^{+-\beta
}(q_2,q_1)}{2\,\,(k_1+k_2)^2} 
$$
\begin{equation}
+n^{+\alpha _1}n^{-\alpha _2}-g^{\alpha _1\alpha _2}-\frac 12n^{+\alpha
_2}n^{-\alpha _1}+t_1\frac{n^{-\alpha _1}\,n^{-\alpha _2}}{
k_1^{-}(k_1^{-}+k_2^{-})}+t_2\frac{n^{+\alpha _1}\,n^{+\alpha _2}}{
k_2^{+}(k_1^{+}+k_2^{+})}\,. 
\end{equation}

One can verify the following gauge property of $A^{\alpha _1\alpha _2}$: 
\begin{equation}
k_1^{\alpha _1}A^{\alpha _1\alpha _2}\,=\frac 12k_2^{\alpha _2}\left( \frac{
k_1^{-}k_2^{+}}{(k_2+q_2)^2}-\frac{k_1^\beta \Gamma ^{+-\beta }(q_2,q_1)}{
(k_1+k_2)^2}\right) , 
\end{equation}
which implies, that the production amplitude $A_{2\rightarrow 4}$ multiplied
by the gluon polarization vectors $e(k_i)$ is gauge invariant. To calculate
the matrix element of $A^{\alpha _1\alpha _2}$ it is convenient to use the
different light-cone gauges for two produced gluons: 
\begin{equation}
e^{-}(k_1)\,=\,0\,,\,\,e^{+}(k_2)\,=\,0. 
\end{equation}
The corresponding polarization vectors can be parametrized by the
two-dimensional vectors $e_{1\perp },\,e_{2\perp }$ as follows 
\begin{equation}
e(k_1)=e_{1\perp }-\frac{(k_1e_{1\perp })}{k_1^{-}}n^{-},\,\,e(k_2)=e_{2
\perp }-\frac{(k_2e_{2\perp })}{k_2^{+}}n^{+}\, 
\end{equation}
and the matrix element of the tensor $A^{\alpha _1\alpha _2}$ can be
expressed in terms of a new tensor $c^{\alpha _1\alpha _2}$ with pure
transverse components according to the definition: 
\begin{equation}
e_{\alpha _1}^{*}(k_1)\,e_{\alpha _2}^{*}(k_2)\,A^{\alpha _1\alpha _2}\equiv
4\,e_{\alpha _1}^{\perp *}(k_1)\,e_{\alpha _2}^{\perp *}(k_2)\,c^{\alpha
_1\alpha _2}\,. 
\end{equation}
Note, that the corresponding tensor $\widetilde{c}^{\alpha _2\alpha _1}$ for 
$A^{\alpha _2\alpha _1}(k_2,k_1)$ is obtained from $c$ by the transmutation
of gluons $1\leftrightarrow 2$ with the simultaneous multiplication by the
matrices

$$
\Omega ^{\alpha _i\alpha _{i^{\prime }}}=g^{\alpha _i\alpha _{i^{\prime
}}}-2\,\frac{k_{i\perp }^{\alpha _i}k_{i\perp }^{\alpha _{i^{\prime }}}}{
k_{i\perp }^2} 
$$
interchanging two light-cone gauges (316).

Using also the reality condition $k^{+}k^{-}=\left| k_i\right| ^2$ for the
produced gluons $i=1,2$ to exclude the light-cone momenta $k_i^{-}$ one can
present the tensor $c^{\alpha _1\alpha _2}$ only in terms of transverse
momenta $\overrightarrow{k_i}\,,\,\overrightarrow{q_i}$ and the relative
Sudakov parameter $x=\frac{k_1^{+}}{k^{+}+k_2^{+}}\,$ [55]:

$$
c^{\alpha _1\alpha _2}=\frac{\delta ^{\alpha _1\alpha _2}}{2Z} 
\overrightarrow{q_1}^2x(1-x)-xk_1^{\alpha _1}\frac{\overrightarrow{q_2}
^2k_1^{\alpha _2}+\overrightarrow{\Delta }^2Q_1^{\alpha _2}-(1-x)^{-1}( 
\overrightarrow{Q_1}^2-\overrightarrow{k_1}^2)k_2^{\alpha _2}}{\kappa 
\overrightarrow{\,k_1}^2} 
$$
$$
-\frac 1{\overrightarrow{k_1}^2}xk_1^{\alpha _1}Q_1^{\alpha _2}+\frac{\Delta
^{\alpha _1}q_1^{\alpha _2}+\delta ^{\alpha _1\alpha _2}\overrightarrow{q_1}
( \overrightarrow{k_1}+x\overrightarrow{q_2})-(1-x)^{-1}q_1^{\alpha
_1}(k_1^{\alpha _2}-x\Delta ^{\alpha _2})}\kappa 
$$
\begin{equation}
+\frac{Q_1^{\alpha _1}Q_1^{\alpha _2}-\frac 12(1-x)(\overrightarrow{Q_1}^2- 
\overrightarrow{k_1}^2)\delta ^{\alpha _1\alpha _2}}t+x\overrightarrow{q_1}
^2\,\frac{k_2^{\alpha _1}k_2^{\alpha _2}+\delta ^{\alpha _1\alpha _2}(1-x) 
\overrightarrow{k_1}\overrightarrow{\Delta }}{\kappa Z}\,, 
\end{equation}
where $\delta ^{\alpha _1\alpha _2}=-g_{\perp }^{\alpha _1\alpha _2}$ is the
Kroniker tensor ($\delta ^{11}=\delta ^{22}=1$) and

$$
\,\overrightarrow{\Delta }\equiv \overrightarrow{q_1}-\overrightarrow{q_2}= 
\overrightarrow{k_1}+\overrightarrow{k_2}\,,\,\overrightarrow{Q_1}= 
\overrightarrow{q_1}-\overrightarrow{k_1}\,,\,\kappa =\frac{(\overrightarrow{
k_1}-x\overrightarrow{\Delta })^2}{x(1-x)}\,, 
$$
\begin{equation}
Z=-x(1-x)(\overrightarrow{\Delta }^2+\kappa )\,,\,\,t=-\frac{( 
\overrightarrow{k_1}-x\overrightarrow{q_1})^2+x(1-x)\overrightarrow{q_1}^2}
x\,. 
\end{equation}

Analogously to the gluon case, the amplitude of the production of the quark
and anti-quark with their momenta $k_1$ and $k_2$ in the central rapidity
region also can be written as a sum of two terms being the matrices in the
spin and colour spaces [55]: 
\begin{equation}
\psi _{c_2c_1}=-g^2\left(
t^{c_1}t^{c_2}b(k_1,k_2)-t^{c_2}t^{c_1}b^T(k_2,k_1)\right) \,. 
\end{equation}
Here $t^c$ are the colour group generators in the fundamental representation
and the expressions for $b(k_1,k_2)$ and $b^T(k_1,k_2)$ are constructed
according to the Feynman rules including the effective vertices: 
\begin{equation}
b(k_1,k_2)\,=\,\gamma ^{-}\,\frac{\hat q_1-\hat k_1}{(q_1-k_1)^2}\,\gamma
^{+}-\frac{\gamma ^\beta \Gamma ^{+-\beta }(q_2,q_1)}{(k_1+k_2)^2}\,, 
\end{equation}
\begin{equation}
b^T(k_1,k_2)\,=\,\gamma ^{+}\,\frac{\hat q_1-\hat k_2}{(q_1-k_2)^2}\,\gamma
^{-}-\frac{\gamma ^\beta \Gamma ^{+-\beta }(q_2,q_1)}{(k_1+k_2)^2}\,. 
\end{equation}
One can calculate the matrix elements of these expressions between the
spinors $u^{(\pm )}(k_1)$ and $v^{(\pm )}(k_2)$ describing the quark and
anti-quark with helicities $\lambda =\pm \frac 12$:

$$
\overline{u}^{(+)}(k_1)b(k_1,k_2)v^{(-)}(k_2)=\frac
12c_{+-}^{*}(k_1)d_{-+}(k_2)\,b^{(+-)}(k_1,k_2)\,, 
$$
$$
\overline{u}^{(-)}(k_1)b^T(k_2,k_1)v^{(-)}(k_2)=\frac
12c_{-+}^{*}(k_1)d_{+-}(k_2)\,b^{(-+)}(k_2,k_1)\,, 
$$
$$
\overline{u}^{(-)}(k_1)b(k_1,k_2)v^{(+)}(k_2)=\frac
12c_{++}^{*}(k_1)d_{--}(k_2)\,b^{(-+)}(k_1,k_2)\,, 
$$
\begin{equation}
\overline{u}^{(-)}(k_1)b^T(k_2,k_1)v^{(+)}(k_2)=\frac
12c_{--}^{*}(k_1)d_{++}(k_2)\,b^{(+-)}(k_2,k_1)\,. 
\end{equation}
The introduced complex wave functions $c$ and $b$ satisfy the
relations

$$
c_{+-}(k_1)=-\frac{k_1^{-}}{k_1^{*}}c_{-+}=-\frac{k_1}{k_1^{+}}
c_{-+}\,,\,d_{+-}(k_2)=-\frac{k_2^{-}}{k_2^{*}}d_{-+}=-\frac{k_2}{k_2^{+}}
d_{-+}\,, 
$$
\begin{equation}
c_{--}(k_1)=-\frac{k_1}{k_1^{-}}c_{++}=-\frac{k_1^{+}}{k_1^{*}}
c_{++}\,,\,d_{--}(k_2)=-\frac{k_2}{k_2^{-}}d_{++}=-\frac{k_2^{+}}{k_2^{*}}
d_{++}\,, 
\end{equation}
and are normalized as follows

\begin{equation}
\mid c_{\pm i}(k_1)\mid ^2=2k_1^{\mp }\,,\,\mid d_{\pm i}(k_2)\mid
^2=2k_2^{\mp }\,. 
\end{equation}
in accordance with the usual normalization condition for the spinors:

\begin{equation}
\overline{u}(k_1)\gamma ^\alpha u(k_1)=2k_1^\alpha \,,\,\overline{v}
(k_2)\gamma ^\alpha v(k_2)=2k_2^\alpha \,. 
\end{equation}
The amplitudes $b$ can be written in the form 
\begin{equation}
b^{(+-)}(k_1,k_2)=(b^{(-+)}(k_1,k_2))^{*}=-4\,\frac{q_1-k_1}{(q_1-k_1)^2}- 
\frac{j^\beta \Gamma ^{+-\beta }(q_2,q_1)}{(k_1+k_2)^2}\,, 
\end{equation}
where the quark current $j$ is given below 
\begin{equation}
j=n+n^{*}\,\frac{k_2^{-}}{k_2^{*}}\,\frac{k_1}{k_1^{-}}-n^{-}\,\frac{k_1}{
k_1^{-}}-n^{+}\,\frac{k_2^{-}}{k_2^{*}}\,. 
\end{equation}
Due to the relations 
$$
n^\alpha k^\alpha =k,\,n^{*\alpha }k^\alpha =k^{*}\,,\,n^{\pm \alpha
}k^\alpha =k^{\pm } 
$$
this current is conserved: 
\begin{equation}
(k_1^\beta +k_2^\beta )\,j^\beta =0\,. 
\end{equation}

\section{Next-to-leading corrections to the BFKL pomeron}

\subsection{Corrections from the gluon and quark pair production}

The imaginary part of the elastic scattering amplitude calculated with the
use of the $s$-channel unitarity condition through the squared production
amplitude $A_{2\rightarrow 4}$ (301, 313) 
contains the infrared divergences at small $%
k_{i\perp }^2$ and $\kappa $. To avoid such divergencies the dimensional
regularization

\begin{equation}
\frac{d^4k}{(2\pi )^4}\,\longrightarrow \mu ^{4-D}\frac{d^Dk}{(2\pi )^D} 
\end{equation}
is used in the gauge theories (where $\mu $ is the normalization point). It
is important, that in the $D$-dimensional space the gluon has $D-2$ degrees
of freedom.

The total cross-section of the gluon production is proportional to the
integral from the squared amplitude $\psi _{d_1d_2c_2c_1}^{\alpha _1\alpha
_2+-}$ summed over all indices. It is expressed in terms of the bilinear
combinations of the tensor $c^{\alpha _1\alpha _2}$ with transverse
components [55]:

\begin{equation}
R\equiv (c^{\alpha _1\alpha _2}(k_1,k_2))^2+(c^{\alpha _1\alpha
_2}(k_2,k_1))^2+c^{\alpha _1\alpha _2}(k_1,k_2)\,
c^{\alpha _2^{\prime }\alpha
_1^{\prime }}(k_2,k_1)\,\Omega ^{\alpha _1\alpha _1^{\prime }}\,\Omega
^{\alpha _2\alpha _2^{\prime }}, 
\end{equation}
where the matrix $\Omega ^{\alpha _i\alpha _{i^{\prime }}}$ interchanges the
left and right gauges:

$$
\Omega ^{\alpha _i\alpha _{i^{\prime }}}=g_{\perp }^{\alpha _i\alpha
_{i^{\prime }}}-2\frac{k_{i\perp }^{\alpha _i}k_{i\perp }^{\alpha
_{i^{\prime }}}}{k_{i\perp }^2}\,. 
$$
In expression (332) we took into account that the colour factor for the 
interference term
is two times smaller than one for the direct contributions. The generalized
BFKL equation describing the pomeron
as a compound state of two reggeized gluons is valid also in the 
next-to-leading approximation. The equation for the total
cross-section of the reggeon scattering 
can be written in the
integral form as follows 
\begin{equation}
\sigma (\overrightarrow{q_1},\,q_1^{+})=\sigma _0(\overrightarrow{q_1}
,q_1^{+})+\int \frac{d\,q_2^{+}}{q_2^{+}}\,\mu ^{4-D}\int d^2\overrightarrow{
q_2}\,K_\delta (\overrightarrow{q_1},\overrightarrow{q_2})\sigma ( 
\overrightarrow{q_2},\,q_2^{+})\,. 
\end{equation}
Here the integration region for the longitudinal momentum $q_2^{+}$ is
restricted from above by the value proportional to $q_1^{+}$: 
\begin{equation}
q_2^{+}\,<\,\delta \,q_1^{+}. 
\end{equation}
The intermediate infinitesimal parameter $\delta >0$ is introduced instead
of the above cut-off $\eta $ to arrange the particles in the groups with
strongly different rapidities. The integral kernel $K_\delta ( 
\overrightarrow{q_1},\overrightarrow{q_2})$ takes into account the
interactions among the particles with approximately equal rapidities. The 
kernel $ K_\delta $ can be calculated in the perturbation theory: 
\begin{equation}
K_\delta (\overrightarrow{q_1},\overrightarrow{q_2})=\sum_{r=1}^{\infty}
\left( \frac{g^2}{2(2\pi )^{D-1}}\right) ^rK_\delta ^{(r)}(\overrightarrow{
q_1},\overrightarrow{q_2}). 
\end{equation}
We remind, that the real contribution to the kernel in the leading
logarithmic approximation is proportional to the square of the effective
vertex: 
\begin{equation}
K_{real}^{(1)}(\overrightarrow{q_1},\overrightarrow{q_2})=-\frac{N_c}{4\, 
\overrightarrow{q_1}^2\,\overrightarrow{q_2}^2}\left( \Gamma ^{+-\beta
}(q_2,q_1)\right) ^2=N_c\,\frac 4{(\overrightarrow{q_1}-\overrightarrow{q_2}
)^2}. 
\end{equation}
The corresponding virtual contribution is determined by the gluon Regge
trajectory $\omega (-\overrightarrow{q}_1^2)$.

The next-to-leading term in $K_\delta $ related with the two gluon
production is given below [55] 
\begin{equation}
K_{gluons}^{(2)}=\int d\kappa \,\frac{d^Dk_1}{\mu ^{D-4}}\,\delta
(k_1^2)\,\delta (k_2^2)\,\frac{16N_c^2R}{\overrightarrow{q_1}^2 
\overrightarrow{q}_2^2}=\frac{16\,N_c^2}{2\overrightarrow{q_1}^2 
\overrightarrow{q_2}^2}\int_\delta ^{1-\delta }\frac{dx}{x(1-x)}\int \frac{
d^{D-2}\overrightarrow{k_1}}{\mu ^{D-4}}\,R\,. 
\end{equation}

The limits in the integral over $x$ correspond to a restriction from above
for the invariant mass $\sqrt{\kappa }$ of the produced gluons. In the
solution of the generalized BFKL equation the dependence from $\delta $
should disappear.

For the physical value $D=4$ of the space-time dimensions one can express $R$
as the following sum:

\begin{equation}
R=R(+-)+R(++) 
\end{equation}
where $R(+-)$ and $R(++)$ are the contributions from the production of two
gluons with the same and opposite helicities correspondingly. 
$$
R(+-)=\frac 12\left( \mid c^{+-}(k_1,k_2)\mid ^2+\mid c^{+-}(k_2,k_1)\mid
^2+Re\,c^{+-}(k_1,k_2)\,c^{-+}(k_2,k_1)\frac{k_1^{*}}{k_1}\frac{k_2}{k_2^{*}}
\right) \,, 
$$
\begin{equation}
R(++)=\frac 12\left( \mid c^{++}(k_1,k_2)\mid ^2+\mid c^{++}(k_2,k_1)\mid
^2+Re\,c^{++}(k_1,k_2)\,c^{++}(k_2,k_1)\frac{k_1^{*}}{k_1}\frac{k_2^{*}}{k_2}
\right) . 
\end{equation}
Here we have used the following relation between the polarization vectors 
(317) in
the right and left gauges $\left[42\right] $: 
\begin{equation}
e_{\perp }^r(k)=-(e_{\perp }^l(k))^{*}\frac k{k^{*}} 
\end{equation}
to express amplitudes (319) in terms of two complex functions $%
c^{+-}(k_1,k_2)$ and $c^{++}(k_1,k_2)$ given below:

$$
c^{+-}(k_1,k_2)\equiv c^{11}+ic^{21}-ic^{12}+c^{22}=\overline{c^{-+}}
(k_1,k_2)=-x\,\frac{q_2\,q_1^{*}}{(k_1-x\,\Delta )\,k_1^{*}}\,, 
$$

$$
c^{++}(k_1,k_2)=c^{11}+ic^{21}+ic^{12}-c^{22}=\overline{c^{--}}(k_1,k_2) 
$$
$$
=-\frac{x\,(Q_1)^2}{\left( (\overrightarrow{k_1}-x\overrightarrow{q_1}
)^2+x(1-x)\overrightarrow{q_1}^2\right) }+\frac{x\,\overrightarrow{q_1}
^2(k_2)^2}{\overrightarrow{\Delta }^2\left( (\overrightarrow{k_1}-x 
\overrightarrow{\Delta })^2+x(1-x)\overrightarrow{\Delta }^2\right) } 
$$
\begin{equation}
-\frac{x(1-x)q_1k_2q_2^{*}}{\Delta ^{*}\,(k_1-x\Delta )k_1^{*}}-\frac{
xq_1^{*}k_1q_2}{\overrightarrow{\Delta }^2(k_1^{*}-x\Delta ^{*})}+\frac{
xq_2^{*}Q_1}{\Delta ^{*}\,k_1^{*}}\,. 
\end{equation}
These expressions were obtained independently also in Ref. [56].

In the Regge regime of small $1-x$ and fixed $k_i,q_i$ amplitudes $c^{\alpha
_1\alpha _2}$ are simplified as follows 
\begin{equation}
c^{+-}(k_1,k_2)\longrightarrow \frac{q_1^{*}\,q_2}{k_1^{*}\,k_2}
\,,\,\,c^{++}(k_1,k_2)\longrightarrow \frac{q_1^{*}}{k_1^{*}}\,\frac{q_1-k_1 
}{q_1^{*}-k_1^{*}}\,\frac{q_2^{*}}{k_2^{*}} 
\end{equation}
and are proportional to the product of the effective vertices $\Gamma
^{+-\beta }$ in the light-cone gauge $\left[ 42\right] $. For $x\rightarrow
0 $ the amplitudes $c^{+-}(k_1,k_2)$ and $c^{++}(k_1,k_2)$ vanish, but for
simultaneously small $k_1$ or $k_1-x\Delta $ one obtains a nonzero
integral contribution because in this region there are poles:

\begin{equation}
c^{++}(k_1,k_2)\rightarrow \frac{q_1q_2^{*}\Delta }{q_1^{*}q_2\Delta ^{*}}
\,c^{+-}(k_1,k_2)\rightarrow -\frac{xq_1q_2^{*}\Delta }{\Delta
^{*}(k_1-x\Delta )k_1^{*}}\,. 
\end{equation}
For large $k_1$ and fixed $q_i\,,\,x$ we obtain 
\begin{equation}
c^{+-}(k_1,k_2)\longrightarrow -x\,\frac{q_1^{*}q_2}{\overrightarrow{k_1}^2}
\,,\,\,c^{++}(k_1,k_2)\longrightarrow -x(1-x)^2\,\frac{q_1q_2}{
\overrightarrow{k_1}^2}-x^3\frac{q_1^{*}q_2^{*}k_1}{(k_1^{*})^3} 
\end{equation}
and therefore the integrals for the cross section of the gluon production do
not contain any ultraviolet divergency.

As it is seen from above formulas, $c^{\alpha _1\alpha _2}$ has infrared
poles at small $\overrightarrow{k_1}$: 
\begin{equation}
c^{+-}(k_1,k_2)\rightarrow \,\frac{q_1^{*}q_2}{\Delta \,k_1^{*}}
\,,\,c^{++}(k_1,k_2)\rightarrow \,\frac{q_1q_2{*}}{\Delta ^{*}\,k_1^{*}}\,\, 
\end{equation}
and at small $\overrightarrow{k_1}-x\,\overrightarrow{\Delta }$: 
\begin{equation}
c^{+-}\rightarrow -\,\frac{q_1^{*}q_2}{\Delta ^{*}(k_1-x\Delta )}
\,,\,\,c^{++}\rightarrow -(1-x)^2\,\frac{q_1\Delta q_2^{*}}{(\Delta
^{*})^2(k_1-x\Delta )}-x^2\frac{q_1^{*}q_2}{\Delta ^{*}(k_1^{*}-x\Delta
^{*}) }\,. 
\end{equation}

Let us return now to the quark-antiquark production. The total cross-section
of this process in accordance with the normalization condition for spinors
is proportional to the integral from the expression: 
\begin{equation}
K_{quarks}^{(2)}=\int_0^1\frac{dx\,\mu ^{4-D}}{x(1-x)}\int d^{D-2} 
\overrightarrow{k_1}\frac{8\,L}{\overrightarrow{q_1}^2\overrightarrow{q_2}^2}
\end{equation}
where we put $\delta =0$ because the integral in $x$ is convergent at $x=0$
and $x=1$.

The expression for $L$ with corresponding colour factors for $D=4$ is given
below: 
\begin{equation}
L=\frac{N_c^2-1}{4N_c}\left( \frac{1-x}x\mid c(k_1,k_2)\mid ^2+\frac
x{1-x}\mid c(k_2,k_1)\mid ^2\right) +\frac 1{2N_c}Re\;c(k_1,k_2)\,c(k_2,k_1) 
\end{equation}
where two equal contributions from two different helicity states of the
quark-anti-quark pair were taken into account. The amplitude $c$ is
determined by the equation  (see (328)):
\begin{equation}
b^{(+-)}(k_1,k_2)=\frac 4{k_1^{*}}\,c(k_1,k_2)\,. 
\end{equation}
and is given below: 
$$
\frac 1x\,c(k_1,k_2)=\frac{(1-x)q_1k_1^{*}-xq_1^{*}k_1+x\overrightarrow{q_1}
^2}{(\overrightarrow{k_1}-x\overrightarrow{q_1})^2+x(1-x)\overrightarrow{q_1}
^2} 
$$
\begin{equation}
-\frac{\overrightarrow{q_1}^2}{\overrightarrow{\Delta }^2}\,\frac{
(1-x)\Delta k_1^{*}-x\Delta ^{*}k_1+x\overrightarrow{\Delta }^2}{( 
\overrightarrow{k_1}-x\overrightarrow{\Delta })^2+x(1-x)\overrightarrow{
\Delta }^2}+\frac{(1-x)q_1q_2^{*}}{\Delta ^{*}(k_1-x\Delta )}-\frac{
xq_1^{*}q_2}{\Delta (k_1^{*}-x\Delta ^{*})}\,. 
\end{equation}

Note, that in the limit of large $N_c$ the quark contribution is smaller
than gluon one and the interference term is suppressed by the factor $%
1/N_c^2 $ in comparison with the direct contribution.

\subsection{Infrared and collinear divergencies}

The gluon and quark production cross-sections contain infrared divergencies
which should be cancelled with the virtual corrections to the multi-Regge
processes. For example from the products of amplitudes $c^{+-}$ we can
calculate $R(+-)$ for $D=4$ and obtain the following expression: 
\begin{equation}
R(+-)=\frac{\overrightarrow{q_1}^2\overrightarrow{q_2}^2}4\left( \frac 1{
\overrightarrow{k_1}^2(\overrightarrow{k_1}-\overrightarrow{\Delta })^2}+ 
\frac{x^2}{\overrightarrow{k_1}^2(\overrightarrow{k_1}-x\overrightarrow{
\Delta })^2}+\frac{(1-x)^2}{(\overrightarrow{k_1}-\overrightarrow{\Delta }
)^2(\overrightarrow{k_1}-x\overrightarrow{\Delta })^2}\right) 
\end{equation}
from which the presence of infrared and collinear singularities is obvious.

As for $R(++)$, we can write it as the sum of singular and regular terms
[55] 
\begin{equation}
R(++)=R_{sing}(++)+R_{reg}(++) 
\end{equation}
where for $R_{sing}(++)$ the following expression is chosen: 
\begin{equation}
R_{sing}(++)=R(+-)+\frac 14\left( r(k_1,x)+r(\Delta -k_1,1-x)\right) . 
\end{equation}
and 
\begin{equation}
r(k_1,x)=\frac{\overrightarrow{q_1}^2\overrightarrow{q_2}^2}{\overrightarrow{
\Delta }^2}\left( x(1-x)-2\right) \frac{2x^2(1-x)\overrightarrow{k_1} 
\overrightarrow{\Delta }}{\overrightarrow{k_1}^2(\overrightarrow{k_1}-x 
\overrightarrow{\Delta })^2}+2\,Re\,\frac{x^3(1-x)^2q_1^2q_2^{*2}\Delta }{
\Delta ^{*2}(k_1-x\Delta )^2k_1}\;. 
\end{equation}
Therefore for the total sum $R_{sing}=R(+-)+R_{sing}(++)$ we have

\begin{equation}
R_{sing}=2\;R(+-)+\frac 14\left( r(k_1,x)+r(\Delta -k_1,1-x)\right) . 
\end{equation}
The term $R_{sing}$ contains all infrared and Regge singularities of $R$ and
is generalized to the case of arbitrary $D$. It decreases rapidly at large $%
k_1$.

One can obtain the following contribution from $R_{sing}$ after its
dimensional regularization (and taking into account $D-4$ non-physical
polarizations for each gluon): 
$$
\frac{\mu ^{4-D}}\pi \int_\delta ^{1-\delta }\frac{d\,x}{x(1-x)}\int
d^{D-2}k_1R_{sing}=\frac 16\,Re\,\frac{q_1^2q_2^{*2}}{\overrightarrow{\Delta 
}^2}+\frac{\overrightarrow{q_1}^2\overrightarrow{q_2}^2}{\overrightarrow{
\Delta }^2}\left( \frac{67}{18}-4\frac{\pi ^2}6\right) 
$$
\begin{equation}
+\frac{\overrightarrow{q_1}^2\overrightarrow{q_2}^2}{\overrightarrow{\Delta }
^2}\frac{2^{4-D}\pi ^{\frac{D-1}2}}{\Gamma (\frac{D-3}2)\sin \,(\pi \frac{%
D-4 }2)}\left| \frac{\overrightarrow{\Delta }^2}{\mu ^2}\right| ^{\frac{D-4}
2}\left( \frac 2{D-4}+4\ln \frac 1\delta \,-\frac{11}6\right) , 
\end{equation}
where it is implied, that the terms of the order of value of $D-4$ should be
omitted. The infrared divergencies at $D\rightarrow 4$ in the above formulas
should be cancelled with the contribution from one-loop corrections to the
Reggeon-Reggeon-particle vertex which will be considered below.

It is important to investigate the region of small $\Delta $: 
\begin{equation}
\left| \Delta \right| \ll \left| q_1\right| \simeq \left| q_2\right| \,. 
\end{equation}
Here the essential integration region corresponds to the soft gluon
transverse momenta: 
\begin{equation}
k_1\sim k_2\sim \Delta \ll q 
\end{equation}
where $q$ means $q_1$ or $q_2$.

The total contribution of the singular and regular terms in the soft region $%
\Delta \rightarrow 0$ can be calculated [55]:

$$
\frac{\mu ^{4-D}}\pi \int_\delta ^{1-\delta }\frac{d\,x}{x(1-x)}\int
d^{D-2}k_1\,R=\frac{\overrightarrow{q}^4}{\overrightarrow{\Delta }^2}\left( 
\frac{\pi \overrightarrow{\Delta }^2}{\mu ^2}\right) ^{\frac{D-4}2}\Gamma
(3-\frac D2)\frac{\Gamma ^2({\frac D2}-1)}{\Gamma (D-2)}\times 
$$
\begin{equation}
\left[ \frac 4{(D-4)^2}+\frac{D-2}{2(D-4)(D-1)}+\frac{2(D-3)}{D-4}\left(
4\ln \frac 1\delta +\psi (1)+\psi ({\frac D2}-1)-2\psi (D-3)\right) \right]
. 
\end{equation}
In this expression three first terms of the expansion in $\epsilon =D-4$
including one of the order of $\epsilon $ are correct. Such accuracy of
calculations is needed because the integration over $\Delta $ in the
generalized BFKL equation leads to the infrared divergency at $\Delta \,=\,0$
for $D\rightarrow 4$. It means, that to obtain the final result after
cancellations of the infrared divergencies one should know the real and
virtual corrections to the BFKL kernel up to terms $\sim \epsilon $.

Note, that it is possible to modify the definition of the singular and
regular parts of $c^{++}(k_1,k_2)$: 
$$
c^{++}(k_1,k_2)=\widetilde{c}_{sing}^{++}(k_1,k_2)+\widetilde{c}
_{reg}^{++}(k_1,k_2)\,, 
$$
$$
\frac 1x\,\widetilde{c}_{sing}^{++}(k_1,k_2)=\frac{q_1^{*}q_2k_1(k_1-x\Delta
)-(2-x)q_1q_2^{*}\Delta k_1+q_1q_2^{*}\,\Delta ^2}{\overrightarrow{\Delta }
^2\left( (\overrightarrow{k_1}-x\overrightarrow{\Delta })^2+x(1-x) 
\overrightarrow{\Delta }^2\right) } 
$$
$$
-\frac{q_1^{*}q_2k_1}{\overrightarrow{\Delta }^2(k_1^{*}-x\Delta ^{*})}
+(2-x) \frac{q_1q_2^{*}}{\Delta ^{*}k_1^{*}}-(1-x)^2\frac{\Delta q_1q_2^{*}}{
\Delta ^{*}(k_1-x\Delta )k_1^{*}}\,, 
$$
$$
\frac 1x\,\widetilde{c}_{reg}^{++}(k_1,k_2)=-\,\frac{Q_1^2}{(\overrightarrow{
k_1}-x\overrightarrow{q_1})^2+x(1-x)\overrightarrow{q_1}^2} 
$$
\begin{equation}
+\frac{q_1^{*}k_1(k_1-x\Delta )-(2-x)q_1\Delta ^{*}k_1+q_1\overrightarrow{
\Delta }^2}{\Delta ^{*}\left( (\overrightarrow{k_1}-x\overrightarrow{\Delta }
)^2+x(1-x)\overrightarrow{\Delta }^2\right) }-\frac{q_2^{*}k_1}{\Delta
^{*}k_1^{*}} 
\end{equation}
in such way to include in the comparatively simple expression for $
\widetilde{c}_{sing}^{++}(k_1,k_2)$ all singular terms without the loss of
its good behaviour at large $k_1$. In the Regge limit $x\rightarrow 1$ and
fixed $k_i$ we obtain

\begin{equation}
\widetilde{c}_{sing}^{++}\rightarrow \frac{q_1q_2^{*}}{k_1^{*}k_2^{*}} 
\end{equation}
and therefore in this region the contributions of the exact amplitude $%
c^{++} $ and approximate one $\widetilde{c}_{sing}^{++}$ to the differential
cross-section also coincide. However, as a result of the singularity at $%
k_1=x\Delta $ after integration over $k_1$ these contributions to the total
cross-section in the Regge limit $x\rightarrow 1$ turn out to be different.

We return now to the quark production amplitude $c(k_1,k_2)$ and present it
in the form analogous to the gluon case: 
\begin{equation}
c(k_1,k_2)=c_{sing}(k_1,k_2)+c_{reg}(k_1,k_2)\,, 
\end{equation}
where the singular and regular terms are chosen as follows

$$
\frac 1xc_{sing}(k_1,k_2)= \frac{(1-x)q_1q_2^{*}}{\Delta ^{*}(k_1-x\Delta )}
- \frac{xq_1^{*}q_2}{\Delta (k_1^{*}-x\Delta ^{*})} -\frac{(1-x)\Delta
q_1q_2^{*}k_1^{*}-x\Delta ^{*}q_1^{*}q_2k_1+x\overrightarrow{\Delta }
^2q_1^{*}q_2}{\overrightarrow{\Delta }^2\left( (\overrightarrow{k_1}-x 
\overrightarrow{\Delta })^2+x(1-x) \overrightarrow{\Delta }^2\right) }, 
$$
\begin{equation}
\frac 1xc_{reg}(k_1,k_2)=\frac{(1-x)q_1k_1^{*}-xq_1^{*}k_1+x\overrightarrow{
q_1}^2}{\left( (\overrightarrow{k_1}-x\overrightarrow{q_1})^2+x(1-x) 
\overrightarrow{q_1}^2\right) }-\frac{(1-x)q_1k_1^{*}-xq_1^{*}k_1+xq_1^{*}
\Delta }{\left( (\overrightarrow{k_1}-x\overrightarrow{\Delta })^2+x(1-x) 
\overrightarrow{\Delta }^2\right) }\,. 
\end{equation}
The term $c_{reg}$ does not lead to any divergency and gives a regular
contribution to the BFKL kernel at the soft region $\Delta \rightarrow 0$.

We obtain the following contributions for the bilinear combinations of $%
c_{sing}$ taking into account the dimensional regularization of the
singularity at $\overrightarrow{k_1}\rightarrow x\overrightarrow{\Delta }$
[55]:

$$
\int_0^1\frac{d\,x}{x^2}\int \frac{d^{D-2}\,k_1}{\mu ^{D-4}\pi }\left|
c_{sing}(k_1,k_2)\right| ^2=\int_0^1\frac{d\,x\,\,}{x(1-x)}\int \frac{
d^{D-2}\,k_1}{\mu ^{D-4}\pi }\,Re\,c_{sing}(k_1,k_2)c_{sing}(k_2,k_1) 
$$
\begin{equation}
=\frac 1{\overrightarrow{\Delta }^2}\left( \frac{\pi \overrightarrow{\Delta }
^2}{\mu ^2}\right) ^{\frac D2-2}\Gamma (3-\frac D2)\frac{\Gamma (\frac
D2)\Gamma (\frac D2)}{\Gamma (D)}\,4\,\left( \frac{6-D}{D-4}\, 
\overrightarrow{q_1}^2\overrightarrow{q_2}^2+\left( \overrightarrow{q_1}, 
\overrightarrow{q_2}\right) ^2\right) \,. 
\end{equation}

The integration over $\Delta $ in the generalized BFKL equation leads to the
infrared divergency at $\Delta =0$ for $D\rightarrow 4$: 
\begin{equation}
-16N_cn_f\pi ^{\frac D2-1}\left( \frac{g^2\,\mu ^{4-D}}{2(2\pi )^{D-1}}
\right) ^2\Gamma (2-\frac D2)\frac{\Gamma (\frac D2)\Gamma (\frac D2)}{
\Gamma (D)}\int d^{D-2}\Delta \left( \overrightarrow{\Delta }^2\right)
^{\frac D2-3}. 
\end{equation}
The quark correction to the RRP vertex expressed in terms of the bare
coupling constant does not give such singular contribution at small $\Delta $
. This divergency is cancelled with the doubled quark contribution to the
gluon Regge trajectory (see below).

\subsection{One-loop corrections to reggeon-particle-particle vertices}

To find next-to-leading corrections to the BFKL equation one should
calculate also the analogous 
corrections to the gluon production amplitudes
in the multi-Regge kinematics [41, 49]. These amplitudes in LLA are
constructed from the RPP and RRP vertices combined by the reggeized gluon
propagators expressed in terms of the Regge trajectory. In the
next-to-leading approximation this structure of the production amplitude
remains and therefore one should calculate one-loop contributions to the
reggeon-particle vertices and two-loop corrections to the gluon Regge
trajectory.

To begin with, we consider the elastic amplitude for the gluon-gluon
scattering. In the Regge model it can be written in the form:

\begin{equation}
M=T_{A^{\prime }A}^c\,\Gamma _{\lambda _{A^{\prime }}\lambda _A}(t)\,\frac
st\,\left[ \left( \frac s{-t}\right) ^{\omega (t)}+\left( \frac{-s}{-t}
\right) ^{\omega (t)}\right] \,T_{B^{\prime }B}^c\,\Gamma _{\lambda
_{B^{\prime }}\lambda _B}(t)\,,\,\,t=q^2, 
\end{equation}
where\thinspace $\Gamma _{\lambda _{A^{\prime }}\lambda _A}(t)$ due to the
parity conservation can be presented as the sum of two terms:

\begin{equation}
\Gamma _{\lambda _{A^{\prime }}\lambda _A}(t)=\Gamma ^{(1)}(t)\,\delta
_{\lambda _{A^{\prime }},\lambda _A}+\Gamma ^{(2)}(t)\,\delta _{\lambda
_{A^{\prime }},-\lambda _A} 
\end{equation}
in the helicity representation or

\begin{equation}
\Gamma _{\alpha ^{\prime }\alpha }(t)=\Gamma ^{(1)}(t)\,(-\delta _{\alpha
^{\prime }\alpha }^{\perp \perp })+\Gamma ^{(2)}(t)\,(\delta _{\alpha
^{\prime }\alpha }^{\perp \perp }-2\frac{q_{\alpha ^{\prime }}^{\perp
}q_\alpha ^{\perp }}{q^2}) 
\end{equation}
in the tensor representation. Here 
\begin{equation}
-\delta _{\alpha ^{\prime }\alpha }^{\perp \perp }=\sum_{\lambda =\pm
}\,e_{\alpha ^{\prime }}^\lambda (p_{A^{\prime }})\,e_\alpha ^{\lambda
*}(p_A)\,,\,\,e_\alpha \,p_B^\alpha =0. 
\end{equation}
is the projector to the physical gluon states in the light-cone gauge. The
vectors $q_\alpha^\perp$ and $q_{\alpha^{\prime}}^\perp$ are orthogonal to
the momenta $k,\, p_B$ and $k^{\prime},\,p_B$ correspondingly . For finding
the next-to-leading correction to $\Gamma ^{(i)}$ it is enough to calculate
in the one-loop approximation the scattering amplitude at high energies. The 
$t$-channel unitarity is convenient for this purpose. In the framework of
this approach the $t$-channel imaginary part of the gluon-gluon scattering
amplitude in the pure gluodynamics is obtained as a product of two Born
amplitudes with summing over the colour and Lorentz indices and integrating
over momenta $-k,\,q+k$ of the intermediate gluons in the octet state. Each
of these amplitudes is presented as a sum of two terms after projecting to
the octet state. For example,

\begin{equation}
A_{AA^{\prime }CC^{\prime }}^{\alpha \alpha ^{\prime }\gamma \gamma ^{\prime
}}(k,q+k,p_A)=g^2(-T_{A^{\prime }A}^d\,T_{B^{\prime }B}^d)\,\left(
A^{(a)}+A^{(n)}\right) _{\alpha \alpha ^{\prime }}^{\gamma \gamma ^{\prime
}}\,, 
\end{equation}
where $A^{(a)}$ is the factorized contribution containing the correct
asymptotic behaviour of the total amplitude:

\begin{equation}
A_{\alpha \alpha ^{\prime }}^{(a)\gamma \gamma ^{\prime }}=\delta _{\alpha
\alpha ^{\prime }}^{\perp \perp }\,\,A^{\gamma \gamma ^{\prime }}(k,k+q,p_A) 
\end{equation}
and $A^{(n)}$ is the non-asymptotic term which is small in the Regge
kinematics. The factor $A^{\gamma \gamma ^{\prime }}(k,k^{\prime },p)$ is
proportional to the effective RPP vertex:

$$
A^{\gamma \gamma ^{\prime }}(k,k^{\prime },p)=-2\,\delta ^{\gamma \gamma
^{\prime }}\,\frac{kp+k^{\prime }p}t+2\left( \frac 1{(k+p)^2}+\frac
2t\right) p^\gamma k^{\gamma ^{\prime }} 
$$
\begin{equation}
+2\left( \frac 1{(k^{\prime }-p)^2}+\frac 2t\right) k^{\prime \,\gamma
}p^{\gamma ^{\prime }}+2\left( \frac 1{(k+p)^2}-\frac 1{(k^{\prime
}-p)^2}\right) p^\gamma p^{\gamma ^{\prime }},\,\,k^{\prime }=k+q\,. 
\end{equation}
This tensor has the simple gauge properties on the mass-shell $k^2=k^{\prime
\,2}=0$:

\begin{equation}
A^{\gamma \gamma ^{\prime }}(k,k^{\prime },p)\,k^\gamma =0\,,\,\,A^{\gamma
\gamma ^{\prime }}(k,k^{\prime },p)\,k^{\prime \,\gamma ^{\prime }}=0 
\end{equation}
and therefore the Faddeev-Popov ghosts are absent
in the $t$-channel imaginary part of the one-loop amplitude. 
Further, the product of two
non-asymptotic terms $A^{(n)}$ does not give any contribution of the order of 
$s$ to the elastic scattering amplitude. The integral contribution from the
product of the terms $A^{(a)}$ is given below in the $D$-dimensional
space-time [41]:

$$
A^{(aa)}{}_{AA^{\prime }BB^{\prime }}^{\alpha \alpha ^{\prime }\beta \beta
^{\prime }}=\frac{g^4}t\,g^4\,\delta _{\perp \perp }^{\alpha \alpha ^{\prime
}}\,\delta _{\perp \perp }^{\beta \beta ^{\prime }}\,N_c\,T_{A^{\prime
}A}^d\,T_{B^{\prime }B}^c\,\frac{2\,s}{(-t)^{2-D/2}}\,\frac{\Gamma
(2-D/2)\,\left[ \Gamma (D/2-1)\right] ^2}{(4\pi )^{D/2}\,\Gamma (D-2)}\times 
$$
\begin{equation}
\left[ \frac{1/2}{D-1}-\frac 4{D-4}-\frac 92+(D-3)\left( \ln \,\frac{(-s)s}{
t^2}\,+2\psi (3-D/2)-4\psi (D/2-2)+2\psi (1)\right) \right] . 
\end{equation}

In an analogous way the contribution from the interference term between $%
A^{(a)}$ and $A^{(n)}$ equals [41]

$$
A^{(na)}{}_{AA^{\prime }BB^{\prime }}^{\alpha \alpha ^{\prime }\beta \beta
^{\prime }}=\,\frac{g^4}t\,\delta _{\perp \perp }^{\beta \beta ^{\prime
}}\,\,N_c\,T_{A^{\prime }A}^d\,T_{B^{\prime }B}^c\,\frac{2\,s}{(-t)^{2-D/2}}
\,\frac{\Gamma (2-D/2)\,\left[ \Gamma (D/2-1)\right] ^2}{(4\pi
)^{D/2}\,\Gamma (D-2)}\times 
$$
\begin{equation}
\left[ \delta _{\perp \perp }^{\alpha \alpha ^{\prime }}\,\left( \frac
1{D-1}+\frac 2{D-4}\right) +\,\frac{q_{\perp }^\alpha q_{\perp }^{\alpha
^{\prime }}}t\,\frac{D-4}{D-1}\right] 
\end{equation}

Using the following regularized expression for the gluon Regge trajectory

\begin{equation}
\omega (t)=g^2\,N_c\,\frac 2{(4\pi )^{D/2}}\,(-t)^{D/2-2}\,\frac{\Gamma
(2-D/2)\,\left[ \Gamma (D/2-1)\right] ^2}{\Gamma (D-3)}\,, 
\end{equation}
we obtain for the unrenormalized form-factors $\Gamma ^{(1)}(t)$ and $\Gamma
^{(2)}(t)$ of the reggeon coupling with gluons [41]:

$$
\Gamma ^{(1)}(t)=g\,+\,g^3\,\frac{N_c}{(-t)^{2-D/2}}\,\frac{\Gamma
(2-D/2)\,\left[ \Gamma (D/2-1)\right] ^2}{(4\pi )^{D/2}\,\Gamma (D-2)}
\,\times 
$$
\begin{equation}
\left[ (D-3)\left( \psi (3-D/2)-2\psi (D/2-2)+\psi (1)\right) -\frac 74- 
\frac{1/4}{D-1}\right] 
\end{equation}
and 
\begin{equation}
\Gamma ^{(2)}(t)=g^3\,\frac{N_c}{(-t)^{2-D/2}}\,\frac{\Gamma (3-D/2)\,\left[
\Gamma (D/2-1)\right] ^2}{(4\pi )^{D/2}\,\Gamma (D-2)}\,\frac 1{D-1} 
\end{equation}
Therefore in the pure gluodynamics in the one-loop approximation the $s$
-channel helicity of colliding particles is not conserved even in the limit $%
D\rightarrow 4$. The contribution to the reggeon-gluon-gluon vertex from the
virtual quarks was calculated in Ref $\left[57\right] $. For $n_f$ massless
quarks the result is

$$
\Delta \Gamma ^{(1)}(t)=2\,n_f\,g^3\,\left( -t\right) ^{D/2-2}\,\frac{\Gamma
(2-D/2)\,\left[ \Gamma (D/2)\right] ^2}{(4\pi )^{D/2}\,\Gamma (D)}\,, 
$$
\begin{equation}
\Delta \Gamma ^{(2)}(t)=-2\,n_f\,g^3\,\left( -t\right) ^{D/2-2}\,\frac{
\Gamma (3-D/2)\,\left[ \Gamma (D/2-1)\right] ^2}{(4\pi )^{D/2}\,\Gamma (D)}
\,. 
\end{equation}
For $D\rightarrow 4$ we obtain for $\Delta \,\Gamma ^{(2)}$:

$$
\Delta \,\Gamma ^{(2)}(t)\rightarrow \Gamma ^{(2)}(t)\,(-\frac{n_f}{N_c})\,. 
$$
The total\thinspace contribution of quarks and gluons to the vertex $\Gamma
^{(2)}$ responsible for the helicity non-conservation is zero for $n_f=N_c=3$
. This fact is a consequence of the super-symmetry. Indeed, in the case of
the super-symmetric Yang-Mills theory the helicity of the gluon is conserved
because it takes place for the gluino belonging to the same super-multiplet.
On the other hand, this conservation in the one-loop approximation appears
as a result of the cancellation between contributions of the virtual gluon
and gluino. The gluino contribution can be obtained from quark one by its
multiplication by the factor $N_c/n_f$ because the gluino belongs to the
adjoint representation and it is invariant under the charge
conjugation.\thinspace It leads to the vanishing of the helicity
non-conserving transition at $n_f=N_c=3$.

The one-loop correction to the reggeon-quark-quark vertex was calculated in
ref. $\left[57\right] $. In the massless case we have: 
$$
\Gamma _q(t)=g+g^3\,\left( -t\right) ^{D/2-2}\,\frac{\Gamma (2-D/2)\,\left[
\Gamma (D/2-1)\right] ^2}{(4\pi )^{D/2}\,\Gamma (D-2)}\left\{ \frac{n_f}2\, 
\frac{2-D}{D-1}+\frac 1{N_c}\left( \frac 32-\frac D2-\frac 2{D-4}\right)
\right. 
$$
\begin{equation}
\left. +N_c\left[ (D-3)\left( \psi (3-\frac D2)-2\psi (\frac D2-2)+\psi
(1)\right) +\frac{1/4}{D-1}-\frac 2{D-4}-\frac 74\right] \right\} . 
\end{equation}
The infrared and ultraviolet divergencies can be extracted easily 
from expressions (317-380) and 
are in the agreement with the general considerations.

\subsection{Loop corrections to the reggeon-reggeon-particle vertex}

The amplitude $M_{2\rightarrow 3}$ for the production of a gluon in the
central rapidity region can be written in the Regge model as follows [41]

$$
M_{2\rightarrow 3}=sT_{A^{\prime }A}^{c_1}\,\Gamma _{\lambda _{A^{\prime
}}\lambda _A}(t_1)\,\frac 1{t_1}\,T_{c_2c_1}^d\,\frac 1{t_2}\,T_{B^{\prime
}B}^{c_2}\,\Gamma _{\lambda _{B^{\prime }}\lambda
_B}\,(t_2)\,f(s,s_1,s_2,q_1^{\perp },q_2^{\perp })\,, 
$$
$$
f=\frac 14\left[ s_1^{\omega _1-\omega _2}+\left( -s_1\right) ^{\omega
_1-\omega _2}\right] \left[ s^{\omega _2}+\left( -s\right) ^{\omega
_2}\right] \,R(t_1,t_2,\overrightarrow{k}_{\perp }^2) 
$$
\begin{equation}
+\frac 14\left[ s_2^{\omega _2-\omega _1}+\left( -s_2\right) ^{\omega
_2-\omega _1}\right] \left[ s^{\omega _1}+\left( -s\right) ^{\omega
_1}\right] \,L(t_1,t_2,\overrightarrow{k}_{\perp }^2) 
\end{equation}
where $j_{1,2}=1+\omega _{1,2}$ (for $\omega _i=\omega (t_i)$) are the gluon
Regge trajectories with the negative signature at the crossing channels $%
t_1,t_2$. It is important, that this amplitude does not have the simultaneous
singularities in the overlapping channels $s_1=2kp_A$ and $s_2=2kp_B$. 
In LLA one
obtains:

\begin{equation}
R(t_1,t_2,\overrightarrow{k}_{\perp }^2)\,+\,L(t_1,t_2,\overrightarrow{k}
_{\perp }^2)\,\rightarrow \,2\,g\,C_\mu (q_2,q_1)\,e_\mu ^{*}(k)\, 
\end{equation}
where $C_\mu $ is the RRP vertex. One can calculate the
discontinuities\thinspace of $M_{2\rightarrow 3}$ in the $s,\,s_1$ and $s_2$
channels for small $g$ with the use of the unitarity condition in the
one-loop approximation in the $D$-dimensional space:

\begin{equation}
R-L=\frac{C_\mu (q_2,q_1)\,e_\mu ^{*}(k)}{\omega _1-\omega _2}\,\frac{
4\,N_cg^3}{(4\pi )^{D/2}}\,\Gamma (3-\frac D2)\,\left[ \ln \frac 1{
\overrightarrow{k}_{\perp }^2}\,-\frac 2{D-4}\right] . 
\end{equation}
Note, that this quantity is real in all physical channels.\thinspace By
adding the analogous contribution of the order of $g^3$ to the sum $R+L$ in
such way to provide correct analytic properties of the production amplitude
in $s,s_1$ and $s_2$ channels we obtain\thinspace for the one-loop
correction to $M_{2\rightarrow 3}$ [41]:

$$
M_{2\rightarrow 3}^{loop}=M_{2\rightarrow 3}^{Born}\,N_c\,\frac{g^2}{(4\pi
)^{D/2}}\,\Gamma (3-\frac D2)\,\frac 12\,\left[ \frac 1{4-D}\,\ln \,\frac{
(-s_1)^2s_1^2(-s_2)^2s_2^2}{(-s)^3s^3\mu ^4}\right. 
$$
\begin{equation}
\left. -\frac 14\left( \ln ^2 \frac{(-s_1)(-s_2)}{-s}+\ln ^2\frac{s_1s_2}{-s}
+\ln ^2\frac{s_1(-s_2)}s+ \ln ^2\frac{(-s_1)s_2}s \right) \right] \,+
\,\Delta M_{2\rightarrow 3}\,, 
\end{equation}
where $\Delta \,$$M_{2\rightarrow 3}$ is an polynomial in $\overrightarrow{
k_{\perp }}^2=s_1s_2/s$. One can reproduce this analytic structure of $%
M_{2\rightarrow 3}$ using the approach based on the unitarity conditions in $%
t_1$ and $t_2$ channels and obtain the following result [41]:

$$
R+L=2g\,C_\mu (q_2,q_1)\,e_\mu ^{*}(k)\,f_1+\left( \frac{p_B}{s_2}-\frac{p_A 
}{s_1}\right) _\mu e_\mu ^{*}(k)\,4g\,f_2\,, 
$$
$$
f_1=\left( 1+c_1\,\frac{N_c\,g^2}{(4\pi )^{D/2}}\,\Gamma (3-\frac D2)\right)
,\,f_2=c_2\,\frac{N_c\,g^2}{(4\pi )^{D/2}}\,\Gamma (3-\frac D2)\,; 
$$
$$
c_1=\frac 12\,\ln {}^2\overrightarrow{k}_{\perp }^2\,-\frac 4{(D-4)^2}+\frac
1{D-4}\left( \frac{11}3+2\,\ln \,(t_1t_2)\right) +\frac 12\,\ln {}^2(t_1t_2) 
$$
$$
-\frac{\overrightarrow{k}_{\perp }^2}3\,\frac{t_1+t_2}{(t_1-t_2)^2}+\frac
16\,\ln \,\frac{t_1}{t_2}\,\left( 11\,\frac{t_1+t_2}{t_1-t_2}\,+\,4 
\overrightarrow{k}_{\perp }^2\frac{t_1t_2}{(t_1-t_2)^3}\right) +\frac{\pi ^2}
2\,; 
$$
\begin{equation}
c_2=\frac 13\ln \frac{t_1}{t_2}\left( 11-\overrightarrow{k}_{\perp }^2\, 
\frac{2\overrightarrow{k}_{\perp }^2+t_1+t_2}{(t_1-t_2)^2}\right) \frac{
t_1t_2}{t_1-t_2}\,-\frac{\overrightarrow{k}_{\perp }^2}6\left( 1+\frac{
t_1+t_2}{(t_1-t_2)^2}\,(2\overrightarrow{k}_{\perp }^2+t_1+t_2)\right) . 
\end{equation}
The contribution to the RRP vertex from the massless quark loop was also
calculated $\left[ 58\right] $:

$$
\frac{N_c}{n_f}\,\Delta \,c_1=\frac{2/3}{D-4}-\frac{\overrightarrow{k}
_{\perp }^2}3\,\frac{t_1+t_2}{(t_1-t_2)^2}+\frac 13\,\ln \,\frac{t_1}{t_2}
\,\left( \frac{t_1+t_2}{t_1-t_2}\,+\,2\overrightarrow{k}_{\perp }^2\frac{
t_1t_2}{(t_1-t_2)^3}\right) ,\,\,\,\frac{N_c}{n_f}\,\Delta c_2= 
$$
\begin{equation}
=\frac 13\ln \frac{t_1}{t_2}\left( -2- \overrightarrow{k}_{\perp }^2\frac{2 
\overrightarrow{k}_{\perp }^2+t_1+t_2}{ (t_1-t_2)^2}\right) \frac{t_1t_2}{%
t_1-t_2}+\frac{\overrightarrow{k}_{\perp }^2}6\left( 1+\frac{t_1+t_2}{%
(t_1-t_2)^2}(2\overrightarrow{k}_{\perp }^2+t_1+t_2)\right) 
\end{equation}
and therefore for the super-symmetric Yang-Mills (where effectively 
$n_f=N_c$)
the total gluon and quark contribution to $c_2$ is significantly
simplified.

\subsection{Two-loop corrections to the gluon Regge trajectory}

For finding the next-to-leading corrections to the gluon Regge trajectory
one can use the unitarity relations in the $t$ or $s$ channels. Initially
the $t$-channel unitarity approach was advocated [49]. 
In the framework of
this approach the universality of the Regge trajectory obtained from the
different high energy processes is obvious. To find the next-to-leading
correction to the gluon trajectory one should calculate the imaginary part
of the elastic scattering amplitude in the two-loop approximation taking
into account the terms of the order of $\ln {}^2s$ and $\ln s$. The
contribution from the two-particle intermediate state in the $t$-channel can
be expressed in terms of the product of two elastic amplitudes. One of these
amplitudes is taken in the Born approximation and another one corresponds to
the helicity conserving part of the above calculated one-loop elastic
amplitude in the Regge kinematics. As for the contribution from the
three-particle intermediate state in the $t$-channel, one should take the
product of
two amplitudes $A_{2\rightarrow 3}$ for the quasi-multi-Regge kinematics for
each of two inelastic amplitudes $A_{2\rightarrow 3}$. Because in the
leading logarithmic approximation the Regge expression for the scattering
amplitude has only two-particle intermediate states, it is enough to
calculate only the asymptotic behaviour of this three-particle 
contribution.

One can use also the $s$-channel unitarity to find the asymptotic behaviour
of the scattering amplitude in the two-loop approximation [59]. In this case
the imaginary parts of the amplitude in the $s$ and $u$ channels should be
calculated with taking into account the terms of the order of $s\,\ln
\,s $ and $s$. Subtracting from the result the known one-loop corrections to
the reggeon-particle-particle vertices, one can calculate the two-loop
corrections to the Regge trajectory. For the case of the gluodynamics it is
given below [59]:

$$
\omega ^{(2)}(t)=\frac{N_c^2\,g^4\,t}{4(2\pi )^{2D-2}}\int \frac{
d^{D-2}q_1\, }{\overrightarrow{q_1}^2}\int \frac{d^{D-2}q_2\,}{
\overrightarrow{q_2}^2}\,f 
$$
$$
f=\frac{\overrightarrow{q}^2}{\left( \overrightarrow{q}-\overrightarrow{q_2}
\right) ^2\left( \overrightarrow{q}-\overrightarrow{q_1}\right) ^2}\,\ln \, 
\frac{\overrightarrow{q}^2}{\left( \overrightarrow{q_2}-\overrightarrow{q_1}
\right) ^2}\,+\frac 2{\left( \overrightarrow{q}-\overrightarrow{q_1}- 
\overrightarrow{q_2}\right) ^2}\,\ln \,\frac{\overrightarrow{q_1}^2}{\left( 
\overrightarrow{q}-\overrightarrow{q_1}\right) ^2} 
$$
$$
+\left( \frac 2{\left( \overrightarrow{q}-\overrightarrow{q_1}- 
\overrightarrow{q_2}\right) ^2}\,-\frac{\overrightarrow{q}^2}{\left( 
\overrightarrow{q}-\overrightarrow{q_2}\right) ^2\left( \overrightarrow{q}- 
\overrightarrow{q_1}\right) ^2}\right) \times 
$$
\begin{equation}
\left( 2\psi (D-3)+\psi (3-\frac D2)-2\psi (\frac D2-2)-\psi (1)+\frac
1{D-3}\,\left( \frac 1{4(D-1)}-\frac 2{D-4}-\frac 14\right) \right) . 
\end{equation}
Note, that the logarithms in $f$ appear as a result of integration over
longitudinal momenta.
 
For the two-loop including the quark loop we obtain the following result
[59]: 
$$
\omega _q^{(2)}(-\overrightarrow{q}^2)=\frac{N_cn_f\pi ^{\frac D2-1}g^4}{
(2\pi )^{2D-2}}\,\Gamma (2-\frac D2)\frac{\Gamma ^2(\frac D2)}{\Gamma (D)}
\int \frac{d^{D-2}q_1\,\overrightarrow{q}^2}{\overrightarrow{q_1}^2\left( 
\overrightarrow{q}-\overrightarrow{q_1}\right) ^2}\left( 2(\frac{
\overrightarrow{q_1}}\mu )^{D-4}-(\frac{\overrightarrow{q}}\mu
)^{D-4}\right) . 
$$

\section{Conclusion}

As it was shown above, the theory of the high-energy scattering in QCD is
based on the BFKL equation summing the leading logarithmic terms in the
perturbation theory. Next-to-leading terms were 
calculated using the $k_\perp$%
-factorization only for the anomalous dimensions describing the transition
of quarks to gluons [60]. Above we expressed the next-to-leading real
contribution to the BFKL kernel in terms of squares of the amplitudes for
the production of gluons and quarks with definite helicities. All infrared
divergencies were extracted from these expressions in an explicit form and
are regularized in the $D$-dimensional space. These divergencies 
cancel with the analogous divergencies from the virtual corrections to the
BFKL equation which were also written above. It is known [55], that the
total next-to-leading corrections to the integral kernel can be expressed in
terms of the dilogarithm integrals and will be available soon. The effective
field theory for the quasi-multi-Regge processes reviewed above can be
used in particular 
for the unitarization program. Because
the effective action is expressed in terms of the Wilson contour integrals, 
one can
attempt to include into consideration also some non-perturbative effects. The
effective field theory can be generalized to the case of the high energy
behaviour of the amplitudes with non-singlet $t$-channel exchanges 
including one with the baryon quantum numbers
and to the small-$x$ asymptotics of the structure functions for
the polarized $ep$ deep-inelastic
scattering [61].

\vspace{1cm} \noindent

{\large {\bf Acknowledgements}}\\

I want to thank Universit\"at Hamburg, DESY-Hamburg and DESY-Zeuthen for the
hospitality during my stay in Germany. I am indebted to the
Alexander-von-Humboldt Foundation for the award, which gave me a possibility
to work on small-$x$ physics problems 
during the last year. The fruitful discussions
with Ya. Ya. Balitsky, J. Bartels, J. Bl\"umlein, V. S. Fadin and E. A.
Kuraev were helpful.

\vspace{1cm} \noindent

{\large {\bf Appendix}}

\vspace{0.5cm} \noindent

There are the following helpful relations for hypergeometric functions: 
$$
F(h,h,2h;x)=\frac{\Gamma (2h)}{\left[ \Gamma (h)\right] ^2}\left( -\ln
(1-x)+2\psi (1)-2\psi (d)-\partial _a-\partial _b-2\partial _c\right)
\,F(h,h,1;1-x)\,, 
$$
$$
x^h\left( \partial _a+\partial _b+2\partial _c\right)
F(h,h,1;1-x)=x^{1-h}\,\left( \partial _a+\partial _b+2\partial _c\right)
F(1-h,1-h,1;1-x)\,, 
$$
$$
x^hF(h,h,1;1-x)=x^{1-h}\,F(1-h,1-h,1;1-x)\,, 
$$
where the derivatives $\partial _a$, $\partial _b$ and $\partial _c$ act on
the corresponding arguments of the function $F$ (122) with the subsequent
substitution of $a,b$ and $c$ by their given values. Using these relations
one can continue analytically $G_{\nu n}(\overrightarrow{\rho _1}, 
\overrightarrow{\rho _2};\overrightarrow{\rho _{1^{^{\prime }}}}, 
\overrightarrow{\rho _{2^{^{\prime }}}})$ to the region near $x=1$: 
$$
c_1^{-1}\,\frac{\left[ \Gamma (h)\Gamma (\widetilde{h})\right] ^2}{\Gamma
(2h)\Gamma (2\widetilde{h})}\,x^{-h}x^{*-\widetilde{h}}G_{\nu n}( 
\overrightarrow{\rho _1},\overrightarrow{\rho _2};\overrightarrow{\rho
_{1^{^{\prime }}}},\overrightarrow{\rho _{2^{^{\prime }}}}) 
$$
$$
=2\pi \cot (\pi h)\,(\ln \left| 1-x\right| ^2-4\psi (1)+\partial _a+\partial
_b+2\partial _c)F(h,h,1;1-x)\,F(\widetilde{h},\widetilde{h},1,1-x) 
$$
$$
+4\left( \psi (h)\psi (\widetilde{h})-\psi (1-h)\psi (1-\widetilde{h}
)\right) F(h,h,1;x)F(\widetilde{h},\widetilde{h},1;x^{*})\,. 
$$

Analogously according to the relations 
$$
F(h,h,2h;x)=\frac{\Gamma (2h)}{\left[ \Gamma (h)\right] ^2}\,(-x)^h\left(
\ln (-x)+2\psi (1)-2\psi (d)-\partial _a-\partial _b-2\partial _c\right)
\,F(h,1-h,1;1)\,, 
$$
$$
\left( \partial _a+\partial _b+2\partial _c\right) F(h,1-h,1;1/x)=\left(
\partial _a+\partial _b+2\partial _c\right) F(1-h,h,1;1/x)\,, 
$$
$$
F(h,1-h,1;1/x)=\,F(1-h,h,1;1/x)\, 
$$
one can continue $G_{\nu n}(\overrightarrow{\rho _1},\overrightarrow{\rho _2}
;\overrightarrow{\rho _{1^{^{\prime }}}},\overrightarrow{\rho _{2^{^{\prime
}}}})$ at large $x$: 
$$
c_1^{-1}\,\frac{\left[ \Gamma (h)\Gamma (\widetilde{h})\right] ^2}{\Gamma
(2h)\Gamma (2\widetilde{h})}\,G_{\nu n}(\overrightarrow{\rho _1}, 
\overrightarrow{\rho _2};\overrightarrow{\rho _{1^{^{\prime }}}}, 
\overrightarrow{\rho _{2^{^{\prime }}}}) 
$$
$$
=-2\pi \cot (\pi h)\left( \ln \left| x\right| ^2+4\psi (1)-\partial
_a-\partial _b-2\partial _c\right) F(h,1-h,1;1/x)\,F(\widetilde{h},1- 
\widetilde{h},1;1/x)\,. 
$$
$$
+4\left( \psi (h)\psi (\widetilde{h})-\psi (1-h)\psi (1-\widetilde{h}
)\right) \,F(h,1-h,1;1/x)\,F(\widetilde{h},1-\widetilde{h},1;1/x)\,. 
$$

\end{document}